\documentclass{aa}  

\usepackage{natbib}
\bibpunct{(}{)}{;}{a}{}{,}
\usepackage{graphicx}
\usepackage{txfonts}
\usepackage{lipsum}
\usepackage{subcaption}        
\usepackage{multirow}           
\usepackage{lscape}            
\usepackage{placeins}          
\usepackage[unicode, colorlinks=True, linkcolor=black, citecolor=blue, urlcolor=blue]{hyperref}

\usepackage{color}

\begin{document}

   \title{Analysis of Io's tidal response as a function of the properties of the partially molten layer}

    \titlerunning{Tidal response and melt in Io's interior}
   \authorrunning{M. Paris et al.} 

   \author{M. Paris\inst{1}\fnmsep\inst{2}\fnmsep\thanks{Corresponding author: matteo.paris@inaf.it}
        \and A. Mura\inst{1}
        \and F. Zambon\inst{1}
        \and A. Genova\inst{3}
        \and F. Tosi\inst{1}
        \and A. Consorzi\inst{4}
        \and G. Mitri\inst{4}
        \and A. Cicchetti\inst{1}
        \and S. Bolton\inst{5}
        \and R. Noschese\inst{1}
        \and G. Piccioni\inst{1}
        \and C. Plainaki\inst{1}
        \and G. Sindoni\inst{6}
        \and R. Sordini\inst{1}
        }

   \institute{Istituto Nazionale di Astrofisica – Istituto di Astrofisica e Planetologia Spaziali (INAF-IAPS), Rome, Italy
   \and Dipartimento di Fisica, Sapienza Università di Roma, Roma, Italy
   \and Dipartimento di Meccanica e Ingegneria Aerospaziale, Sapienza Università di Roma, Roma, Italy
   \and Dipartimento di Ingegneria e Geologia, Università d’Annunzio, Pescara, Italy
   \and Southwest Research Institute, San Antonio, TX, USA
   \and Agenzia Spaziale Italiana, Roma, Italy}


  \abstract
    {Io’s internal heat, generated by Jupiter-driven tidal dissipation and the Laplace resonance, partially melts the mantle. However, the resulting melt fraction, depth, and spatial distribution of dissipation remain poorly constrained.}
    {We use Io’s tidal response to constrain its internal structure, focusing on partial melt distribution and dissipation mechanisms. We link the observed tidal deformation to the mantle physical state through a parametric approach accounting for melting onset depth and latent heat of fusion.}
    {We modeled Io as a three-layer body with a fluid core, a viscoelastic mantle, and an elastic lithosphere. The degree-2 potential Love number $k_2$ was computed by solving spheroidal oscillation equations with an adapted California Planetary Geophysics Code (\texttt{CPGC}). We iteratively updated the mantle properties, viscosity, shear modulus, and the Andrade parameter ($\beta$) based on the local melt fraction ($\phi(r)$). Moving beyond state-of-the-art models, we explicitly incorporated mantle compressibility into our framework.}
    {Within the adopted rheological framework, reproducing the observed $\Re(k_2)$ requires melt fractions below the rheologically critical melt fraction. Our analysis indicates that a limited mantle enhancement in tidal dissipation emerges naturally, although the deep mantle is the primary region of tidal heating. The presence of melt reduces the effective viscosity and increases anelasticity, enhancing tidal dissipation in the upper mantle. Incompressible models provide conservative upper bounds on the melt fraction, whereas compressible models yield slightly higher values of $\Re(k_2)$, strengthening this interpretation. An a posteriori mass-flux analysis showed that melt percolation capacity exceeds the thermodynamic production rate, indicating efficient drainage. The reference Andrade parameter $\beta_0$ strongly affects the imaginary components of the Love numbers ($k_2$, $h_2$, $l_2$) and the predicted libration amplitude.}
    {These combined constraints support a heterogeneous, partially molten mantle characterized by a "magmatic sponge" structure rather than a global magma ocean. The proposed framework links Io’s internal structure and tidal dissipation to recent observations from the Juno mission.}

 \keywords{Planets and satellites: individual: Io --
          Planets and satellites: interiors --
          Planets and satellites: physical evolution --
          Planets and satellites: composition --
          Methods: numerical}

   \maketitle
   \nolinenumbers
   \let\linenumbers\nolinenumbers

\section{Introduction}
Jupiter’s moon Io, the most volcanically active body in the Solar System~\citep{1994JGR....9917095V}, exhibits intense internal heating driven by tidal friction~\citep{1979Sci...203..892P}. While tidal dissipation is widely accepted as the main mechanism generating this extraordinary volcanic activity, the spatial distribution and depth of heat generation remain unresolved~\citep{2021BAAS...53d.178K}. The tidal response of Io is described by the Love number $k_2$~\citep{https://doi.org/10.1029/2007JE002908}, whose real part describes the in-phase deformation induced by Jupiter’s gravitational potential, whereas the imaginary part represents the out-of-phase anelastic response associated with internal energy dissipation.
The study of the tidal response of a planetary body is a crucial tool, as the deformation under external gravitational forces encodes valuable information about its internal structure, rheology, and temperature distribution and the state of partial melting. Among all bodies in the Solar System, Io represents a unique natural laboratory for studying tidal dissipation. Its extreme internal heating and vigorous volcanism are directly sustained by tidal interactions with Jupiter and the other Galilean satellites, making it an ideal target to test geophysical models against observations.

Despite decades of observations and modeling, the internal state of Io’s mantle remains highly uncertain~\citep{2001JGR...10632963A, 2004jpsm.book..281S, 2007iag..book...89M, 2010Icar..209..651B}. 
A primary debate regarding this subject concerns whether tidal dissipation occurs within a solid viscoelastic mantle containing a partially molten layer~\citep{1988Icar...75..187S, Bierson, 2020Icar..33513299S, 2021A&A...650A..72K} or is instead distributed within a global magma ocean~\citep{2011Sci...332.1186K, 2015ApJS..218...22T,2024GeoRL..5107869A}. The presence, thickness, and composition of such a partially molten zone have deep implications for Io’s thermal balance, magnetic induction response, and long-term orbital evolution.

To rigorously address this debate, the rheological description of the mantle material is critical. While simple Maxwell models have traditionally been used to approximate planetary interiors~\citep{ROSS1985391, 2003JGRE..108.5096M, 2015ApJS..218...22T}, they fail to account for the transient anelastic response that dominates dissipation at tidal frequencies in high-temperature silicates~\citep{https://doi.org/10.1029/2010JE003664, 2012ApJ...746..150E}. Consequently, we employed the Andrade rheological model, an empirical formulation that successfully reproduces the transient creep behavior of mantle materials observed in laboratory experiments~\citep{2004JGRB..109.6201J, JACKSON2010151, Bierson, park_ios_2025}.
Since its application to tidal heating studies, several versions of the Andrade rheology have been proposed~\citep{bierson2024impact}. Among these, we chose the variant developed by~\cite{JACKSON2010151}, which involves, together with the classical $\alpha$ and $\beta$ parameters, a modified tidal period that accounts for a temperature dependence in the material's response (more details are provided in Appendix \ref{andrade}). In this version, the complex compliance $J(\omega)$ is defined as\begin{equation}
    J(\omega) = \frac{1}{\mu_U} - i\frac{1}{\omega_P\eta} + \beta\cdot (i\omega_P)^{-\alpha}\Gamma(\alpha+1)\,,
    \label{eq:andrade_compliance}
\end{equation}
where $\mu_U$ is the unrelaxed shear modulus, $\eta$ is the viscosity, $\alpha$ is a dimensionless exponent that sets the power-law scaling of the Andrade anelastic term, and $\beta$ is a generalized compliance parameter scaling the intensity of the transient response, which depends on material composition, grain-size, and temperature. Finally, $\Gamma$ is the Euler gamma function. The pseudo-frequency $\omega_P$ reads as\begin{equation}
     \omega_P=2\pi \left\{P\exp\biggl[\frac{E_b}{R_g}\biggl(\frac{1}{T}-\frac{1}{T_r}\biggr)\biggr]\right\}^{-1}\,,
\end{equation}
with $P$ indicating the tidal period, $E_b$ the activation energy, and $R_g$ the gas constant, while $T_r$ is a reference temperature. This model has already been employed in similar works (e.g., \citet{Bierson} and~\citet{park_ios_2025}).
The selection of the material parameters is guided by experimental constraints on ultramafic mineralogies likely characterizing Io's mantle. Laboratory studies on olivine aggregates and dunite at high temperatures suggest that $\alpha$ typically falls in the range of $0.2$ -- $0.4$, reflecting the mechanics of grain-boundary sliding~\citep{2002JGRB..107.2360J}. Accordingly, we adopted a fixed characteristic value of $\alpha \approx 0.3$~\citep{park_ios_2025} and explored $\beta$ values within the range of $10^{-13}$ -- $10^{-12}\,\mathrm{Pa^{-1}s^{-0.3}}$~\citep{2002JGRB..107.2360J, 2004JGRB..109.6201J,park_ios_2025}.
This approach allowed us to explore the rheological parameter space effectively, linking the macroscopic tidal response observed by spacecraft to the microphysical properties of the mantle rock.

Recent spacecraft observations, including those from the Juno mission~\citep{park_ios_2025}, have provided improved constraints on Io’s global tidal response, quantified by the degree-2 Love number $k_2$. The measured value of $k_2$ encodes crucial information about the rheology and internal layering of the satellite. However, a robust interpretation requires frameworks that self-consistently couple solid-state rheology with the effects of partial melting~\citep{1988Icar...75..187S, moore_thermal_2001, 2003JGRE..108.5096M, 2007Icar..192..491K, Bierson, 2020Icar..33513299S, 2021A&A...650A..72K, 2022Icar..37314737K, Miyazaki_2022, 2024GeoRL..5107869A, 2025NatCo..16.6798V}.
Current estimates of $k_2$ remain compatible with a wide range of interior models~\citep{2024GeoRL..5107869A}, leaving open the key question of which melt fraction distribution, $\phi(r)$, is required to reconcile these models with Juno observations.
Similarly, attempts to constrain Io’s interior structure using the spatial distribution of volcanic hot spots observed at the surface remain inconclusive. Recent analyses of narrow-band near-infrared observations, including M-band ($4.5$ -- $5.0$ $\mu$m) measurements, show that the observed hot-spot distribution cannot uniquely distinguish between different internal heating or structural models of Io~\citep{2025FrASS..1268185T}.

In this study, we address these open questions by modeling Io’s tidal response using a layered viscoelastic structure, described in detail in Sect. \ref{melt}. The sensitivity of the tidal Love number $k_2$ to the physical parameters of a partially molten mantle is systematically investigated in Sect. \ref{res}. Specifically, we explore variations in the melt fraction $\phi(r)$, its onset depth, and the resulting changes in viscosity and shear modulus~\citep{Bierson}. This approach allowed us to identify the parameter combinations that reproduce the $k_2$ value inferred from Juno observations. Finally, in Sect. \ref{disc}, we provide a quantitative estimate for the melt fraction required to satisfy observational constraints and discuss the implications for Io's internal dynamics and thermal state.

\section{Methods}\label{melt}
\subsection{Overview and parametric study}
We modeled Io’s tidal response by computing the degree-2 Love number $k_2$ as a function of two mantle parameters: the radial position at which melting begins ($R_{\phi0}$) and the latent heat of fusion ($L$) of the constituent materials. These parameters control the extent and efficiency of tidal dissipation within Io's mantle, which, in turn, determines the satellite's global deformation and energy release.
The onset radius $R_{\phi0}$ defines the depth at which partial melting occurs, influencing whether tidal energy is primarily dissipated in the upper or lower mantle. Shallow melting (higher $R_{\phi0}$) enhances dissipation near the surface, whereas deeper melting confines it to the lower mantle. The latent heat of fusion $L$ governs the degree to which the mantle material melts under tidal stresses, affecting both the local melt fraction and the effective rheology. Since the exact composition of Io's mantle remains uncertain (though likely dominated by olivine-rich silicates;~\citealt{Hamilton}), varying $L$ allows for the exploration of different potential mineralogical assemblages.
We explored the sensitivity of $k_2$ to the onset and intensity of partial melting by systematically varying $R_{\phi0}$ and $L$ within physically plausible ranges~\citep{LESHER2015113}.
This parametric approach allowed us to estimate the melt fraction within the partially molten layer and identify spherically symmetric 1D interior configurations that reproduce the $k_2$ value observed by Juno~\citep{park_ios_2025}.
Unlike previous studies, which often assumed a fixed latent heat or predefined dissipation profiles, we systematically explored a range of $L$ values. Rather than imposing the depth of heating a priori, our coupled rheological model allows the localized enhancement in shallow mantle dissipation to emerge naturally from the melt structure.

Specifically, with the elastic lithosphere thickness fixed at $50\,\mathrm{km}$~\citep{park_ios_2025} (assuming coincidence of the compositional crust and mechanical lithosphere), we varied the melting onset radius ($R_{\phi0}$) between $1340\,\mathrm{km}$ and $1680\,\mathrm{km}$, and $L$ from $2\times10^5$ to $8\times10^5\,\mathrm{J\,kg^{-1}}$. Furthermore, we evaluated the impact of mantle compressibility~\citep{2021A&A...650A..72K, Tobie} on Io's tidal response, which allowed us to further refine the required upper limits for the inferred melt fraction. This framework provides a comprehensive view of how interior configurations dictate Io's tidal response, laying the foundation for the numerical modeling described in the following sections.

Moreover, this versatile methodology is well suited to interpret data from ongoing and future missions, such as Europa Clipper~\citep{2020NatCo..11.1311H} and JUICE~\citep{2013P&SS...78....1G}, and can be extended to other icy satellites, including Europa and Ganymede. The model is also capable of deriving other crucial tidal parameters, including the Love numbers $h_2$ and $l_2$~\citep{Bierson}, and the libration amplitude~\citep{Comstock2003, 2013Icar..223..621N, 2019JGRE..124.2198R, 2019Icar..321..126R, https://doi.org/10.1029/2020JE006473}. These additional observables can further constrain the internal structure and dynamics, improving the understanding of tidally active worlds.

\subsection{Numerical approach}

We derived the Love number $k_2$ by solving the spheroidal oscillation equations~\citep{Takeuchi} for an initial three-layer structure: a fluid core, a viscoelastic mantle, and an elastic lithosphere. The calculations used an adapted version of the California Planetary Geophysics Code (\texttt{CPGC}; \citealt{ermakov_california_2024}) specifically designed to iteratively link tidal dissipation with mantle rheology.
We computed tidal dissipation from the radial and tangential displacement and stress fields~\citep{beuthe_spatial_2013}, yielding the energy converted into heat within each layer. This computed dissipation, in turn, altered the local rheological properties (viscosity, shear modulus, and the Andrade parameter $\beta$).
This established the critical feedback loop between tidal forcing and mantle rheology that we solved with the iterative approach.
With this study, we extend previous work (e.g.,~\citealt{Bierson, park_ios_2025}) by doing the following:
\begin{itemize}
    \item Systematically varying the latent heat of fusion $L$ from $2\times10^5$ to $8\times10^5\,\mathrm{J\,kg^{-1}}$ to explore a range of thermal states and possible melt fractions~\citep{LESHER2015113}.
    \item Exploring the rheological parameter space without imposing the depth of dissipation a priori. Although the system shows a deep-mantle dissipation (e.g.,~\citealt{1988Icar...75..187S, 2018AJ....156..207R}), coupling the Andrade parameter $\beta$ directly to the local melt fraction allows our framework to naturally capture an additional localized enhancement of dissipation in the upper mantle (e.g.,~\citealt{1990Icar...85..309R, 2013E&PSL.361..272H, 2024NatAs...8...94D, 2024ComEE...5..340M, 2025PSJ.....6...84P}).
    \item Evaluating the impact of mantle compressibility~\citep{2021A&A...650A..72K,Tobie} on Io's tidal response, which allowed us to further refine the required upper limits for the inferred melt fraction.
    \item Performing a targeted sensitivity analysis on the reference Andrade parameter ($\beta_0$) to reconcile the modeled imaginary part of $k_2$ with the precise $1$$\sigma$ Juno confidence ~\citep{park_ios_2025}.
    \item Conducting a comprehensive mass-flux balance analysis that compares thermodynamic melt production~\citep{1979Sci...203..892P,https://doi.org/10.1029/GL008i004p00313} with melt percolation capacity~\citep{10.1093/petrology/25.3.713}, physically validating the long-term stability of the inferred "magmatic sponge" structure~\citep{Miyazaki_2022}.
\end{itemize}

This expanded methodology provides a robust, physically grounded framework linking Io’s internal structure, partial melting, and tidal response. It allowed us to quantitatively assess how dynamic assumptions about mantle rheology and magmatic transport influence the observed $k_2$ and the overall efficiency of tidal energy dissipation.

\subsection{Internal structure and rheology}\label{internal_structure}
\begin{figure}[!ht]
\resizebox{\hsize}{!}{
\includegraphics{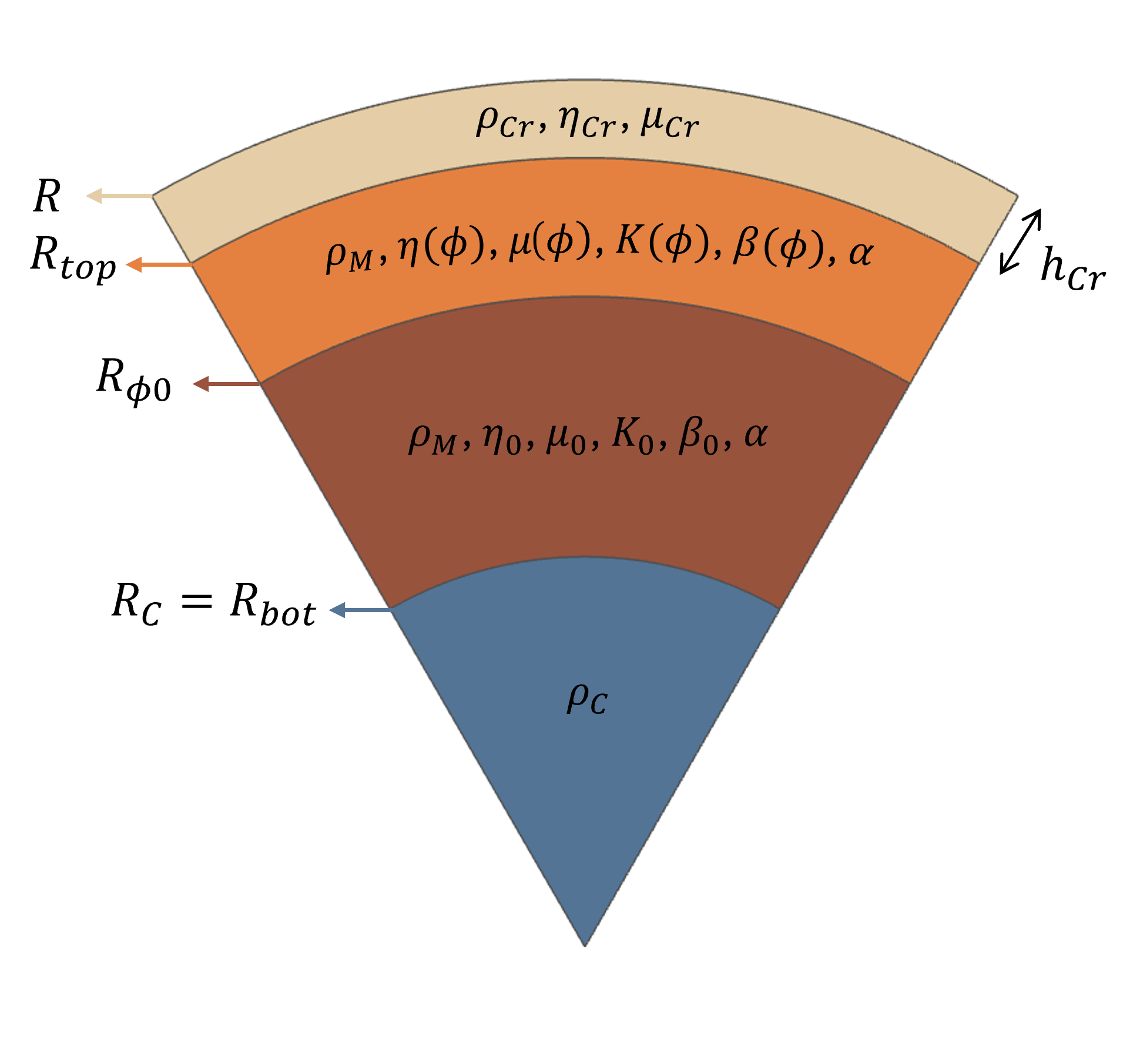}}
    \caption{Schematic representation of the Io interior model. Radial boundaries indicate the core radius ($R_C = R_{\text{bot}}$), the mantle-lithosphere boundary ($R_{\text{top}}$), and the total radius ($R$), which are held constant as fixed structural constraints, while the onset of partial melting ($R_{\phi0}$) acts as a free parameter. The thickness of the lithosphere is denoted by $h_{\text{Cr}}$. Physical and rheological parameters ($\rho$, $\eta$, $\mu$, $K$, $\beta$, $\alpha$) are indicated for each structural domain, depending explicitly on the local melt fraction ($\phi$) within the partially molten layer.}
    \label{fig_models_Io}
\end{figure}

At the beginning of the iterative process, we modeled Io as a three-layer body:
\begin{itemize}
    \item Fluid core: Io likely hosts a fully liquid core due to its high mantle temperature~\citep{1997tiph.conf..345S,2004jpsm.book..281S, 2011Sci...332.1186K}.
    \item Mantle: It follows the Andrade rheology ($\alpha = 0.3$) with a reference Andrade parameter $\beta_0$ between $10^{-13}$ and $10^{-12}\,\mathrm{Pa^{-1}s^{-0.3}}$~\citep{Bierson, park_ios_2025}.
    \item Lithosphere: We modeled it as an elastic layer that modulates surface tidal bulges~\citep{1998Sci...279.1514S, park_ios_2025}, assuming that the compositional crust and the mechanical lithosphere coincide.
\end{itemize}
The structural configuration of the interior model is schematically illustrated in Fig.~\ref{fig_models_Io}. Throughout all numerical simulations, we held the core radius ($R_C$), the mantle-lithosphere boundary ($R_{\text{top}}$), and the total planetary radius ($R$) constant as fixed structural constraints. Consequently, the lithospheric thickness ($h_{\text{Cr}}$) remained constant. Conversely, we treated the radial position marking the onset of partial melting within the mantle ($R_{\phi0}$) as a free parameter and varied it systematically across the different model configurations to explore the resulting tidal response.

The baseline rheological parameters used for this parametric analysis are reported in Table \ref{tab1}.
\begin{table*}[ht]
\caption{Baseline rheological parameters for the three-layer model.}
\centering
\begin{tabular}{l c c c c c}
\hline\hline
Layer & Rheology & Thickness & Density & Viscosity $\eta_0$ & Shear modulus $\mu_0$ \\
 & & ($\mathrm{km}$) & ($\mathrm{kg\;m^{-3}}$) & ($\mathrm{Pa\;s}$) & ($\mathrm{GPa}$) \\
\hline
Core & Fluid & 950 & 5150 & -- & -- \\
Mantle & Andrade & 820 & 3259 & $10^{21}$ & 40 \\
Lithosphere & Elastic & 50 & 3259 & $10^{25}$ & 40 \\
\hline
\end{tabular}
\label{tab1}
\end{table*}
The values of the rheological parameters agree with~\citet{park_ios_2025}, while the initial values of $\beta_0$ were chosen to remain consistent with the literature~\citep{Bierson, park_ios_2025}.
Regarding the reference mantle viscosity, our sensitivity analysis confirmed the findings of~\citet{park_ios_2025}: the tidal response remains robust across a broad range of viscosities, showing significant deviations only for values $\leq 10^{15}\,\mathrm{Pa\,s}$.
We adopted a mantle shear modulus of $40\,\mathrm{GPa}$, corresponding to the upper limit expected for partially molten olivine aggregates~\citep{park_ios_2025}.
To evaluate the robustness of this baseline shear modulus, we performed a preliminary sensitivity analysis by varying $\mu_0$ between $20\,\mathrm{GPa}$ and $60\,\mathrm{GPa}$ within our parametric framework. Decreasing $\mu_0$ to $20\,\mathrm{GPa}$ increased the overall elastic compliance of the mantle, which systematically drove the real part of the Love number, $\Re(k_2)$, upward beyond the $3$$\sigma$ uncertainty threshold of the Juno spacecraft constraints, while lowering its imaginary counterpart, $|\Im(k_2)|$. Conversely, increasing $\mu_0$ to $60\,\mathrm{GPa}$ stiffened the interior profile, resulting in a reduction of $\Re(k_2)$ but causing a significant increase in $|\Im(k_2)|$ that systematically exceeded the upper $3$$\sigma$ observational bounds. Among the tested values, $\mu_0 = 40\,\mathrm{GPa}$ was the only baseline shear modulus that simultaneously satisfied both the tidal amplitude and dissipation constraints within the established statistical confidence intervals.
Furthermore, the elastic lithosphere thickness was constrained by mechanical stability requirements; following~\citet{2016NatGe...9..429B},~\citet{2023ASSL..468.....L},~\citet{Gyalay2024}, and~\citet{park_ios_2025}, the selected value represents the minimum thickness necessary to support the observed topographic relief (i.e., mountains) on Io's surface.
In addition to the upper lithospheric boundary, the core radius ($R_C$) defines the lower boundary condition of the viscoelastic mantle domain, representing a structural constraint inferred from Io's bulk density and moment of inertia factors~\citep[e.g.,][]{park_ios_2025}. Within a forward parametric architecture, variations in the core radius shift the volumetric proportion of the highly dissipative mantle. Increasing $R_C$ reduces the overall mantle volume while placing the core-mantle boundary (CMB) closer to the surface, which typically amplifies the elastic compliance and modulates the baseline tidal amplitude $\Re(k_2)$. However, sensitivity tests within canonical planetary ranges indicate that while the core size sets the foundational baseline for the deformation, its impact on the complex Love number is secondary compared to the orders-of-magnitude variations driven by the thermal state, $\beta_0$, and the melt distribution above $R_{\phi0}$~\citep{Bierson, park_ios_2025}. The nominal core radius chosen here represents a self-consistent structural constraint compatible with recent gravity science measurements~\citep{park_ios_2025}.

To capture the depth-dependent effects of partial melting, the model implemented an iterative refinement scheme. The mantle began as a single viscoelastic layer and was progressively subdivided based on the calculated dissipation, converging to a final, high-resolution grid of 67 layers. This final resolution allowed a smooth representation of variations in viscosity, shear modulus, and $\beta$ as a function of the local melt fraction (see Fig.~\ref{fig_profile}).

At the end of the iterative process, the mantle was resolved into sub-layers with a final thickness of $7\,\mathrm{km}$ in the upper mantle (radial range $1380-1770\,\mathrm{km}$) and $40\,\mathrm{km}$ in the lower mantle ($980-1380\,\mathrm{km}$). Consequently, in the final configuration, the value of $R_{\phi0}$ represents the inner radius of the deepest sub-layer where the local melt fraction ($\phi$) exceeds $1\%$ (see Fig. \ref{fig_models_Io}). From this radius outward, the melt fraction remains above this threshold continuously up to the base of the lithosphere ($1770\,\mathrm{km}$). For instance, if the model converges to an $R_{\phi0}$ of $1604\,\mathrm{km}$ (within the upper mantle grid), this indicates that the first partially molten sub-layer spans $1604\,\mathrm{km}$ to $1611\,\mathrm{km}$, while the total partially molten zone extends continuously from $1604\,\mathrm{km}$ up to the lithospheric boundary at $1770\,\mathrm{km}$ (yielding a total thickness of $166\,\mathrm{km}$).

\subsection{Melt fraction and iterative procedure}\label{melt_and_iterative}
The iterative procedure is illustrated schematically in Fig.~\ref{loop}.
First, we solved the spheroidal oscillation equations to obtain the initial radial profile of the volumetric tidal heating rate, $Q_T(r)$, and the corresponding complex Love number, $k_2$. We then calculated the local melt fraction $\phi(r)$ from $Q_T(r)$ following~\citet{Bierson}, linking internal energy dissipation to the partial melt fraction at each depth. Regions with a higher $\phi(r)$ are rheologically weaker (softer) and more dissipative, thereby modifying the local tidal response.
Based on this newly computed $\phi(r)$ profile, the mantle discretization was refined. Concurrently, the rheological parameters of the mantle, such as viscosity, shear modulus, and the Andrade parameter $\beta$, were updated following~\citet{Bierson}, as shown in the equations below.
\begin{linenomath*}
\begin{equation}
\begin{split}
    \eta(\phi)&=\eta_0e^{-\alpha_{melt}\phi},\\
    \mu(\phi)&=\mu_0[1+c\phi]^{-1},\\
    \beta(\phi)&=\beta_0e^{n_{\beta}\phi}.  
\end{split}
\label{eq1}
\end{equation}
\end{linenomath*}
Here, $\eta_0$, $\mu_0$, and $\beta_0$ denote the initial (reference) viscosity, shear modulus, and Andrade parameter, respectively, assigned to the three-layer model.
The terms $\alpha_{melt}$, $c$, and $n_{\beta}$ are experimental parameters set following~\citet{Bierson}: $\alpha_{melt}=26$~\citep{MEI2002491, SCOTT2006177}, $c=67/15$~\citep{1980JGR....85.5173M}, and $n_{\beta}=20$~\citep{1980JGR....85.5173M}.

To derive the melt fraction $\phi(r)$ from the volumetric tidal heating rate $Q_T(r)$, we used a custom Python script to integrate the 1D equations formulated by~\citet{moore_thermal_2001}. This approach assumes that the melt distribution is determined by a balance between melt production and melt transport (via percolation). The integration parameters were initially set following~\citet{Bierson} (see our Table \ref{tab:parameters}). We also explored alternative parameter values, including different choices for the velocity scale $\gamma$. The results are largely insensitive to these variations, justifying the retention of the~\citet{Bierson} parameter set (see Figs.~\ref{figS4} and \ref{figS6}).
An initial melt fraction of $1\%$ was assumed; lower initial values do not significantly affect the final results but increase computational time, which justifies this choice (see Fig.~\ref{figS8}). As noted by~\citet{Bierson}, the equations are invalid for melt fractions exceeding $\sim30\%$. Furthermore, values above $\sim20\%$ are thought to represent an unstable layer likely to form a continuous magma ocean~\citep{Miyazaki_2022}.

Thus, this methodology captures partially molten states up to the onset of extensive melting.
With the updated mantle structure, we solved the spheroidal oscillation equations again to obtain a new $Q_T(r)$ profile and a new $k_2$. The resulting $Q_T(r)$ was used to recalculate the melt fraction $\phi(r)$. Based on this new $\phi(r)$ profile, the mantle rheology and grid discretization were updated, establishing a self-consistent feedback loop between tidal dissipation and material properties. The iteration continued until the variations in the real and imaginary components of $k_2$ fell simultaneously below the threshold of $10^{-6}$ between successive steps (see Figs.~\ref{figS10} and \ref{figS12}). The final converged models featured a mantle resolved into 67 layers, from which we extracted the final $k_2$ values.
We performed this entire iterative procedure for a grid of $R_{\phi0}$ and $L$ values. This parametric study explores how the depth (controlled by $R_{\phi0}$) and intensity (controlled by $L$) of mantle melting influence the tidal response. This approach allowed us to identify interior models consistent with the observational constraints from the Juno mission~\citep{park_ios_2025}.

\subsection{Rheological parameterization}

To capture the thermomechanical feedback within Io's interior without imposing the depth of dissipation a priori, we implemented a dynamic parameterization for the Andrade parameter $\beta$. Specifically, it evolves as a continuous function of the local melt fraction ($\phi(r)$). This physically grounded framework allowed the mantle rheology to respond directly to the internal melt structure, naturally enhancing anelastic dissipation in regions with higher melt fractions~\citep{2004JGRB..109.6201J,Bierson}.
By directly linking the rheological properties to the thermal state, this approach allowed us to self-consistently compute the complex Love number $k_2$ and infer the corresponding melt distribution. The resulting depth-dependent profile of tidal dissipation is discussed in Sect. \ref{res} and Sect. \ref{disc}.

The structural robustness of this continuous formulation relies on specific microphysical assumptions regarding the distribution of the liquid phase. The adopted exponential scaling given in Eq. \ref{eq1} implicitly presumes a texturally equilibrated and macroscopically isotropic distribution of melt along grain boundaries, acting primarily via melt-enhanced grain boundary sliding~\citep{2004JGRB..109.6201J}. However, this continuous formulation represents an idealized configuration that does not account for complex geodynamic processes such as local melt segregation or the formation of localized shear bands, which could introduce nonnegligible anelastic anisotropy under high-frequency tidal forcing.

Furthermore, the predictive behavior of this dynamic framework is inherently coupled to both the chosen viscoelastic framework and its baseline parameters. The initial Andrade parameter $\beta_0$ acts as the primary control governing the trade-off between elastic compliance and fluid-like relaxation, heavily dictating the physical characteristics of the mantle models required to match the observed $k_2$ values. However, both $\beta_0$ and the scaling parameters themselves carry significant experimental and microphysical uncertainties~\citep{Walterova2023, bierson2024impact}.
 
Beyond parametric variations within a single formulation, the tidal response is also dependent on the choice of the rheological model itself. Recent comparative studies demonstrate that the differences between distinct formulations, such as the Burgers or Sundberg-Cooper frameworks, can be substantial under planetary tidal frequencies (e.g., \citealp{Tobie} for icy satellites, and \citealp{2018ApJ...857...98R} specifically for Io's extreme environment). These alternative rheologies prescribe different frequency-dependent complex compliances because they represent the spectrum of anelastic relaxation differently. Specifically, the Andrade rheology describes a broad power-law transient-creep response, whereas the Burgers and Sundberg-Cooper formulations introduce additional relaxation mechanisms with characteristic timescales and potentially an additional dissipation peak. Depending on the position of the forcing frequency relative to these characteristic relaxation times, alternative models may predict different real and imaginary components of the tidal response at Io's orbital frequency. Consequently, the specific melt fractions required to match the observed $k_2$ values are intrinsically conditional on the adopted Andrade formulation. Determining whether another constitutive framework would shift the inferred solutions above or below the rheologically critical melt fraction (RCMF) requires explicit comparative calculations.

\subsection{Melt production and migration budget}\label{meltprod}

To assess a posteriori the physical plausibility of the modeled melt fraction ($\phi$) and the potential existence of a global magma ocean, we evaluated the mass-flux balance within the partially molten mantle shell defined by $R_{\text{bot}} \le r \le R_{\text{top}}$, where $R_{\text{bot}}$ represents the CMB and $R_{\text{top}}$ corresponds to the base of the lithosphere. The CMB radius ($R_{\text{bot}}$) was set to $950\,\mathrm{km}$ to ensure consistency with the core dimensions adopted by~\citet{park_ios_2025}. Meanwhile, the upper mantle boundary ($R_{\text{top}}$) corresponds to the base of the lithosphere, which was fixed at a radius of $1770\,\mathrm{km}$ throughout the simulations. The analysis distinguishes between the thermodynamic rate of melt generation ($\dot{M}_{\text{gen}}$) and the melt percolation capacity ($\dot{M}_{\text{migr}}$).

The melt generation rate is constrained by the total tidal dissipation power ($P_{\text{tot}}$) integrated over the shell volume. Assuming a steady-state thermal equilibrium where latent heat absorption balances tidal heating~\citep{1979Sci...203..892P, https://doi.org/10.1029/GL008i004p00313}, $\dot{M}_{\text{gen}}$ is given by\begin{equation}
    \dot{M}_{\text{gen}} = \frac{\int_{R_{\text{bot}}}^{R_{\text{top}}} 4\pi r^2 Q_T(r) \, dr}{L}\,,
    \label{eq:melt_generation}
\end{equation}
where $Q_T(r)$ is the volumetric tidal heating rate and $L$ is the latent heat of fusion.

In contrast, the migration rate represents the maximum melt flux that can be extracted through the top of the shell ($R_{\text{top}}$) through buoyancy-driven porous flow. Following Darcy's law for two-phase flow~\citep{10.1093/petrology/25.3.713}, the melt percolation capacity is defined as\begin{equation}
    \dot{M}_{\text{migr}} = 4\pi R_{\text{top}}^2 \cdot \rho_{\text{melt}} \cdot \gamma \cdot \phi^m\,,
    \label{eq:melt_migration}
\end{equation}
where $\gamma$ is the reference velocity scale, $m$ is the permeability exponent, $\rho_{\text{melt}}$ is the melt density, and $\phi$ is explicitly evaluated as the local melt fraction at the upper boundary of the shell ($r = R_{\text{top}}$). This comparison serves as an idealized quasi-steady, radially one-dimensional mass-balance test. It implicitly assumes a continuous vertically interconnected porous network through which all melt generated within the shell must ultimately migrate and cross $R_{\text{top}}$. Consequently, this formulation neglects compaction effects, melt network geometry variations, localized refreezing, temporary storage, lateral transport, and channelized or dike-assisted migration. Furthermore, because the baseline transport rate is assumed to scale cubically with porosity ($m = 3$), the resulting mass flux is highly sensitive to uncertainties in $\phi(R_{\text{top}})$, $\gamma$, and the chosen permeability exponent $m$. The adopted numerical values for these parameters are listed in Table~\ref{tab:parameters}.

Comparing $\dot{M}_{\text{gen}}$ and $\dot{M}_{\text{migr}}$ provides an a posteriori consistency check for the modeled partially molten mantle configurations:
\begin{itemize}
    \item Accumulation-prone regime ($\dot{M}_{\text{gen}} > \dot{M}_{\text{migr}}$): If the total thermodynamic melt production integrated over the mantle column exceeds the estimated transport rate at the upper boundary, steady porous drainage through $R_{\text{top}}$ is insufficient. Under the adopted assumptions, this indicates that temporary melt storage, localized porosity increases, or alternative transport mechanisms (such as diking) would be required.
    
    \item Near-balance regime ($\dot{M}_{\text{gen}} \approx \dot{M}_{\text{migr}}$): When the column-integrated production rate approximates the estimated extraction rate at $R_{\text{top}}$, the system operates near its modeled fluid-dynamic limit. This state represents a threshold where the assumed porous network is maximally utilized to transport the upward-fluxing magma.
    
    \item Drainage-capable regime ($\dot{M}_{\text{gen}} < \dot{M}_{\text{migr}}$): If the estimated migration rate at the base of the lithosphere exceeds the total deeper thermodynamic production, the inferred porosity profile is compatible with the assumed porous-drainage model. While this does not independently prove the absence of localized accumulation, it demonstrates that the interior structure is physically consistent with a drainage-capable magmatic sponge configuration.
\end{itemize}

\subsection{Framework overview}

To ensure the robustness of our numerical implementation, we compared our computed $k_2$ values with the observed estimates~\citep{park_ios_2025} and independently verified them using the \texttt{ALMA3} code~\citep{melini_computing_2022}. Because \texttt{ALMA3} does not currently account for mantle compressibility, we restricted this code-to-code comparison to incompressible mantle configurations. However, the agreement found in this regime confirms that our computed $k_2$ trends are robust and not artifacts of a specific numerical implementation. This successful benchmark (see Appendix \ref{app1}) validates our core computational framework, providing strong confidence in our extension of the modified \texttt{CPGC} to fully compressible models.

Building on this validated foundation, the methodology we developed integrates several key components to self-consistently model Io's thermomechanical state:
\begin{itemize}
    \item A layered viscoelastic internal structure, evaluating both incompressible and compressible mantle configurations.    \item Iterative updates of the mantle rheology (including viscosity, shear modulus, and the Andrade parameter $\beta$) dynamically coupled to the local melt fraction ($\phi(r)$).    \item A parametric exploration of the latent heat of fusion ($L$) and the melting onset radius ($R_{\phi0}$).    \item A targeted sensitivity analysis on the reference Andrade parameter ($\beta_0$) to reconcile the imaginary part of the Love number with observational constraints.    \item An a posteriori mass-flux balance analysis comparing thermodynamic melt production with magmatic percolation capacity.
\end{itemize}

This integrated approach provides a rigorous framework for linking Io's interior structure, partial melting, and tidal dissipation with the observed $k_2$, allowing us to identify mantle configurations strictly compatible with Juno data. By explicitly coupling the rheology to the local melt fraction and incorporating magmatic transport constraints, our methodology expands upon previous studies to offer a self-consistent assessment of Io's interior dynamics. For reference, we provide a complete summary of all physical and numerical parameters adopted in our simulations in Table \ref{tab:parameters}.

\section{Results}\label{res}

In this section we present the results of our parametric investigation into Io's tidal response. We systematically examine how the melting onset radius ($R_{\phi0}$), the latent heat of fusion ($L$), and the initial Andrade parameter ($\beta_0$) govern the degree-2 Love number ($k_2$), the volumetric tidal heating rate $Q_T(r)$, and the radial melt fraction profile $\phi(r)$. 
By explicitly coupling the Andrade parameter $\beta$ to the local melt fraction, we demonstrate how this physically grounded rheology naturally generates a localized shallow-mantle enhancement in dissipation, superimposing it upon the deep-mantle heating regime.

The results are organized into two main cases: A) an incompressible mantle with a fixed initial Andrade parameter $\beta_0$ and B) a compressible mantle, exploring $\beta_0$ variations for models consistent with the observed $\Re(k_2)$.
For each case, the real ($\Re(k_2)$) and imaginary ($|\Im(k_2)|$) parts of $k_2$ are presented, alongside radial profiles of the volumetric heating rate ($Q_T(r)$) and the melt fraction ($\phi(r)$). Within the adopted Andrade rheological framework, these results provide quantitative constraints on the 1D melt distribution and mantle rheology, forming the basis for the discussion of Io’s interior structure and tidal heating processes.

\subsection{Incompressible mantle framework with fixed initial \texorpdfstring{$\beta_0$}{beta0}}\label{Incompressible}

This investigation focuses on three key parameters that influence the tidal Love number $k_2$: the radial position at which mantle melting begins ($R_{\phi0}$), the latent heat of fusion ($L$), and the initial Andrade parameter ($\beta_0$). In Case A, we establish our baseline physical framework by treating the mantle as strictly incompressible (i.e., assuming an infinite bulk modulus) and fixing the initial reference $\beta_0$ value at $10^{-12}\,\mathrm{Pa^{-1}s^{-0.3}}$.

The incompressible assumption confines tidal dissipation entirely to shear deformation, consistent with previous literature~\citep{1988Icar...75..187S, 2013E&PSL.361..272H, Bierson, 2024GeoRL..5107869A, park_ios_2025, 2025NatCo..16.6798V}. Furthermore, the chosen $\beta_0$ value aligns with~\citet{park_ios_2025}, who showed that it yields a $k_2$ estimate within $1$$\sigma$ of the Juno measurement, even for a non-iterative three-layer model where rheological properties are not updated based on the local melt fraction.

With this fixed initial $\beta_0$ value, we systematically investigated the variation in $k_2$ as a function of $R_{\phi0}$ and $L$. We subsequently examined the corresponding radial profiles of the volumetric tidal heating rate ($Q_T(r)$) and the melt fraction ($\phi(r)$), with a strict focus on interior configurations consistent with Juno measurements. By explicitly coupling the rheology to the melt distribution, this analysis directly links Io's macroscopic tidal response to its internal thermomechanical state. We note, however, that the adopted 1D modeling framework inherently yields spherically averaged, radial profiles. Consequently, while providing robust depth-dependent constraints, this approach cannot resolve the full 3D spatial heterogeneity of tidal heating and partial melt.

\begin{figure}[!ht]
\resizebox{\hsize}{!}
    {\includegraphics{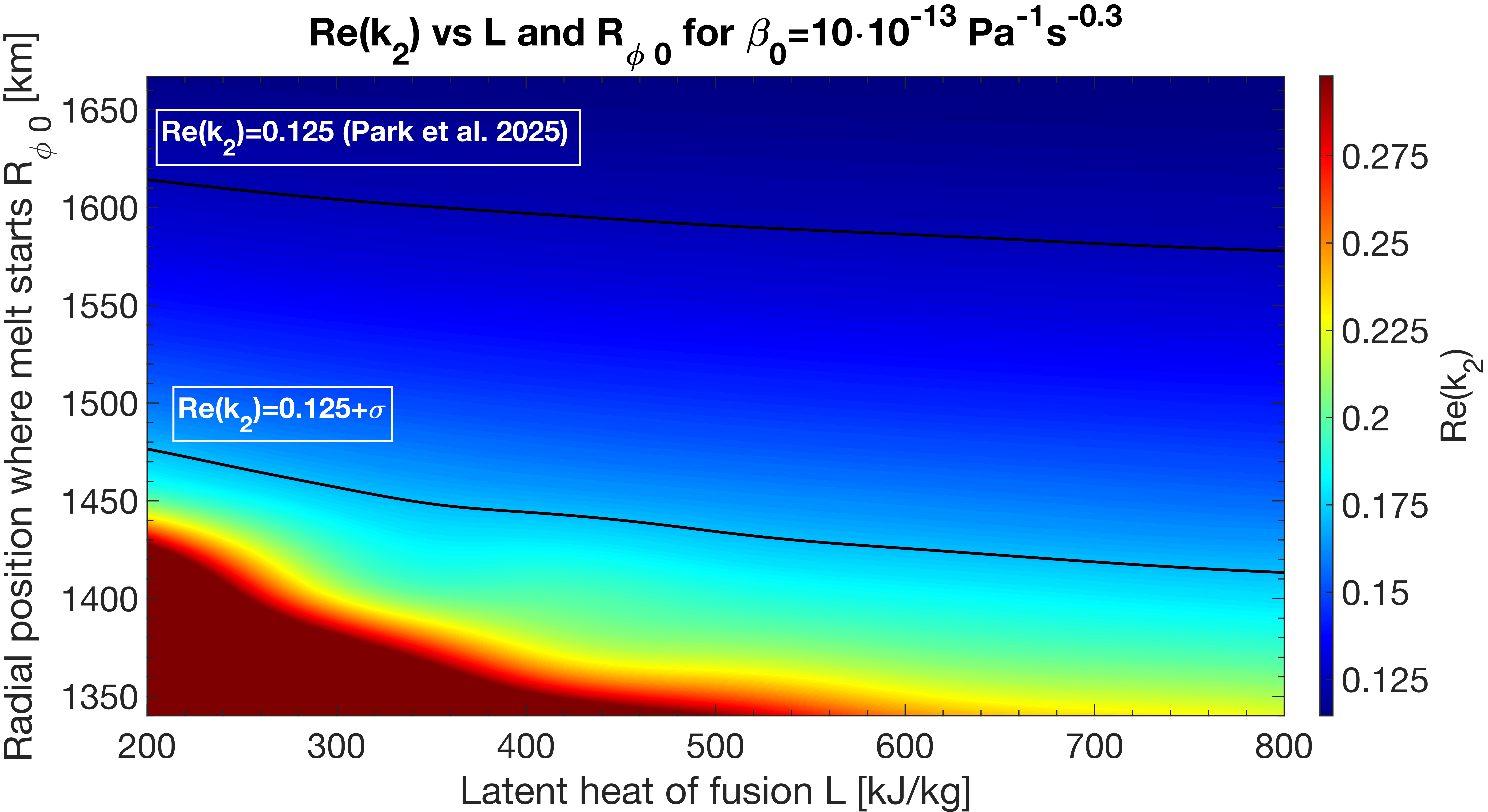}}
    \hspace{0.6 mm}
\resizebox{\hsize}{!}
    { \includegraphics{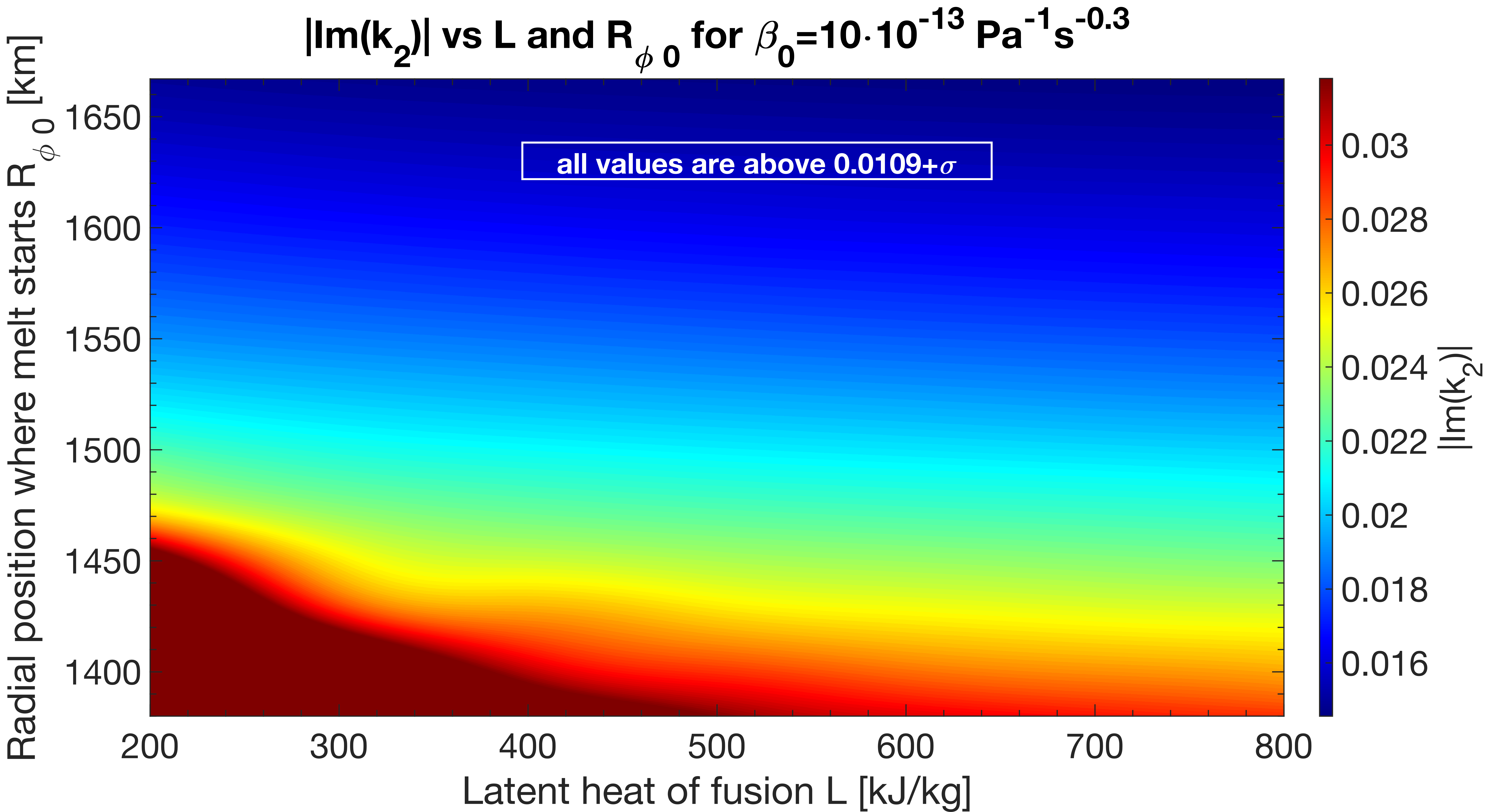}}
    \caption{Love number $k_2$ as a function of melting radius ($R_{\phi 0}$) and latent heat of fusion ($L$). In the top panel, the estimated value of $\Re(k_2)$ from~\citet{park_ios_2025} ($0.125\pm0.047$) is shown as a reference. In the bottom panel, no reference is shown for $|\Im(k_2)|$, as all values exceed $1$$\sigma$, i.e., $0.0163$.}
    \label{fig5}
\end{figure}

As formulated in Eq. \ref{eq1}, our model allows the Andrade parameter $\beta$ to vary dynamically across the mantle sub-layers in response to the local melt fraction. Figure~\ref{fig5} presents the resulting tidal response and internal dissipation profiles derived for Case A. The real part of the Love number, $\Re(k_2)$, matches observational constraints across a wide range of $R_{\phi0}$ and $L$ values. The imaginary part, $|\Im(k_2)|$, lies outside the strictly $1$$\sigma$ confidence interval, but remains robustly within $2$$\sigma$ of the observed value.

\begin{figure}[!ht]
    \centering
    \includegraphics[width=0.75\hsize]{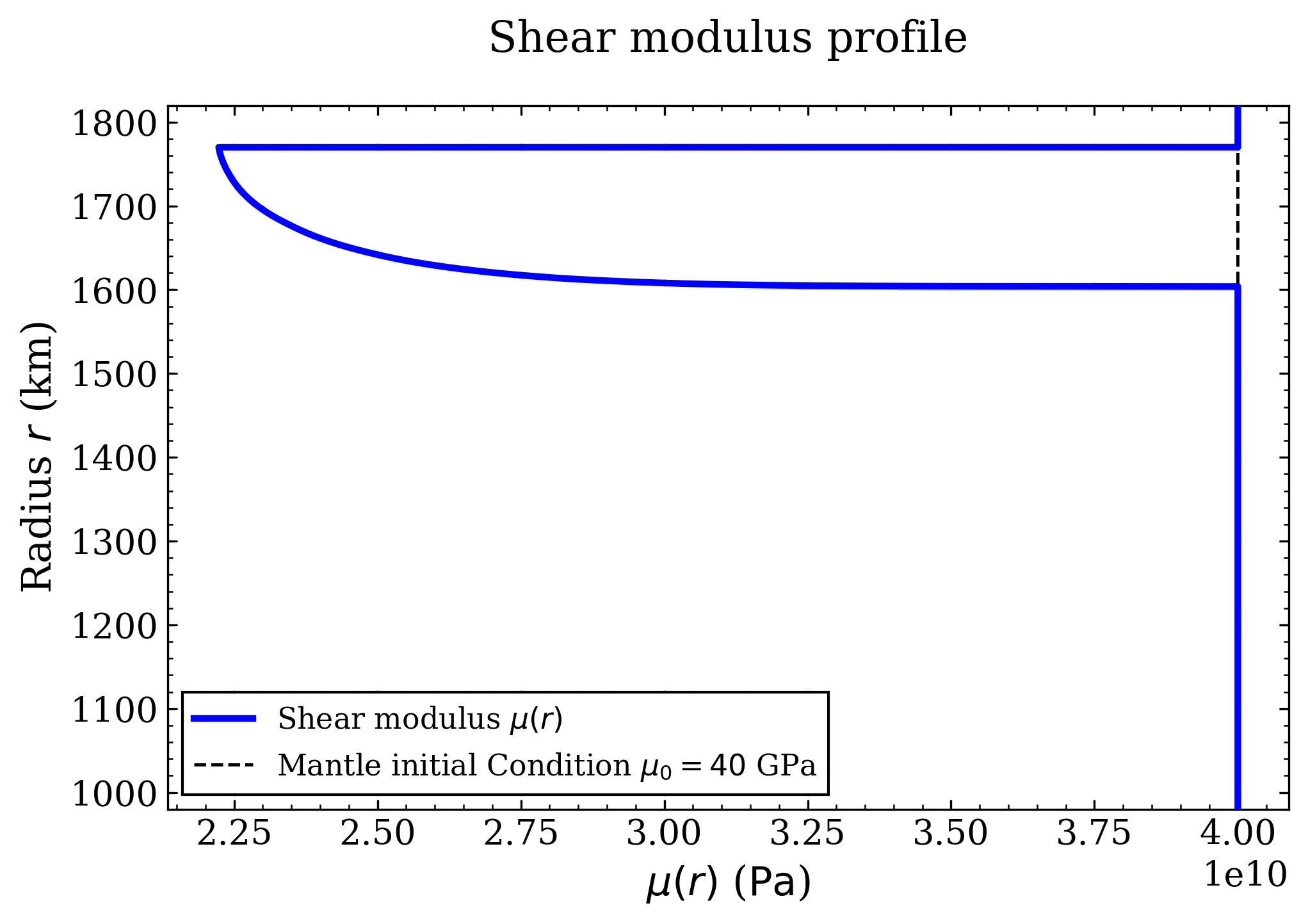}  
    \includegraphics[width=0.75\hsize]{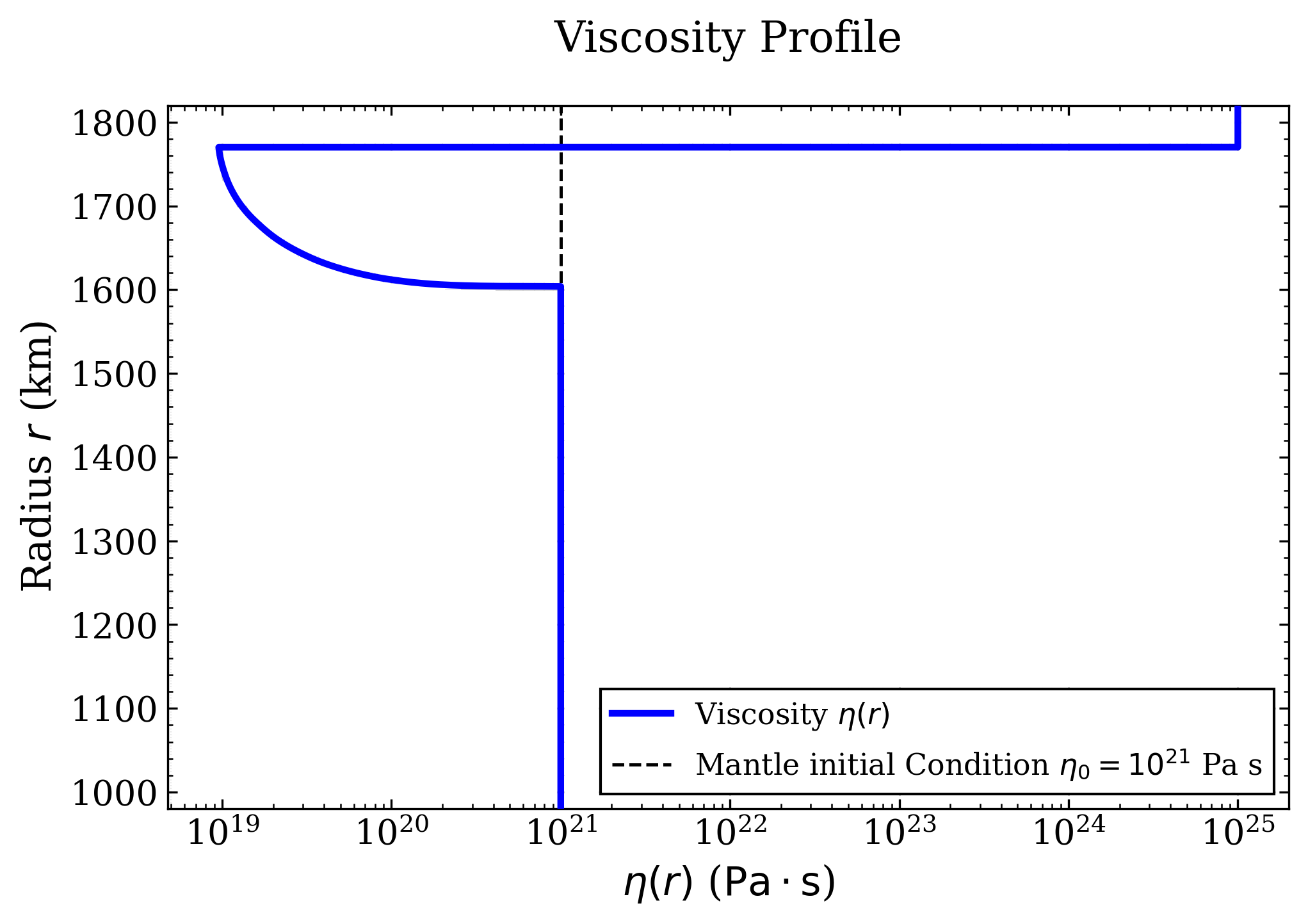}    
    \includegraphics[width=0.75\hsize]{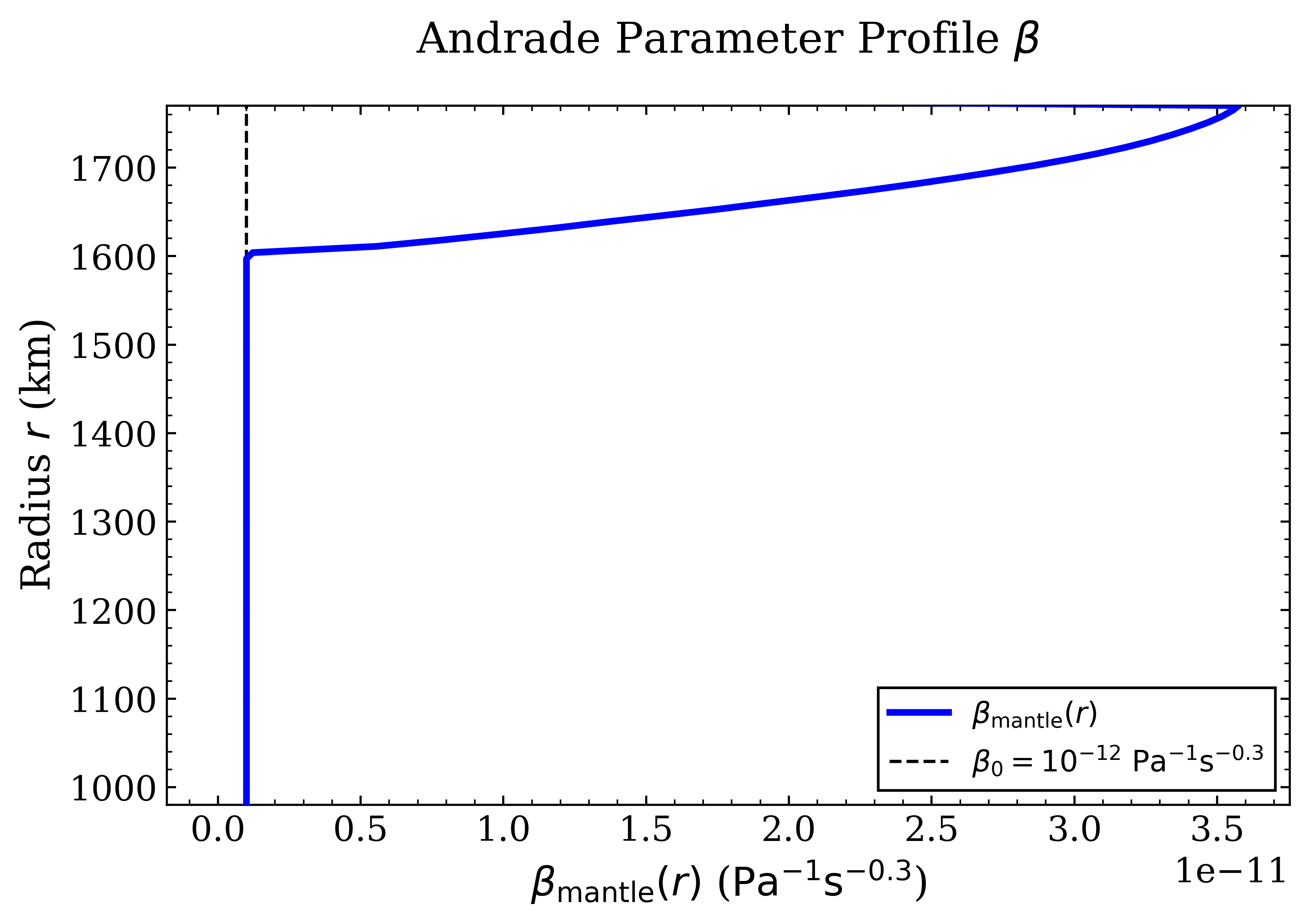}
    
    \caption{Radial profiles of mantle viscosity, shear modulus, and the Andrade parameter ($\beta$) for the representative interior configuration matching the observational constraint $\Re(k_2) = 0.125$. This specific model is parameterized by a melting onset radius $R_{\phi0} = 1604\,\mathrm{km}$ and a latent heat of fusion $L = 3\times 10^{5}\,\mathrm{J\,kg^{-1}}$. The depth-dependent variations across the mantle reflect the self-consistent thermomechanical feedback, wherein the rheological properties are dynamically coupled to the local melt fraction.}
    \label{fig_profile}
\end{figure}

Multiple combinations of $R_{\phi0}$ and $L$ yield a $\Re(k_2)$ within $1$$\sigma$ of the Juno measurement~\citep{park_ios_2025} (see our Fig.~\ref{fig5}). This agreement is lost, however, if the partially molten layer becomes too thick; specifically, when $R_{\phi_0} < 1400\,\mathrm{km}$ (i.e., the layer exceeds half the mantle thickness), the $\Re(k_2)$ values fall outside the $1$$\sigma$ range. The imaginary part, $|\Im(k_2)|$, is consistently higher than the estimate by~\citet{park_ios_2025} and closer to that of~\citet{2009Natur.459..957L}. These values lie outside the $1$$\sigma$ confidence interval, but remain within $2$$\sigma$. A specific benchmark case that reproduces the observed  $\Re(k_2)\approx 0.125$ is found in $R_{\phi0} = 1604\,\mathrm{km}$ and $L = 3\times10^{5}\,\mathrm{J\,kg^{-1}}$ (see Fig.~\ref{figS2});
for this configuration, Fig.~\ref{fig_profile} illustrates the resulting radial profiles of key mantle properties, namely viscosity, shear modulus, and the Andrade parameter $\beta$.

\begin{figure}[!ht]
\resizebox{\hsize}{!}{
\includegraphics{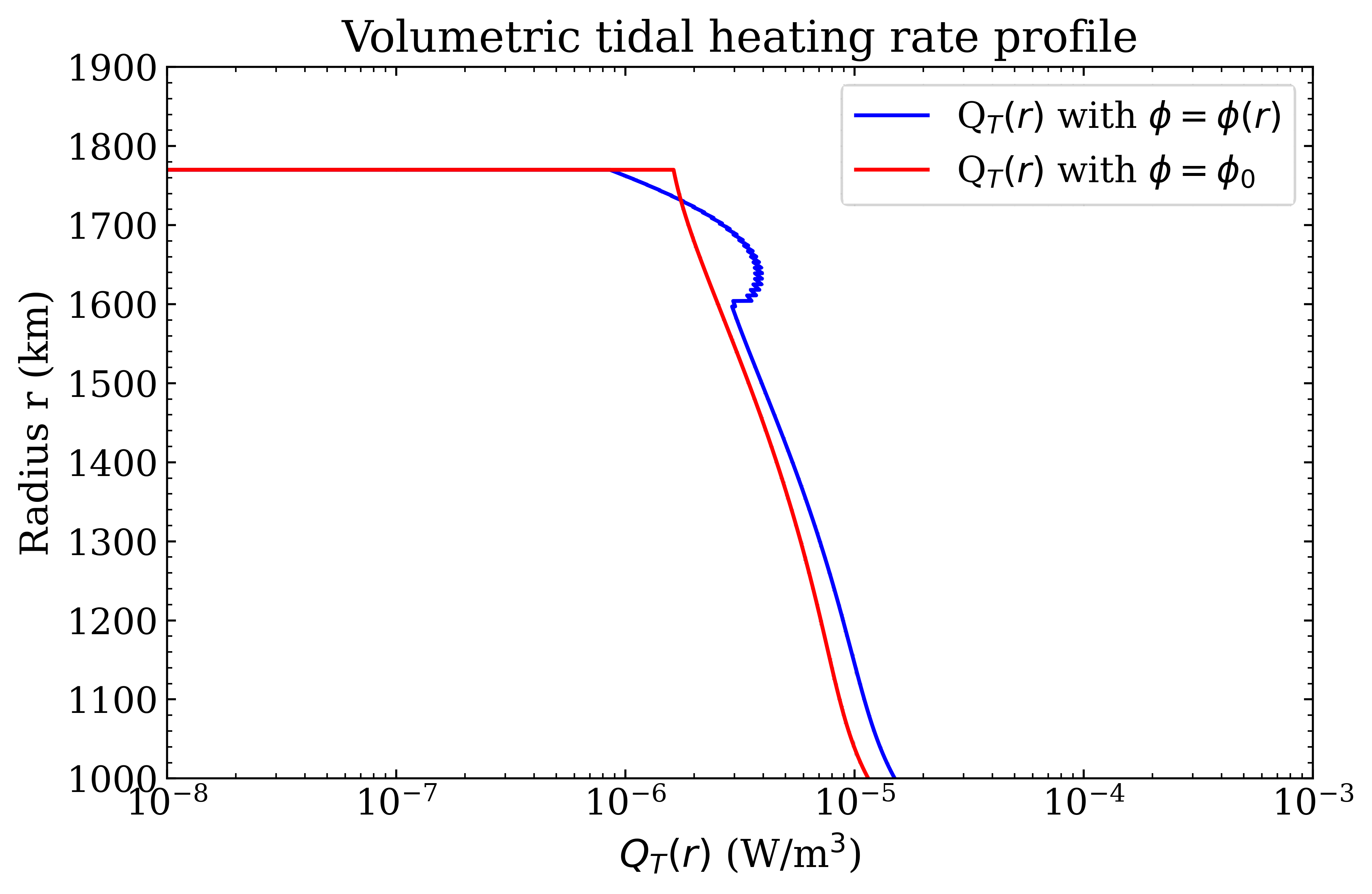}}
\caption{Radial profile of the volumetric tidal heating rate, $Q_T(r)$, for the benchmark case ($R_{\phi0} = 1604\,\mathrm{km}$, $L = 3\times10^{5}\,\mathrm{J\,kg^{-1}}$) that matches the observed  $\Re(k_2)\approx 0.125$~\citep{park_ios_2025}. The red curve shows the initial (first iteration) profile; the blue curve represents the final self-consistently converged profile. By dynamically coupling the rheology to the local melt fraction, the model illustrates the natural emergence of an additional localized enhancement of tidal heating in the upper mantle, superimposing it upon the deep-mantle heating regime.}
\label{fig7}
\end{figure}

\begin{figure}[!ht]
\resizebox{\hsize}{!}
{\includegraphics{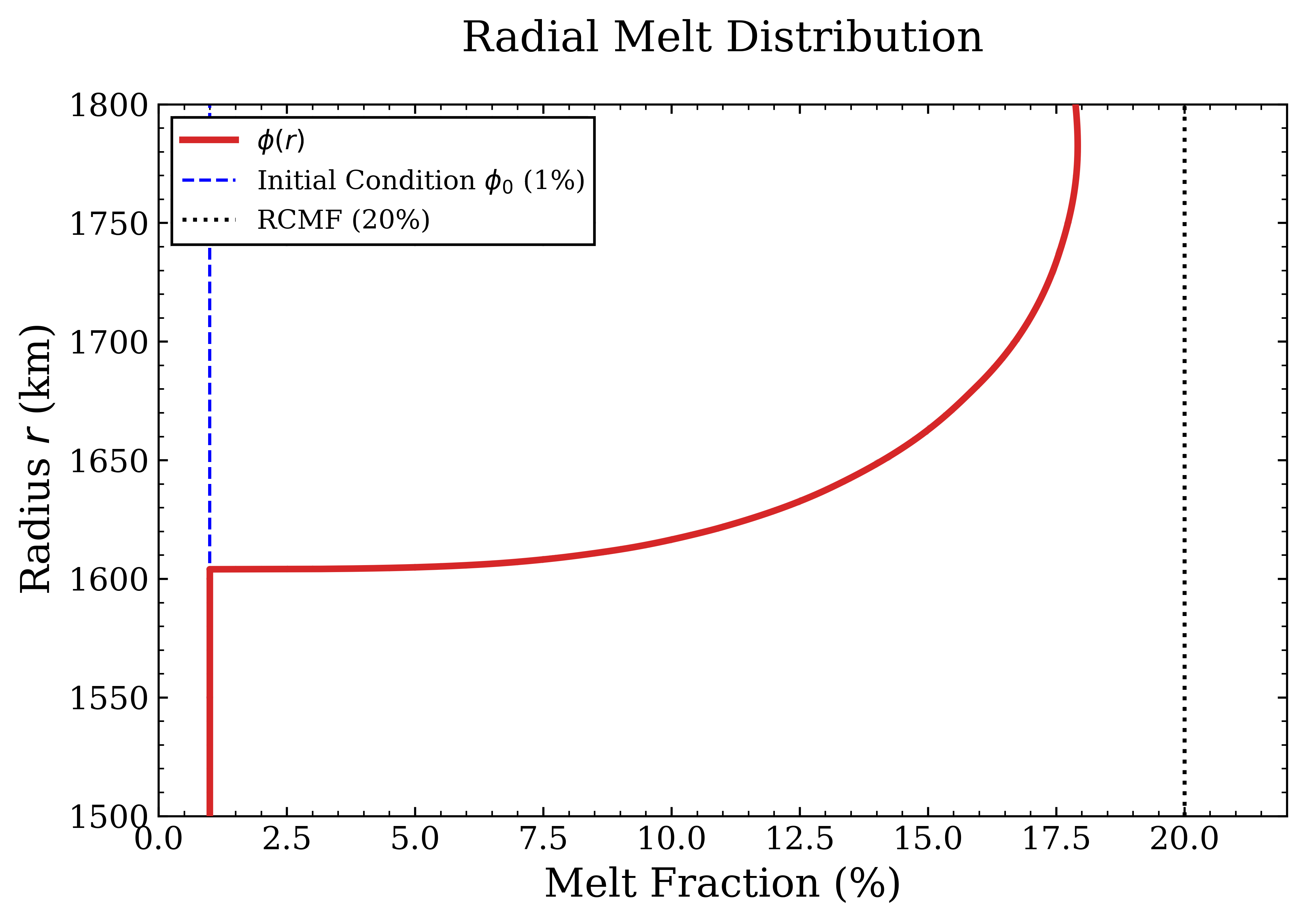}}
\caption{Final radial profile of the melt fraction, $\phi(r)$, for the same benchmark model shown in Fig.~\ref{fig7} ($R_{\phi0} = 1604\,\mathrm{km}$, $L = 3\times10^{5}\,\mathrm{J\,kg^{-1}}$, and $\beta_0=10^{-12}\,\mathrm{Pa^{-1}s^{-0.3}}$). The melt fraction remains below the RCMF associated with the formation of a laterally uniform magma layer~\citep{Miyazaki_2022}. An initial melt fraction of $1\%$ was assumed; lower initial values did not significantly affect the results but increased computational time, justifying the choice of $1\%$.}
\label{fig8}
\end{figure}

\begin{figure}[!ht]
\resizebox{\hsize}{!}
{\includegraphics{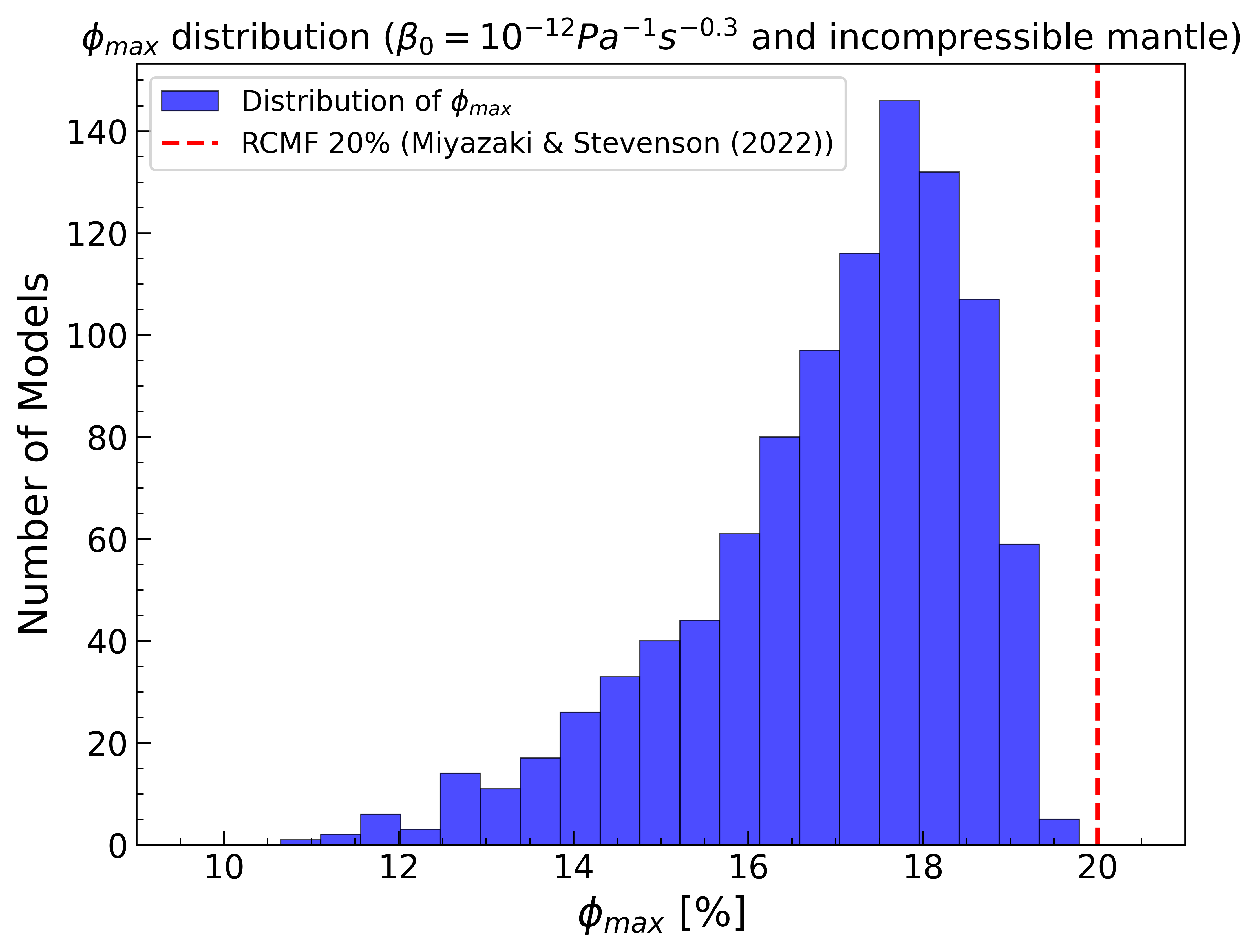}}
\caption{Distribution of peak melt fractions ($\phi_{max}$) for models consistent with Juno observations. The figure displays the maximum melt fraction obtained for all incompressible model configurations that reproduce the real part of the Love number, $\Re(k_2)$, within the $1$$\sigma$ confidence interval estimated by~\citet{park_ios_2025}. The vertical red dashed line indicates the RCMF of $20\%$~\citep{Miyazaki_2022}. All valid models remain below the RCMF threshold, supporting a subcritical partially molten mantle.}
\label{figS14}
\end{figure}

The corresponding radial profiles for the volumetric tidal heating rate ($Q_T(r)$) and melt fraction ($\phi(r)$) for this model are shown in Figs.~\ref{fig7} and~\ref{fig8}, respectively.
As shown in Figs.~\ref{fig7} and \ref{fig8}, alongside the primary deep-mantle heating, there is an additional localized enhancement of tidal heating in the upper mantle because the presence of partial melt locally decreases the effective viscosity and increases anelastic dissipation. The benchmark model that reproduces the value of $\Re(k_2)$ estimated by~\citet{park_ios_2025} exhibits a melt fraction of approximately $18\%$. This finding is consistent across all 1D models that yield $\Re(k_2)$ values within $1$$\sigma$ of the Juno observation; such models maintain melt fractions below the critical threshold (see Fig.~\ref{figS14}).

Therefore, within our 1D spherically averaged framework, only models with subcritical melt fractions (i.e., $\phi(r) <$ RCMF) successfully reproduce the Juno data. This result indicates that agreement with the Juno observations can be achieved without requiring a global magma ocean in Io's mantle.

For the benchmark case, an a posteriori mass-flux analysis (see Sect.~\ref{meltprod}) yields a column-integrated thermodynamic melt production rate of $\dot{M}^{\text{max}}_{\text{gen}} \approx 8.01 \times 10^7 \, \text{kg s}^{-1}$. In comparison, the maximum melt percolation capacity calculated using the local melt fraction at the upper boundary ($\phi(R_{\text{top}})$) is $\dot{M}^{\text{max}}_{\text{migr}} \approx 3.75 \times 10^8 \, \text{kg s}^{-1}$.

This comparison reveals that $\dot{M}^{\text{max}}_{\text{migr}} > \dot{M}^{\text{max}}_{\text{gen}}$ by a factor of approximately $4.68$. This inequality serves as an internal consistency check validating the fluid-dynamics plausibility of the inferred configuration, while the geodynamic implications of the resulting melt transport regime are discussed in detail in Sect.~\ref{comparison}.

The calculated value for $\dot{M}_{\text{gen}}^{\text{max}}$ is consistent with independent empirical constraints. Scaling the localized volumetric effusion rates documented by~\citet{https://doi.org/10.1029/2025JE008940} yields an integrated global volcanic mass discharge range of $\sim 1.4 \times 10^7$ to $1.4 \times 10^8\,\mathrm{kg\,s^{-1}}$. The model-derived steady-state production falls within this observationally constrained interval. A comprehensive description of the scaling formulation and a comparative analysis against transient eruptive fluxes ~\citep{Mura_Synchronized}  are provided in Sect.~\ref{comparison}.

\begin{figure}[!ht]
\resizebox{\hsize}{!}
{\includegraphics{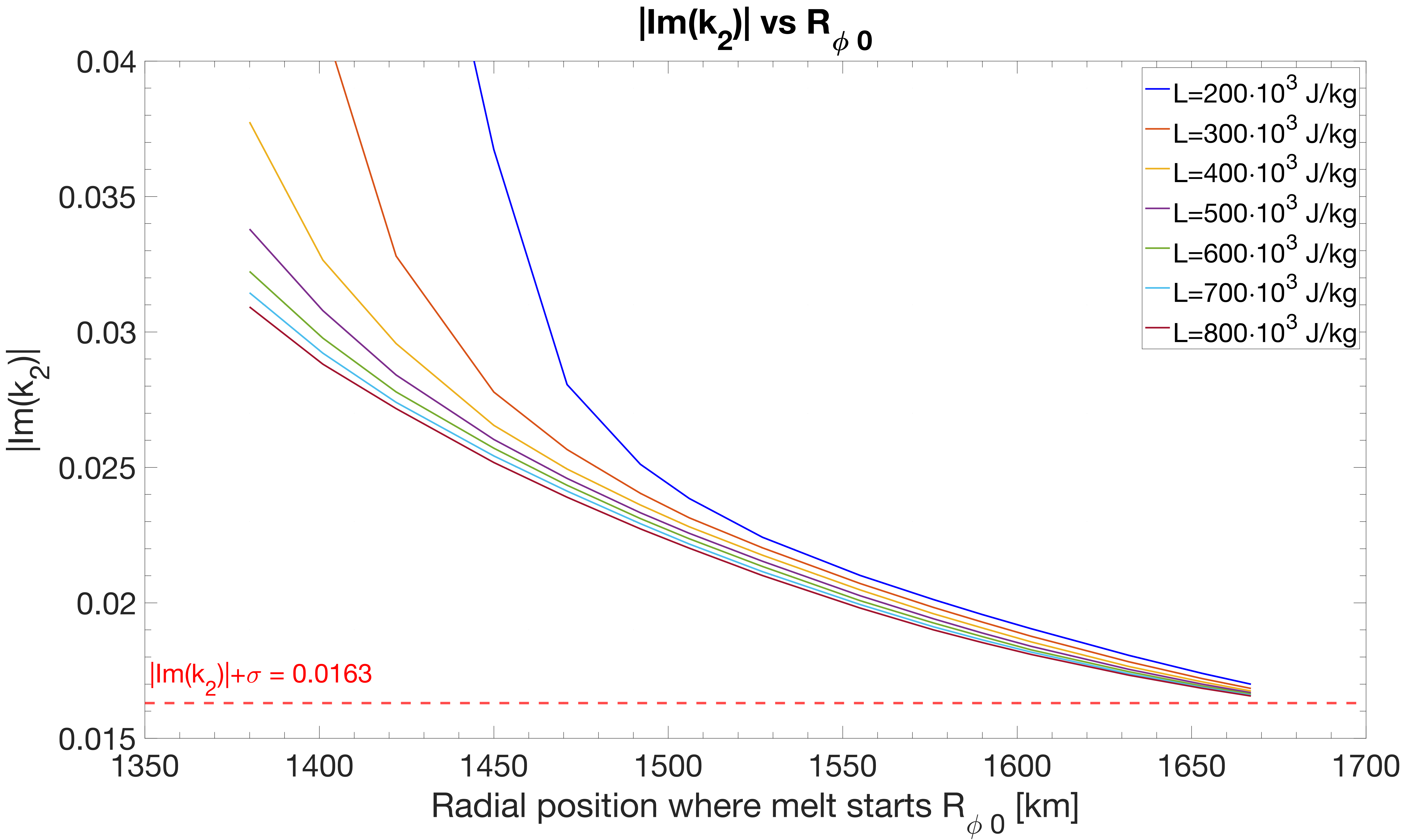}}
    \caption{Magnitude of the imaginary Love number, $|\Im(k_2)|$, as a function of melting onset radius ($R_{\phi0}$) for different latent heat ($L$) values. This figure shows how $|\Im(k_2)|$ decreases with $R_{\phi0}$.}
    \label{fig9}
\end{figure}

Because the internal thermal state and melt distribution directly govern the satellite's rheological response to tidal forcing, we subsequently extend our analysis to further evaluate the tidal dissipation. Figure~\ref{fig9} illustrates the trends in the magnitude of the imaginary Love number, $|\Im(k_2)|$, which is observed to decrease as the melting onset radius ($R_{\phi0}$) increases.

For the reference $\beta_0$ value of $10^{-12}\,\mathrm{Pa^{-1}s^{-0.3}}$, the computed $|\Im(k_2)|$ values lie within $2$$\sigma$ of the observational constraints. Although broadly consistent, this discrepancy motivates a dedicated sensitivity analysis of $k_2$ with respect to the initial Andrade parameter, $\beta_0$. This analysis is necessary to better constrain Io’s interior structure using the Juno observations~\citep{park_ios_2025} and to achieve closer agreement with the observed imaginary component.

Indeed, to further reconcile the imaginary component of $k_2$ with observations, we performed a sensitivity analysis on the initial Andrade parameter, $\beta_0$. The previous simulations (using $\beta_0 = 10^{-12}\,\mathrm{Pa^{-1}s^{-0.3}}$) yielded $|\Im(k_2)|$ values that are consistently higher than the $1$$\sigma$ estimate from~\citet{park_ios_2025}. In this regime, dissipation is positively correlated with $\beta_0$; therefore, to reduce $|\Im(k_2)|$ to the observed value, we focused the analysis on smaller parameters, exploring the range $10^{-13}$-$10^{-12}\,\mathrm{Pa^{-1}s^{-0.3}}$. This investigation varied $\beta_0$ and $L$ while holding $R_{\phi0}$ fixed at the values previously found to be consistent with the observed $\Re(k_2)$. 

Exploring $\beta_0$ values significantly larger than $10^{-10}\,\mathrm{Pa^{-1}s^{-0.3}}$ was not considered. Although such values might also reduce $|\Im(k_2)|$ by moving past the dissipation peak~\citep{Tobie}, they would simultaneously increase $\Re(k_2)$ significantly, placing it well outside the $3$$\sigma$ confidence interval reported by~\citet{park_ios_2025}.

As shown in Appendix \ref{sensitivity_study}, the agreement with the observed $|\Im(k_2)|$ can be achieved by lowering the initial Andrade parameter $\beta_0$ (i.e., $\beta_0 < 10^{-12}\,\mathrm{Pa^{-1}s^{-0.3}}$). However, this adjustment produces a simultaneous reduction in $\Re(k_2)$. Consequently, these models (those that successfully match $|\Im(k_2)|$) correspond to a lower mantle melt fraction compared to the reference models that used $\beta_0 = 10^{-12}\,\mathrm{Pa^{-1}s^{-0.3}}$.

The $\beta_0 = 10^{-12}\,\mathrm{Pa^{-1}s^{-0.3}}$ case can therefore be interpreted as providing an upper limit on the melt fraction required to be consistent with the Juno observations. This sensitivity analysis (Fig.~\ref{fig11}) also demonstrated that the initial Andrade parameter $\beta_0$ is the primary factor controlling the tidal response. 
Because $\beta_0$ dictates the scaling of the viscoelastic dissipation, the systematic clustering of Juno-compatible models just below the RCMF threshold must be presented as a consequence of this adopted rheological parameterization rather than a model-independent physical result.
The $k_2$ values cluster tightly based on $\beta_0$, showing a much weaker dependence on latent heat of fusion $L$. 
This analysis thus identified the models (for a fixed $R_{\phi0}$) that fall within the $1$$\sigma$ uncertainty range of the Juno observations.

\subsection{Compressible mantle (fixed \texorpdfstring{$R_{\phi_0}$}{Rphi0})}\label{compressible}
To test the effect of compressibility, we repeated the analysis using the benchmark $R_{\phi0}$ values identified in the incompressible cases. In this specific case, the radial position was fixed at $R_{\phi0} = 1604\,\mathrm{km}$. These values correspond to the models that reproduced the observed $\Re(k_2)$ using reference $\beta_0 = 10^{-12}\,\mathrm{Pa^{-1}s^{-0.3}}$.

The sensitivity of the Love number $k_2$ to the latent heat of fusion ($L$) and the initial Andrade parameter ($\beta_0$) was then investigated for a compressible mantle. In Case B, the bulk modulus $K$ was assumed to be finite (rather than infinite) and calculated using Eq. \ref{eq2}:
\begin{linenomath*}
\begin{equation}
K(\phi)=\lambda+\frac{2}{3}\mu(\phi)\,,
\label{eq2}
\end{equation}
\end{linenomath*}
where $\lambda$ is the Lamé coefficient and $\mu(\phi)$ is the melt-dependent shear modulus. The value of $\lambda$ was set to $223.\bar{3}\,\mathrm{GPa}$, a reference value for typical mantle materials~\citep{Tobie}.
Because the shear modulus ($\mu$) is iteratively updated based on the melt fraction, the bulk modulus ($K$) also becomes melt-dependent in this scenario. 

Since $\mu(\phi)$ is a complex quantity in our formulation, the bulk modulus $K(\phi)$ mathematically inherits a melt-dependent imaginary component. We explicitly clarify, however, that we did not evaluate the volumetric heating associated with this imaginary component. Our formulation does not incorporate an independent bulk viscosity, nor does it compute explicit dissipative volumetric heating terms (i.e., the volumetric dissipation power is set to zero). Instead, the physical dissipation mechanism remains entirely shear controlled through the Andrade framework. This distinguishes our approach from geodynamic formulations where bulk or compaction-driven volumetric dissipation channels are actively modeled~\citep{beuthe_spatial_2013, 2021A&A...650A..72K}. In our framework, the introduction of finite compressibility increases the total mechanical compliance of the mantle, which modifies the radial and tangential displacement and stress fields and shifts the overall deformation state. Ultimately, this modification of the elastic response results in a slightly higher $k_2$ value and enhanced shear heating for the same $L$ and $\beta_0$ parameters compared to the incompressible baseline.
Consequently, reproducing the $\Re(k_2)$ value reported by~\citet{park_ios_2025} requires a lower melt fraction in the compressible scenario. Incompressible models (e.g., with $\beta_0 = 10^{-12}\,\mathrm{Pa^{-1}s^{-0.3}}$) can therefore be interpreted as providing a conservative upper limit on the melt fraction expected within Io’s partially molten mantle.

For the compressible mantle case, the tidal Love numbers $h_2$ and $l_2$, and the libration amplitude were also computed. The values of these parameters were specifically examined for the models found to be consistent with the $k_2$ value estimated by~\citet{park_ios_2025}, yielding explicit numerical ranges for each observable (Table~\ref{tab:predictions}). The libration amplitude, in particular, offers an additional potential constraint on the internal structure of the mantle and lithosphere~\citep{https://doi.org/10.1029/2020JE006473}.

\begin{figure}[!ht]
\resizebox{\hsize}{!}
{\includegraphics{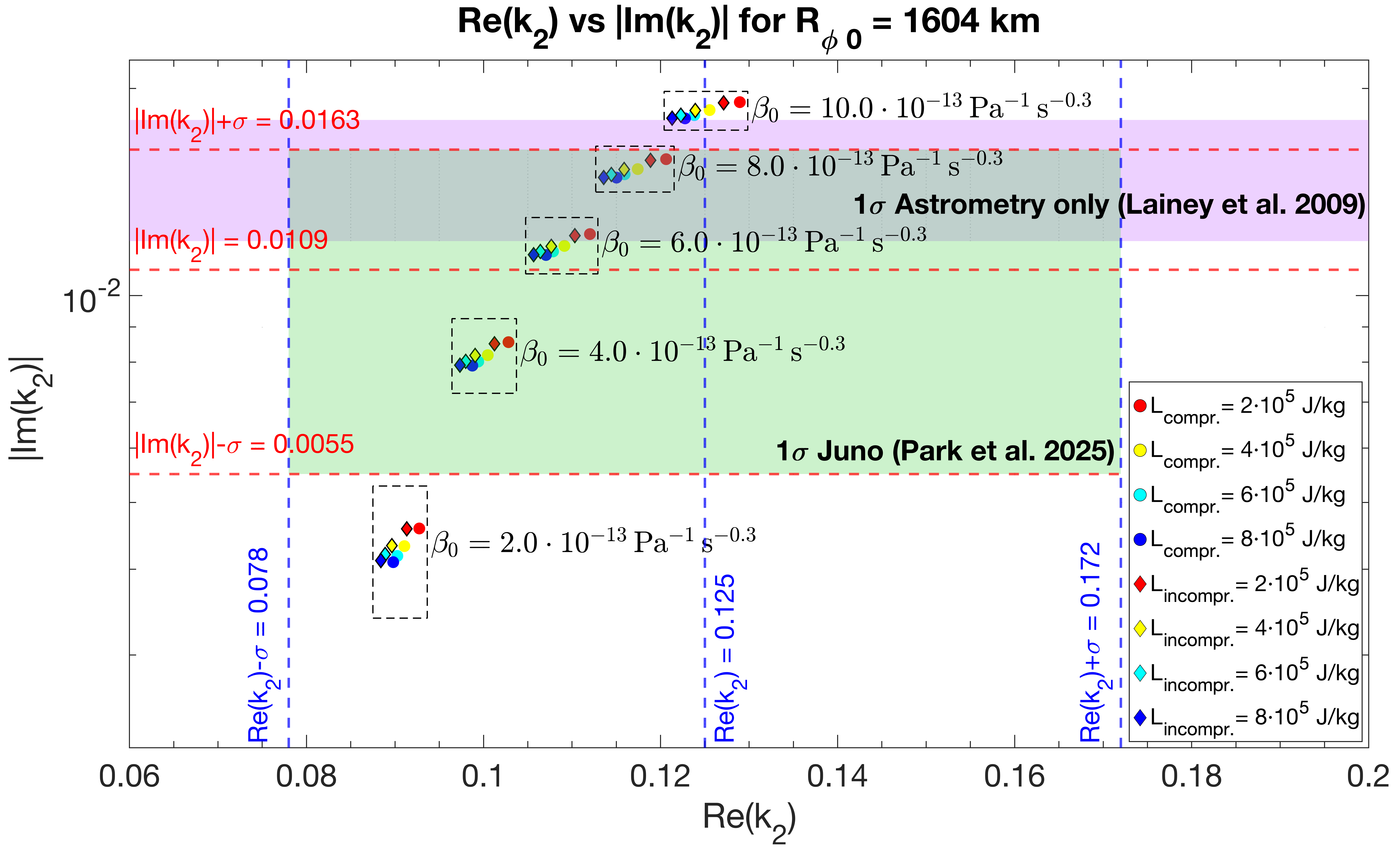}}
\caption{Variation in $\Re(k_2)$ versus $|\Im(k_2)|$. The melting onset radius is fixed at $R_{\phi0} = 1604\,\mathrm{km}$, the value that reproduces the observed $\Re(k_2)$ in the reference incompressible model (i.e., for $\beta_0=10^{-12}\,\mathrm{Pa^{-1}s^{-0.3}}$). Models are grouped by the initial $\beta_0$ value and color coded by latent heat ($L$). The green shaded region represents the $1$$\sigma$ confidence interval of the Juno measurements~\citep{park_ios_2025}, while the purple region corresponds to the $|\Im(k_2)|$ range estimated by~\citet{2009Natur.459..957L}. Circles denote the compressible models; diamonds (for comparison) denote the incompressible models.}
\label{fig13}
\end{figure}

For this specific compressible scenario, the melting onset radius was fixed at $R_{\phi0} = 1604\,\mathrm{km}$, the value previously found to reproduce the observed $\Re(k_2)$. Figure~\ref{fig13} shows the resulting $k_2$ variation as a function of the initial Andrade parameter $\beta_0$ (varied over $10^{-13}$ -- $10^{-12}\,\mathrm{Pa^{-1}s^{-0.3}}$) and the latent heat of fusion $L$ ($2\times10^5$ -- $8\times10^{5}\,\mathrm{J\,kg^{-1}}$). The compressible models consistently yield slightly higher $k_2$ values for the same $L$ and $\beta_0$. The plot thus allows for the identification of models consistent with both estimates, highlighting the effect of mantle compressibility on the Love number.

The y-axis ($|\Im(k_2)|$) is logarithmic, as the values are highly sensitive to the initial $\beta_0$ and span several orders of magnitude. By contrast, the x-axis ($\Re(k_2)$) is linear, as this component is less sensitive to $\beta_0$ variations. Due to this mixed log-linear scaling, the plot is intended only to highlight which models fall within the $1$$\sigma$ confidence interval. It cannot be used to infer which model is "closest" or "most likely," because the vertical (log) and horizontal (linear) distances from the reference value are not directly comparable.

Table \ref{tab:combined_models_4} summarizes the parameter values for models (from Fig.~\ref{fig13}) that fall within the $1$$\sigma$ observational constraints. The parameters listed are the latent heat of fusion ($L$) and the initial Andrade parameter ($\beta_0$), because $R_{\phi0}$ was fixed at $1604\,\mathrm{km}$ for this whole scenario.

\begin{table}[!ht]
\caption{Model parameters ($L$, $\beta_0$) for the compressible scenario (fixed $R_{\phi0} = 1604\,\mathrm{km}$).}
\centering
\begin{tabular}{c c l}
\hline\hline
$\beta_0$ ($10^{-13}\,\mathrm{Pa^{-1}s^{-0.3}}$) & $L$ ($10^{5}\,\mathrm{J\,kg^{-1}}$) & Satisfied Constraints \\
\hline
4 & 2--8 & Juno \\
\hline
\multirow{2}{*}{6} & 2 & Juno + Lainey \\
 & 3--8 & Juno \\
\hline
8 & 2--8 & Juno + Lainey \\
\hline
\end{tabular}
\tablefoot{The last column indicates satisfied observational constraints: "Juno" refers to the $1$$\sigma$ interval for $\Re(k_2)$ and $|\Im(k_2)|$ \protect\citep{park_ios_2025}; "Lainey" refers to the $|\Im(k_2)|$ range by \protect\citet{2009Natur.459..957L}.}
\label{tab:combined_models_4}
\end{table}

\begin{figure}[!ht]
\resizebox{\hsize}{!}
{\includegraphics{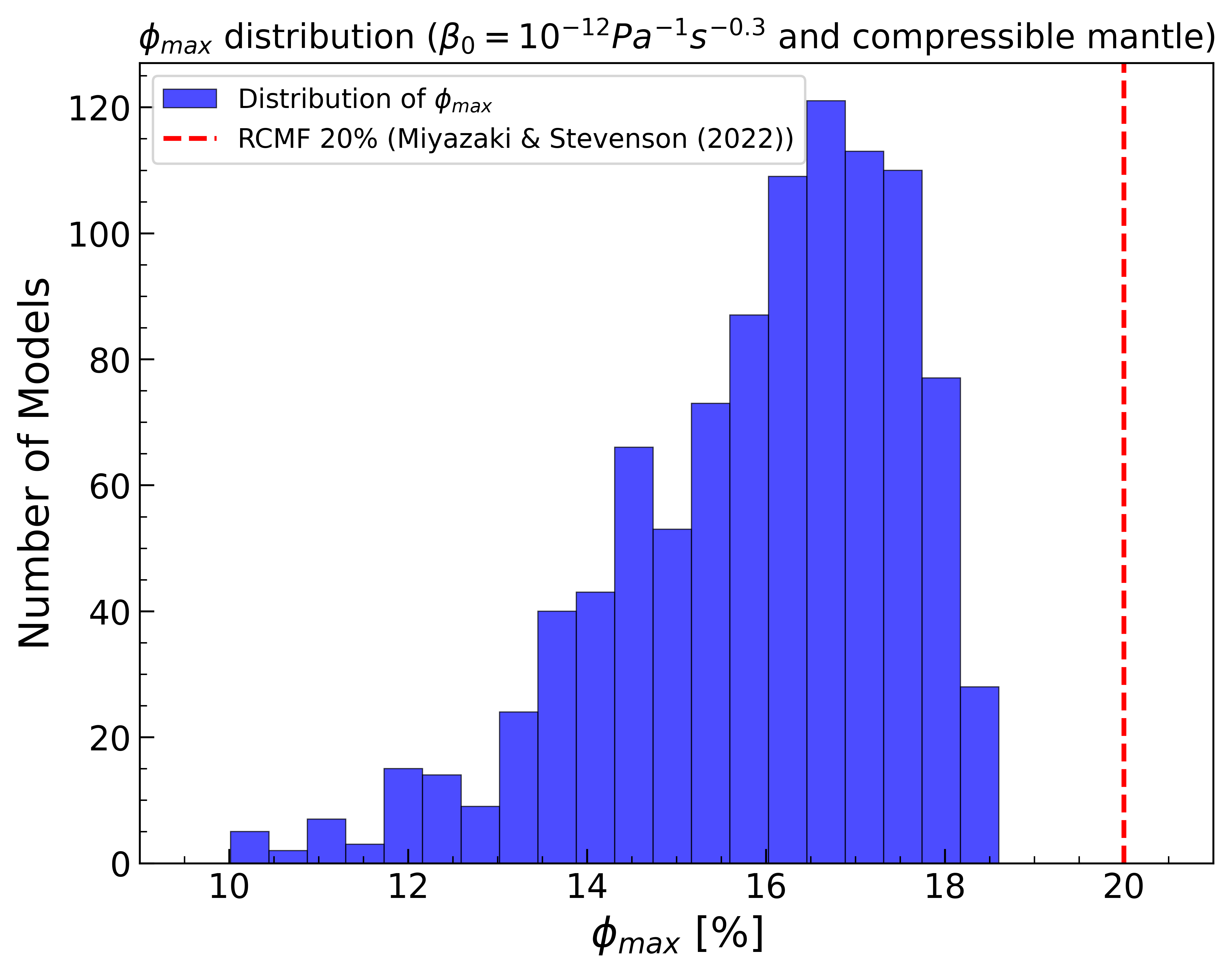}}
\caption{Distribution of peak melt fractions $\phi_{max}$ for 1D models consistent with Juno observations. The figure displays the maximum melt fraction obtained for all compressible model configurations with $\beta_0=10^{-12}\,\mathrm{Pa^{-1}s^{-0.3}}$ that reproduce the real part of the Love number, $\Re(k_2)$, within the $1$$\sigma$ confidence interval estimated by~\citet{park_ios_2025}. The vertical red dashed line indicates the RCMF of $20\%$~\citep{Miyazaki_2022}.}
\label{fig_distribution}
\end{figure}

Even in the compressible case, all 1D models (for different values of $R_{\phi0}$, $L$ and $\beta_0$) within the $1$$\sigma$ uncertainty of the Juno-derived $\Re(k_2)$ value exhibit melt fractions below the RCMF (see Fig.~\ref{fig_distribution} for $\beta_0$ equal to $10^{-12}\,\mathrm{Pa^{-1}s^{-0.3}}$). This result reinforces the main conclusion: Within the adopted rheological framework, the $k_2$ value reported by~\citet{park_ios_2025} is incompatible with a globally uniform magma layer, regardless of the compressibility assumption. The differences between the compressible and incompressible models are minimal, confirming that showing all the results for different values of $R_{\phi0}$ is unnecessary. However, the comparison shows that the melt fractions derived from the incompressible models serve as a conservative upper limit.

Quantitatively, by isolating the subset of models consistent with the 1$\sigma$ observational constraints of $k_2$~\citep{park_ios_2025}, we established explicit and self-consistent bounds for both the real and imaginary components of the Love numbers $h_2$ and $l_2$, as well as for the libration amplitude. These predicted ranges, derived for the compressible scenario with the melting onset fixed at $R_{\phi0} = 1604\,\mathrm{km}$, are summarized in Table~\ref{tab:predictions}.\\
\indent For these Juno-compatible configurations, the real part $\Re(h_2)$, which dictates the radial surface displacement amplitude, is constrained within the range $0.15$--$0.23$, while its dissipative imaginary component $|\Im(h_2)|$ spans from $0.010$ to $0.030$. Similarly, the transverse displacement parameters are bounded, with $\Re(l_2)$ ranging from $0.030$ to $0.065$ and $|\Im(l_2)|$ restricted to the interval $0.0025$--$0.012$. Concurrently, leveraging the empirical bounds of $k_2$, the predicted value of the libration amplitude is resolved within a bracket of $258$--$267\,\mathrm{m}$.\\
\indent The detailed behavioral trends of these parameters plotted against the complex Love number $k_2$, illustrating the clustering of models parameterized by the initial $\beta_0$ and latent heat of melting ($L$), are provided in Appendix \ref{app:companion_parameters} (Figs.~\ref{fig_h2_2_app}, \ref{fig_A_2_app}, and \ref{fig_l2_2_app}).

\begin{table}[!ht]
\caption{Predicted ranges for the companion Love numbers and longitudinal libration amplitude.}
\centering
\footnotesize 
\begin{tabular}{l c c}
\hline\hline
Observable & Component & Range (Juno 1$\sigma$) \\
\hline
\multirow{2}{*}{$h_2$} & $\Re(h_2)$ & $0.15$--$0.23$ \\
                                  & $|\Im(h_2)|$ & $0.010$--$0.030$ \\
\hline
\multirow{2}{*}{$l_2$} & $\Re(l_2)$ & $0.030$--$0.065$ \\
                                      & $|\Im(l_2)|$ & $0.0025$--$0.012$ \\
\hline
Libration Amp. & $A_{libration}$ & $258$--$267\,\mathrm{m}$ \\
\hline
\end{tabular}
\tablefoot{The values were constrained by the 1$\sigma$ Juno $k_2$ observations~\citep{park_ios_2025}. The reported values are derived from the compressible scenario with a fixed melting onset boundary at $R_{\phi0} = 1604\,\mathrm{km}$.}
\label{tab:predictions}
\end{table}

\section{Discussion}\label{disc}

\subsection{Mantle parameters controlling Io’s tidal response}

This parametric study shows that Io’s tidal response, quantified by the Love number $k_2$, is primarily controlled by three mantle parameters: the melting onset radius $R_{\phi_0}$ (i.e., the radius at which the melt fraction exceeds $1\%$), the latent heat of fusion $L$, and the reference Andrade parameter $\beta_0$. The dependence of $k_2$ on these parameters is examined in two configurations: (A) an incompressible mantle with a fixed $\beta_0$; (B) a compressible mantle in which $\beta_0$ is varied for models consistent with the observed $\Re(k_2)$~\citep{park_ios_2025}.

We evaluated each interior configuration by dynamically coupling the Andrade parameter ($\beta$) to the local melt fraction. Rather than confining tidal heating to a single dominant layer, the framework yields models with a localized enhancement of dissipation in the upper mantle, while deep-mantle heating remains active. The primary objective is to identify models consistent with the observational constraints reported by~\citet{park_ios_2025} and to determine the corresponding melt fractions implied by these models.

\subsection{Modeling approach and physical interpretation}\label{modeling_approach}
The model transforms an initial three-layer structure into a configuration featuring a finer radial mantle discretization (i.e., 67 sub-layers). Throughout this process, mantle properties (viscosity, shear modulus, and Andrade parameter $\beta$) are iteratively updated as a function of the local melt fraction. This establishes a self-consistent coupling between rheology and tidal heating, as detailed in Sect.~\ref{melt}. This setup represents an extension of the foundational 1D framework proposed by~\citet{Bierson}, which investigated steady-state tidal heating balanced by melt segregation under an Andrade rheology. While we adopt their core approach regarding the iterative rheological update, the melt-fraction calculation, and the melt-transport formalism, the present formulation advances their methodology with the addition of dynamically varying Andrade parameters and a compressible formulation.

We treated the model parameters as deterministic simulation inputs and did not associate them with formal uncertainties; instead, we explored their influence through a systematic parametric analysis.
The resulting 1D radial profile of the melt fraction does not resolve detailed spatial variations in melt distribution or lateral magma migration.
Nevertheless, a robust interpretation remains possible: As long as the melt fraction $\phi(r)$ remains below the RCMF of $\sim20\%$~\citep{Miyazaki_2022}, i.e., the threshold above which the partially molten layer is expected to become gravitationally unstable and segregate, the models exclude the presence of a laterally continuous global magma ocean.

In particular, we evaluated mantle properties within a shell extending from the CMB ($R_{\text{bot}} = 950\,\mathrm{km}$,~\citealt{park_ios_2025}) to a fixed upper limit of $R_{\text{top}} = 1770\,\mathrm{km}$, a geometric boundary dictated by a rigid, $50\,\mathrm{km}$ thick lithosphere~\citep{2016NatGe...9..429B, Gyalay2024, park_ios_2025}. For the benchmark scenario, the model resolves a distinct melt-rich region expanding from a radius of $1604\,\mathrm{km}$ up to $R_{\text{top}}$, yielding a $166\,\mathrm{km}$ thick asthenospheric layer where the melt fraction consistently exceeds $1\%$ and reaches a peak value of $18\%$.

The maintenance of a subcritical melt state ($\phi < \phi_{\text{RCMF}}$) throughout this $166\,\mathrm{km}$ thick asthenosphere justifies the description of this layer as a "magmatic sponge," a terminology formalized by~\citet{Miyazaki_2022}. Rather than denoting a mechanically decoupled fluid layer or a disconnected melt network, the magmatic sponge represents a texturally equilibrated, porous silicate matrix where grains retain continuous structural contact and finite shear strength. This configuration sustains high viscoelastic tidal dissipation via the Andrade creep function~\citep{Bierson, 2021A&A...650A..72K}, while simultaneously maintaining an interconnected interstitial pore space that accommodates upward buoyant Darcy flow, as verified by our mass-flux analysis.\\
\indent This structural configuration provides an alternative to the asthenospheric models invoked to explain Galileo magnetometry data.~\citet{2011Sci...332.1186K} proposed a global subsurface magma ocean with a thickness exceeding $50\,\mathrm{km}$ and an elevated fluid fraction ($>$$20$-$30\%$) to satisfy planetary electrical conductivity constraints. However, as demonstrated by~\citet{Miyazaki_2022}, high-porosity liquid layers are susceptible to fluid-dynamic instabilities and are prone to rapid mechanical compaction and segregation. Our results show that a vertically extensive, $166\,\mathrm{km}$ thick melt-rich zone with a subcritical peak porosity of $18\%$ avoids mechanical collapse while remaining structurally continuous. Laboratory and theoretical developments~\citep[e.g.,][]{Pommier2008, 2011CoMP..162..637N, 2014Natur.509...81S} support this perspective, demonstrating that strong planetary magnetic induction signals can be successfully generated within subcritical frameworks ($\phi \le 20\%$), provided that dissolved volatile species sufficiently enhance the magma’s electrical conductivity within the interconnected pore network.\\
\indent The radial localization of tidal heating aligns with classic formulations~\citep{1988Icar...75..187S, moore_thermal_2001, 2003JGRE..108.5096M} that confined sub-lithospheric dissipation within an asthenosphere of several tens to a few hundred kilometers. This spatial scale is closely consistent with more recent multi-dimensional numerical frameworks; for instance, parametric studies by~\citet{2020Icar..33513299S} and~\citet{2022Icar..37314737K} systematically explore asthenospheric thicknesses ranging from $50$ up to $200\,\mathrm{km}$. Our benchmark estimate of $166\,\mathrm{km}$ falls squarely within this dynamically viable parameter space. While some modern formulations argue that matching Io's total heat budget requires near- or super-critical melt fractions ($\phi > 25\text{--}30\%$) if dissipation is primarily confined to a highly melted asthenosphere~\citep{2021A&A...650A..72K, 2022Icar..37314737K}, alternative approaches, including the deeper mantle-dominated regimes of~\citet{2022Icar..37314737K} and models invoking lateral rheological variations~\citep{2025NatCo..16.6798V}, confirm that partially molten interiors can satisfy the energy requirements under subcritical melt fractions ($\phi \le 20\%$). In contrast to end-member liquid magma oceans~\citep{2024GeoRL..5107869A}, our benchmark scenario yields a stable $166\,\mathrm{km}$ thick magmatic sponge with an $18\%$ peak melt fraction, demonstrating that coupled viscoelastic solid-frame deformation and porous drainage can sustain steady-state tidal responses without requiring a mechanically decoupled fluid layer or super-critical melt states.\\
\indent These conclusions are conditional on the adopted Andrade rheology and melt-dependent parameterization. Alternative rheological formulations, such as Burgers or Sundberg-Cooper models, may shift the admissible melt fractions and should be explored in future work. Nevertheless, within the present framework, the Juno-compatible solutions consistently remain below the RCMF.

\subsection{Comparison with Juno observations}\label{comparison}
In the case of an incompressible mantle with fixed $\beta_0 = 10^{-12}\,\mathrm{Pa^{-1}s^{-0.3}}$ (see Sect.~\ref{Incompressible}), the dynamically coupled framework yields interior configurations consistent with the observed $\Re(k_2)$ reported by~\citet{park_ios_2025}. In the benchmark case, the solution converges toward a peak melt fraction of $\sim18\%$, and all models consistent with the Juno-derived constraint~\citep{park_ios_2025} maintain melt fractions below the critical threshold (see Fig.~\ref{figS14}). These values remain below the RCMF, supporting a partially molten "magmatic sponge" structure. However, a significant discrepancy remains, as the corresponding $|\Im(k_2)|$ values systematically exceed the $1$$\sigma$ observational constraint.

This discrepancy in $|\Im(k_2)|$ motivated a further sensitivity analysis on the initial Andrade parameter $\beta_0$, while keeping $R_{\phi0}$ fixed. This analysis identifies models consistent with both $\Re(k_2)$ and $|\Im(k_2)|$ within the $1$$\sigma$ Juno confidence interval. These solutions correspond to melt fractions smaller than those obtained in the reference case with fixed $\beta_0$. This result implies that the melt fraction derived from the $\beta_0 = 10^{-12}\,\mathrm{Pa^{-1}s^{-0.3}}$ scenario represents a conservative upper limit compatible with the Juno-derived constraints, further supporting the absence of a global magma ocean.

This conclusion is further supported by the compressible mantle case (see Sect.~\ref{compressible}). For equivalent parameter values, compressible models yield slightly higher $k_2$ values, due to the increased mechanical compliance of the medium, implying that a lower melt fraction is required to match the observations reported by~\citet{park_ios_2025} (see our Fig.~\ref{fig_distribution}).
Quantitatively, when restricting the analysis to the models consistent with the 1$\sigma$ Juno measurement~\citep{park_ios_2025}, the additional tidal parameters, namely the Love numbers $h_2$, $l_2$, as well as the libration amplitude, are constrained within narrow, self-consistent intervals. As summarized in Table~\ref{tab:predictions}, the Love number $h_2$ is constrained to $0.15$--$0.23$ for its real component and $0.010$--$0.030$ for its dissipative imaginary part. Similarly, the $l_2$ parameter is bounded, with $\Re(l_2)$ ranging from $0.030$ to $0.065$ and $|\Im(l_2)|$ restricted to the interval $0.0025$--$0.012$. Concurrently, by leveraging the 1$\sigma$ empirical bounds of both the real and imaginary components of $k_2$, the predicted surface amplitude of the forced longitudinal libration is resolved within a bracket of $258$--$267\,\mathrm{m}$.
Importantly, this predicted interval exhibits a good agreement with the expected libration amplitude computed via the interior models of~\citet{https://doi.org/10.1029/2020JE006473} for scenarios without a global magma ocean. This agreement demonstrates that the independent viscoelastic framework, when constrained by Juno observations, consistently reproduces the geodetic response expected for a mechanically coupled lithosphere-mantle system.

To further evaluate the implications of our model, the a posteriori mass-flux balance provides a critical internal consistency check by comparing the column-integrated thermodynamic melt production rate ($\dot{M}_{\text{gen}}$) with the local extraction capacity at the base of the lithosphere ($\dot{M}_{\text{migr}}$).
We conducted this assessment for the benchmark incompressible models.
The analysis yields $\dot{M}^{\text{max}}_{\text{gen}} \approx 8.01 \times 10^7 \, \text{kg s}^{-1}$ and $\dot{M}^{\text{max}}_{\text{migr}} \approx 3.75 \times 10^8 \, \text{kg s}^{-1}$, indicating that $\dot{M}^{\text{max}}_{\text{migr}}$ exceeds $\dot{M}^{\text{max}}_{\text{gen}}$ by a factor of $\sim4.68$. This result indicates that the system operates in a transport-efficient regime in which permeability is sufficient to extract the generated melt. Consequently, the long-term magmatic flux is energy-limited rather than transport-limited, independent of the spatial distribution of tidal heating. This constraint prevents significant melt accumulation and suggests that the steady-state melt fraction ($\phi_{\text{eq}}$) required to balance production is likely lower than the modeled value. Such a balance supports the interpretation of a stable "magmatic sponge" structure.

 The model-derived thermodynamic melt production rates ($\sim$$10^7\text{--}10^8\,\mathrm{kg\,s^{-1}}$) are consistent with independent observational constraints inferred from the Jovian InfraRed Auroral Mapper (JIRAM) instrument aboard the Juno spacecraft. To explicitly clarify how the mass-flux scaling is performed, we integrated the local volumetric effusion rates ($V_{\text{local}} \approx 50\text{--}500\,\mathrm{m^3\,s^{-1}}$) documented by~\citet{https://doi.org/10.1029/2025JE008940} for long-lived Promethean flow fields into a global planetary mass budget ($\dot{M}_{\text{global}}$) according to the relation
\begin{linenomath*}
\begin{equation}
\dot{M}_{\text{global}} = V_{\text{local}} \cdot N \cdot \rho_{\text{melt}}\,,
\label{eq_mass_flux}
\end{equation}
\end{linenomath*}
where we adopted $N = 100$ to capture the order of magnitude of the global volcanic output, representing a nominal population of persistently active, long-lived volcanic centers~\citep[e.g.,][]{Lopes2004}. Although comprehensive planetary inventories compiled from Galileo and Juno/JIRAM observations identify a higher absolute number of thermal anomalies, ranging from approximately $152$~\citep{Lopes2004} up to $325$ distinct hot spots~\citep{2023GeoRL..5000597Z, 2024NatAs...8...94D, 2025PSJ.....6...84P}, a significant portion of these features represents low-power or transient activity. Concurrently, we adopted a baseline of $N = 100$ as a representative nominal count for major steady-state eruptive centers characterized by continuous, long-term silicate effusion. We took the reference basaltic magma density as $\rho_{\text{melt}} = 2800\,\mathrm{kg\,m^{-3}}$.

Substituting the observational limits from the JIRAM data into this scaling formulation yielded a global empirical mass discharge interval of $\sim 1.4 \times 10^7$ to $1.4 \times 10^8\,\mathrm{kg\,s^{-1}}$. The steady-state production of $8.01 \times 10^7\,\mathrm{kg\,s^{-1}}$ calculated from our models falls near the center of this observationally constrained range. 
Furthermore, these estimates are highly consistent with the synchronized eruption frameworks of~\citet{Mura_Synchronized} for the largest localized event observed on Io, where the flux of a single event ($\sim 10^8\,\mathrm{kg\,s^{-1}}$) was found to be comparable to the total global production predicted by the model.

\subsection{Parameter selection and robustness}
We adopted the parameters linking the tidal heating rate $Q_T(r)$ to the local melt fraction $\phi(r)$ from~\citet{moore_thermal_2001}, consistent with the formulation used by~\citet{Bierson} (see our Sect.~\ref{melt_and_iterative} and Table~\ref{tab:parameters}). The parameters governing the rheological dependence on $\phi(r)$ were taken from the same studies to ensure consistency with previous modeling approaches~\citep{moore_thermal_2001, 2003JGRE..108.5096M,Bierson}. A sensitivity analysis confirmed that varying these parameters within physically reasonable ranges did not significantly affect the main results. The adopted parameter set was therefore retained as a representative and physically consistent configuration, supporting the robustness of the conclusions regarding Io's tidal response (see Figs. \ref{figS4} to \ref{figS12}).

The predicted $k_2$ values were also cross-validated by using~\texttt{ALMA3}~\citep{melini_computing_2022} as an independent tidal solver for tidal Love numbers. As~\texttt{ALMA3} is restricted to incompressible media, this comparison was performed for the incompressible mantle configurations. The agreement between the two approaches confirms that the computed trends in $k_2$ are robust and not artifacts of the numerical implementation, thereby validating the solver performance across the investigated rheological parameterizations.

\section{Conclusions}
This study represents a parametric analysis of Io’s tidal response aimed at constraining the degree of partial melting and the efficiency of tidal energy dissipation within its mantle. Using a three-layer interior model coupled with an iterative melt-dependent rheology, we investigated the sensitivity of the degree-2 Love number $k_2$ to the onset depth of melting ($R_{\phi0}$), the latent heat of fusion ($L$), and the reference Andrade parameter ($\beta_0$). Both incompressible and compressible mantle configurations were analyzed within a physically grounded framework in which the Andrade parameter ($\beta$) is dynamically coupled to the local melt fraction. This approach allowed the mantle rheology to evolve self-consistently with the internal melt structure~\citep{Bierson}. The resulting solutions indicate that a limited mantle enhancement in tidal dissipation emerges naturally, while the deep mantle remains the primary region of tidal heating.

Our results show that under this rheological framework, spherically symmetric 1D interior configurations reproducing the $k_2$ value inferred from Juno observations~\citep{park_ios_2025} consistently require melt fractions strictly below the RCMF.
This conclusion holds across the explored parameter space independent of the specific latent heat, melting onset radius, or mantle compressibility. Although the integrated $k_2$ value alone cannot uniquely constrain the depth distribution of tidal dissipation, the physically consistent framework adopted here systematically converges toward subcritical melt fractions. These results indicate that Io's mantle is partially molten but does not host a laterally continuous global magma ocean. Instead, the inferred melt distribution is consistent with a heterogeneous "magmatic sponge" structure, where the models converge toward a stable $\sim$$150\text{--}166\,\mathrm{km}$ thick asthenospheric layer located immediately beneath a rigid $50\,\mathrm{km}$ thick lithosphere, reaching a peak porosity of $18\%$ within the reference benchmark scenario.

We emphasize that the recovery of subcritical melt fractions is reliant on the adopted Andrade rheology and melt-dependent scaling rather than a model-independent diagnostic. Testing alternative rheological formulations will be necessary to assess the generality of this result.

The physical plausibility of the inferred melt fractions was evaluated via an a posteriori mass-flux balance analysis comparing column-integrated thermodynamic melt production with the local extraction capacity at the upper boundary. Across all Juno-compatible interior configurations, the maximum magmatic migration capacity exceeds the production rate by a factor of a few. Rather than constituting independent evidence against a global magma ocean, this balance serves as an internal consistency check demonstrating that under the assumption of a vertically interconnected and continuous porous network, the steady-state mantle operates in a transport-efficient, energy-limited regime. Within this framework, efficient melt drainage prevents runaway fluid accumulation at the base of the lithosphere, supporting the long-term sustainability of a partially molten subcritical mantle structure.

A sensitivity analysis of the initial Andrade parameter $\beta_0$ showed that this parameter primarily controls the imaginary component of the Love number and the associated tidal dissipation. Models consistent with both the real and imaginary components of $k_2$ within the $1$$\sigma$ Juno confidence interval correspond to melt fractions lower than those obtained for the reference $\beta_0 = 10^{-12}\,\mathrm{Pa^{-1}s^{-0.3}}$ configuration. This reference case therefore represents a conservative upper limit on the melt fraction compatible with the observations. The inclusion of mantle compressibility slightly increases $k_2$ for equivalent parameter values, further reducing the melt fraction required to reproduce the Juno constraint. Additional tidal observables, including the Love numbers $h_2$, $l_2$, and the forced libration amplitude, were computed for models consistent with the $k_2$ constraint, providing additional predictions for Io’s mechanical response.

Future extensions of this framework will include the computation of induced magnetic fields associated with the inferred melt distributions, enabling direct comparison with Galileo and Juno magnetic measurements~\citep{2011Sci...332.1186K, 2018JGRA..123.9286B}. More generally, the methodology developed here provides a quantitative framework for linking internal structure, partial melting, and tidal dissipation in tidally active planetary bodies, with potential applications to other satellites of the outer Solar System.

\section*{Data availability}
Numerical simulations were performed with in-house software written in the general-purpose Python programming language (https://www.python.org). This software is an adapted version of \texttt{CPGC}~\citep{ermakov_california_2024} that iteratively links tidal dissipation to mantle rheology. Figures were produced with custom MATLAB scripts that read the text files containing the tidal quantities of interest. The corresponding dataset is available in the Zenodo repository (https://zenodo.org/records/18877424) of~\citet{paris_2026_18877424}.

\begin{acknowledgements}
     JIRAM was built and supported by the Italian Space Agency (ASI), and is currently supported by the Italian National Institute for Astrophysics (INAF) through a dedicated contract. This work was funded both by ASI, through the Addendum N. 2016-23-H.3-2023 to the ASI-INAF Agreement N. 2016-23-H.0, and by INAF funds (F.O. 1.05.04.87.01). AG acknowledges funding from the Italian Space Agency (ASI) grant n. 2023-60-HH.0.
\end{acknowledgements}

\bibliographystyle{aa}
\bibliography{bibliography}

@PROCEEDINGS{2023ASSL..468.....L,
        title = "{Io: A New View of Jupiter's Moon}",
    booktitle = {Io: A New View of Jupiter's Moon},
         year = 2023,
       editor = {{Lopes}, Rosaly M.~C. and {de Kleer}, Katherine and {Keane}, James Tuttle},
       series = {{ASSL}},
       volume = {468},
        month = jan,
          doi = {10.1007/978-3-031-25670-7},
       adsurl = {https://ui.adsabs.harvard.edu/abs/2023ASSL..468.....L}
}

@article{bierson2024impact,
  title="{The impact of rheology model choices on tidal heating studies}",
  author={Bierson, Carver J},
  journal={\icarus},
  volume={414},
  pages={116026},
  year={2024},
  publisher={Elsevier}
}

@INPROCEEDINGS{2021BAAS...53d.178K,
       author = {{Keane}, James and {Ahern}, Alexandra A. and {Bagenal}, Fran and {Barr Mlinar}, Amy C. and {Basu}, Ko and {Becerra}, Patricio and {Bertrand}, Tanguy and {Beyer}, Ross A. and {Bierson}, Carver J. and {Bland}, Michael T. and {Breuer}, Doris and {Davies}, Ashley G. and {de Kleer}, Katherine and {de Pater}, Imke and {DellaGiustina}, Daniella N. and {Denk}, Tilmann and {Echevarria}, Ariana and {Elder}, Catherine M. and {Feaga}, Lori M. and {Grava}, Cesare and {Gregg}, Patricia M. and {Gregg}, Tracy K.~P. and {Hamilton}, Christopher W. and {Harris}, Camilla D.~K. and {Harris}, Walter M. and {Hay}, Hamish C.~F.~C. and {Hendrix}, Amanda R. and {H{\"o}rst}, Sarah M. and {Huang}, Rowan and {Hughes}, Andr{\'e}a C.~G. and {Jessup}, Kandis Lea and {Jia}, Xianzhe and {Jozwiak}, Lauren M. and {Keane}, James T. and {Kerber}, Laura and {Kestay}, Laszlo P. and {Khurana}, Krishan K. and {Kiefer}, Walter and {Kirchoff}, Michelle R. and {Kite}, Edwin S. and {Klaiber}, Lea and {Klima}, Rachel L. and {Kling}, Corbin L. and {Lainey}, Valery J. and {Lopes}, Rosaly M.~C. and {Lucchetti}, Alice and {Mandt}, Kathleen E. and {Matsuyama}, Isamu and {McCarthy}, Christine and {McEwen}, Alfred S. and {McGrath}, Melissa A. and {Mont{\'e}si}, Laurent G.~J. and {Moses}, Julieanne I. and {Moullet}, Arielle and {N{\'e}non}, Quentin and {Neumann}, Gregory A. and {Neveu}, Marc F. and {Nimmo}, Francis and {Noonan}, John W. and {Pajola}, Maurizio and {Panning}, Mark P. and {Park}, Ryan S. and {Pommier}, Anne and {Quick}, Lynnae C. and {Radebaugh}, Jani and {Rathbun}, Julie A. and {Retherford}, Kurt D. and {Roberts}, James H. and {Roussos}, Elias and {Schenk}, Paul M. and {Schneider}, Nick M. and {Schools}, Joe W. and {Sood}, Rohan and {Spencer}, John R. and {Spencer}, Dan C. and {Steinbr{\"u}gge}, Gregor and {Sulaiman}, Ali H. and {Sutton}, Sarah S. and {Trinh}, Antony and {Tsang}, Constantine C.~C. and {Vertesi}, Janet and {Vorburger}, Audrey and {Westlake}, Joseph H. and {Williams}, David A.},
        title = "{The Science Case for Io Exploration}",
    booktitle = {Bulletin of the American Astronomical Society},
         year = 2021,
       volume = {53},
        month = may,
          eid = {178},
        pages = {178},
          doi = {10.3847/25c2cfeb.f844ca0e},
       adsurl = {https://ui.adsabs.harvard.edu/abs/2021BAAS...53d.178K}
}

@ARTICLE{Hamilton,
       author = {{Breuer}, Doris and {Hamilton}, Christopher W. and {Khurana}, Krishan},
        title = "{The Internal Structure of Io}",
      journal = {Eleme},
         year = 2022,
        month = dec,
       volume = {18},
       number = {6},
        pages = {385-390},
          doi = {10.2138/gselements.18.6.385},
       adsurl = {https://ui.adsabs.harvard.edu/abs/2022Eleme..18..385B}
}

@ARTICLE{Takeuchi,
       author = {{Takeuchi}, Hitoshi and {Saito}, Masanori},
        title = "{Seismic Surface Waves}",
      journal = {MComP},
         year = 1972,
        month = jan,
       volume = {11},
        pages = {217-295},
          doi = {10.1016/B978-0-12-460811-5.50010-6},
       adsurl = {https://ui.adsabs.harvard.edu/abs/1972MCPAR..11..217T}
}

@ARTICLE{Bierson,
author = {Bierson, C. J. and Nimmo, F.},
title = {A test for Io's magma ocean: Modeling tidal dissipation with a partially molten mantle},
journal = {JGRE},
volume = {121},
number = {11},
pages = {2211-2224},
doi = {https://doi.org/10.1002/2016JE005005},
url = {https://agupubs.onlinelibrary.wiley.com/doi/abs/10.1002/2016JE005005},
eprint = {https://agupubs.onlinelibrary.wiley.com/doi/pdf/10.1002/2016JE005005},
year = {2016}
}

@ARTICLE{park_ios_2025,
       author = {{Park}, R.~S. and {Jacobson}, R.~A. and {Gomez Casajus}, L. and {Nimmo}, F. and {Ermakov}, A.~I. and {Keane}, J.~T. and {McKinnon}, W.~B. and {Stevenson}, D.~J. and {Akiba}, R. and {Idini}, B. and {Buccino}, D.~R. and {Magnanini}, A. and {Parisi}, M. and {Tortora}, P. and {Zannoni}, M. and {Mura}, A. and {Durante}, D. and {Iess}, L. and {Connerney}, J.~E.~P. and {Levin}, S.~M. and {Bolton}, S.~J.},
        title = "{Io's tidal response precludes a shallow magma ocean}",
      journal = {Nat},
         year = 2025,
        month = feb,
       volume = {638},
       number = {8049},
        pages = {69-73},
          doi = {10.1038/s41586-024-08442-5},
       adsurl = {https://ui.adsabs.harvard.edu/abs/2025Natur.638...69P}
}

@ARTICLE{moore_thermal_2001,
	title = {The {Thermal} {State} of {Io}},
	volume = {154},
	issn = {0019-1035},
	url = {https://www.sciencedirect.com/science/article/pii/S0019103501967399},
	doi = {https://doi.org/10.1006/icar.2001.6739},
	number = {2},
	journal = {\icarus},
	author = {Moore, William B.},
	year = {2001},
	pages = {548--550},
}

@ARTICLE{Tobie,
       author = {{Tobie}, G. and {Auclair-Desrotour}, P. and {B{\v{e}}hounkov{\'a}}, M. and {Kervazo}, M. and {Sou{\v{c}}ek}, O. and {Kalousov{\'a}}, K.},
        title = "{Tidal Deformation and Dissipation Processes in Icy Worlds}",
      journal = {\ssr},
         year = 2025,
        month = feb,
       volume = {221},
       number = {1},
          eid = {6},
        pages = {6},
          doi = {10.1007/s11214-025-01136-y},
       adsurl = {https://ui.adsabs.harvard.edu/abs/2025SSRv..221....6T}
}

@ARTICLE{beuthe_spatial_2013,
	title = {Spatial patterns of tidal heating},
	volume = {223},
	issn = {0019-1035},
	url = {https://www.sciencedirect.com/science/article/pii/S0019103512004745},
	doi = {https://doi.org/10.1016/j.icarus.2012.11.020},
	number = {1},
	journal = {\icarus},
	author = {Beuthe, Mikael},
	year = {2013},
	pages = {308--329},
}

@ARTICLE{melini_computing_2022,
	title = {On computing viscoelastic {Love} numbers for general planetary models: the {ALMA3} code},
	volume = {231},
	issn = {0956-540X},
	url = {https://doi.org/10.1093/gji/ggac263},
	doi = {10.1093/gji/ggac263},
	number = {3},
	journal = {GeoJI},
	author = {Melini, D and Saliby, C and Spada, G},
	month = jul,
	year = {2022},
	note = {},
	pages = {1502--1517},
}

@ARTICLE{https://doi.org/10.1029/2007JE002908,
author = {Efroimsky, Michael and Lainey, Valéry},
title = {Physics of bodily tides in terrestrial planets and the appropriate scales of dynamical evolution},
journal = {JGRE},
volume = {112},
number = {E12},
pages = {},
doi = {https://doi.org/10.1029/2007JE002908},
url = {https://agupubs.onlinelibrary.wiley.com/doi/abs/10.1029/2007JE002908},
eprint = {},
year = {2007}
}

@misc{ermakov_california_2024,
	title = {California {Planetary} {Geophysics} {Code}},
	url = {https://doi.org/10.5281/zenodo.14029354},
	publisher = {Zenodo},
	author = {Ermakov, Anton and Akiba, Ryunosuke},
	month = nov,
	year = {2024},
	doi = {10.5281/zenodo.14029354},
}

@ARTICLE{Miyazaki_2022,
doi = {10.3847/PSJ/ac9cd1},
url = {https://doi.org/10.3847/PSJ/ac9cd1},
year = {2022},
month = {nov},
publisher = {The American Astronomical Society},
volume = {3},
number = {11},
pages = {256},
author = {Miyazaki, Yoshinori and Stevenson, David J.},
title = {A Subsurface Magma Ocean on Io: Exploring the Steady State of Partially Molten Planetary Bodies},
journal = {PSJ}
}

@ARTICLE{2009Natur.459..957L,
       author = {{Lainey}, Val{\'e}ry and {Arlot}, Jean-Eudes and {Karatekin}, {\"O}zg{\"u}r and {van Hoolst}, Tim},
        title = "{Strong tidal dissipation in Io and Jupiter from astrometric observations}",
      journal = {Nat},
         year = 2009,
        month = jun,
       volume = {459},
       number = {7249},
        pages = {957-959},
          doi = {10.1038/nature08108},
       adsurl = {https://ui.adsabs.harvard.edu/abs/2009Natur.459..957L}
}

@ARTICLE{2015ApJS..218...22T,
       author = {{Tyler}, Robert H. and {Henning}, Wade G. and {Hamilton}, Christopher W.},
        title = "{Tidal Heating in a Magma Ocean within Jupiter's Moon Io}",
      journal = {ApJS},
         year = 2015,
        month = jun,
       volume = {218},
       number = {2},
          eid = {22},
        pages = {22},
          doi = {10.1088/0067-0049/218/2/22},
       adsurl = {https://ui.adsabs.harvard.edu/abs/2015ApJS..218...22T}
}

@ARTICLE{https://doi.org/10.1029/2020JE006473,
author = {Van Hoolst, Tim and Baland, Rose-Marie and Trinh, Antony and Yseboodt, Marie and Nimmo, Francis},
title = {The Librations, Tides, and Interior Structure of Io},
journal = {JGRE},
volume = {125},
number = {8},
pages = {e2020JE006473},
doi = {https://doi.org/10.1029/2020JE006473},
url = {https://agupubs.onlinelibrary.wiley.com/doi/abs/10.1029/2020JE006473},
eprint = {https://agupubs.onlinelibrary.wiley.com/doi/pdf/10.1029/2020JE006473},
year = {2020}
}

@ARTICLE{1994JGR....9917095V,
       author = {{Veeder}, Glenn J. and {Matson}, Dennis L. and {Johnson}, Torrence V. and {Blaney}, Diana L. and {Goguen}, Jay D.},
        title = "{Io's heat flow from infrared radiometry: 1983-1993}",
      journal = {JGRE},
         year = 1994,
        month = aug,
       volume = {99},
       number = {E8},
        pages = {17095-17162},
          doi = {10.1029/94JE00637},
       adsurl = {https://ui.adsabs.harvard.edu/abs/1994JGR....9917095V}
}

@ARTICLE{2024NatAs...8...94D,
       author = {{Davies}, Ashley Gerard and {Perry}, Jason E. and {Williams}, David A. and {Nelson}, David M.},
        title = "{Io's polar volcanic thermal emission indicative of magma ocean and shallow tidal heating models}",
      journal = {NatAs},
         year = 2024,
        month = jan,
       volume = {8},
       number = {1},
        pages = {94-100},
          doi = {10.1038/s41550-023-02123-5},
archivePrefix = {arXiv},
       eprint = {2310.12382},
 primaryClass = {astro-ph.EP},
       adsurl = {https://ui.adsabs.harvard.edu/abs/2024NatAs...8...94D}
}

@ARTICLE{2023GeoRL..5000597Z,
       author = {{Zambon}, F. and {Mura}, A. and {Lopes}, R.~M.~C. and {Rathbun}, J. and {Tosi}, F. and {Sordini}, R. and {Noschese}, R. and {Ciarniello}, M. and {Cicchetti}, A. and {Adriani}, A. and {Agostini}, L. and {Filacchione}, G. and {Grassi}, D. and {Piccioni}, G. and {Plainaki}, C. and {Sindoni}, G. and {Turrini}, D. and {Brooks}, S. and {Hansen-Koharcheck}, C. and {Bolton}, S.},
        title = "{Io Hot Spot Distribution Detected by Juno/JIRAM}",
      journal = {GeoRL},
         year = 2023,
        month = jan,
       volume = {50},
       number = {1},
          eid = {e2022GL100597},
        pages = {e2022GL100597},
          doi = {10.1029/2022GL100597},
       adsurl = {https://ui.adsabs.harvard.edu/abs/2023GeoRL..5000597Z}
}

@ARTICLE{2024ComEE...5..340M,
       author = {{Mura}, Alessandro and {Tosi}, Federico and {Zambon}, Francesca and {Lopes}, Rosaly M.~C. and {Mouginis-Mark}, Peter J. and {Becker}, Heidi and {Filacchione}, Gianrico and {Migliorini}, Alessandra and {Hansen}, Candice. J. and {Adriani}, Alberto and {Altieri}, Francesca and {Bolton}, Scott and {Cicchetti}, Andrea and {Di Mico}, Elisa and {Grassi}, Davide and {Noschese}, Raffaella and {Moirano}, Alessandro and {Pettine}, Madeline and {Piccioni}, Giuseppe and {Plainaki}, Christina and {Rathbun}, Julie and {Sordini}, Roberto and {Sindoni}, Giuseppe},
        title = "{Hot rings on Io observed by Juno/JIRAM}",
      journal = {ComEE},
         year = 2024,
        month = jun,
       volume = {5},
       number = {1},
          eid = {340},
        pages = {340},
          doi = {10.1038/s43247-024-01486-5},
       adsurl = {https://ui.adsabs.harvard.edu/abs/2024ComEE...5..340M}
}

@ARTICLE{2007Icar..192..491K,
       author = {{Keszthelyi}, Laszlo and {Jaeger}, Windy and {Milazzo}, Moses and {Radebaugh}, Jani and {Davies}, Ashley Gerard and {Mitchell}, Karl L.},
        title = "{New estimates for Io eruption temperatures: Implications for the interior}",
      journal = {\icarus},
         year = 2007,
        month = dec,
       volume = {192},
       number = {2},
        pages = {491-502},
          doi = {10.1016/j.icarus.2007.07.008},
       adsurl = {https://ui.adsabs.harvard.edu/abs/2007Icar..192..491K}
}

@ARTICLE{2011Sci...332.1186K,
       author = {{Khurana}, Krishan K. and {Jia}, Xianzhe and {Kivelson}, Margaret G. and {Nimmo}, Francis and {Schubert}, Gerald and {Russell}, Christopher T.},
        title = "{Evidence of a Global Magma Ocean in Io{\textquoteright}s Interior}",
      journal = {Sci},
         year = 2011,
        month = jun,
       volume = {332},
       number = {6034},
        pages = {1186},
          doi = {10.1126/science.1201425},
       adsurl = {https://ui.adsabs.harvard.edu/abs/2011Sci...332.1186K}
}

@ARTICLE{2020Icar..33513299S,
       author = {{Steinke}, T. and {Hu}, H. and {H{\"o}ning}, D. and {van der Wal}, W. and {Vermeersen}, B.},
        title = "{Tidally induced lateral variations of Io's interior}",
      journal = {\icarus},
         year = 2020,
        month = jan,
       volume = {335},
          eid = {113299},
        pages = {113299},
          doi = {10.1016/j.icarus.2019.05.001},
       adsurl = {https://ui.adsabs.harvard.edu/abs/2020Icar..33513299S}
}

@ARTICLE{2013E&PSL.361..272H,
       author = {{Hamilton}, Christopher W. and {Beggan}, Ciar{\'a}n D. and {Still}, Susanne and {Beuthe}, Mikael and {Lopes}, Rosaly M.~C. and {Williams}, David A. and {Radebaugh}, Jani and {Wright}, William},
        title = "{Spatial distribution of volcanoes on Io: Implications for tidal heating and magma ascent}",
      journal = {E\&SS},
         year = 2013,
        month = jan,
       volume = {361},
        pages = {272-286},
          doi = {10.1016/j.epsl.2012.10.032},
       adsurl = {https://ui.adsabs.harvard.edu/abs/2013E&PSL.361..272H}
}

@ARTICLE{2025PSJ.....6...84P,
       author = {{Perry}, Jason E. and {Davies}, Ashley Gerard and {Williams}, David A. and {Nelson}, David M.},
        title = "{Hot Spot Detections and Volcanic Changes on Io during the Juno Epoch: Orbits PJ5 to PJ55}",
      journal = {PSJ},
         year = 2025,
        month = apr,
       volume = {6},
       number = {4},
          eid = {84},
        pages = {84},
          doi = {10.3847/PSJ/adbae3},
       adsurl = {https://ui.adsabs.harvard.edu/abs/2025PSJ.....6...84P}
}

@ARTICLE{1988Icar...75..187S,
       author = {{Segatz}, M. and {Spohn}, T. and {Ross}, M.~N. and {Schubert}, G.},
        title = "{Tidal dissipation, surface heat flow, and figure of viscoelastic models of Io}",
      journal = {\icarus},
         year = 1988,
        month = aug,
       volume = {75},
       number = {2},
        pages = {187-206},
          doi = {10.1016/0019-1035(88)90001-2},
       adsurl = {https://ui.adsabs.harvard.edu/abs/1988Icar...75..187S}
}

@ARTICLE{1990Icar...85..309R,
       author = {{Ross}, M.~N. and {Schubert}, G. and {Spohn}, T. and {Gaskell}, R.~W.},
        title = "{Internal structure of Io and the global distribution of its topography}",
      journal = {\icarus},
         year = 1990,
        month = jun,
       volume = {85},
       number = {2},
        pages = {309-325},
          doi = {10.1016/0019-1035(90)90119-T},
       adsurl = {https://ui.adsabs.harvard.edu/abs/1990Icar...85..309R}
}

@ARTICLE{2018AJ....156..207R,
       author = {{Rathbun}, Julie A. and {Lopes}, Rosaly M.~C. and {Spencer}, John R.},
        title = "{The Global Distribution of Active Ionian Volcanoes and Implications for Tidal Heating Models}",
      journal = {\apj},
         year = 2018,
        month = nov,
       volume = {156},
       number = {5},
          eid = {207},
        pages = {207},
          doi = {10.3847/1538-3881/aae370},
archivePrefix = {arXiv},
       eprint = {1810.00725},
 primaryClass = {astro-ph.EP},
       adsurl = {https://ui.adsabs.harvard.edu/abs/2018AJ....156..207R}
}

@ARTICLE{2001JGR...10632963A,
       author = {{Anderson}, John D. and {Jacobson}, Robert A. and {Lau}, Eunice L. and {Moore}, William B. and {Schubert}, Gerald},
        title = "{Io's gravity field and interior structure}",
      journal = {JGRE},
         year = 2001,
        month = dec,
       volume = {106},
       number = {E12},
        pages = {32963-32970},
          doi = {10.1029/2000JE001367},
       adsurl = {https://ui.adsabs.harvard.edu/abs/2001JGR...10632963A}
}

@INCOLLECTION{2007iag..book...89M,
       author = {{Moore}, William B. and {Schubert}, Gerald and {Anderson}, John D. and {Spencer}, John R.},
        title = "{The interior of Io}",
    booktitle = {Io After Galileo: A New View of Jupiter's Volcanic Moon},
         year = 2007,
    publisher = {Springer},
       editor = {{Lopes}, Rosaly M.~C. and {Spencer}, John R.},
        pages = {89},
          doi = {10.1007/978-3-540-48841-5_5},
       adsurl = {https://ui.adsabs.harvard.edu/abs/2007iag..book...89M}
}

@INCOLLECTION{2004jpsm.book..281S,
       author = {{Schubert}, G. and {Anderson}, J.~D. and {Spohn}, T. and {McKinnon}, W.~B.},
        title = "{Interior composition, structure and dynamics of the Galilean satellites}",
    booktitle = {Jupiter. The Planet, Satellites and Magnetosphere},
         year = 2004,
       editor = {{Bagenal}, Fran and {Dowling}, Timothy E. and {McKinnon}, William B.},
       publisher = {Cambridge University Press},
       volume = {1},
        pages = {281-306},
       adsurl = {https://ui.adsabs.harvard.edu/abs/2004jpsm.book..281S}
}

@ARTICLE{2010Icar..209..651B,
       author = {{Baland}, Rose-Marie and {Van Hoolst}, Tim},
        title = "{Librations of the Galilean satellites: The influence of global internal liquid layers}",
      journal = {\icarus},
         year = 2010,
        month = oct,
       volume = {209},
       number = {2},
        pages = {651-664},
          doi = {10.1016/j.icarus.2010.04.004},
       adsurl = {https://ui.adsabs.harvard.edu/abs/2010Icar..209..651B}
}

@INPROCEEDINGS{1997tiph.conf..345S,
       author = {{Spohn}, T.},
        title = "{Tides of Io}",
    booktitle = {Tidal Phenomena.},
         year = 1997,
       editor = {{Wilhelm}, Helmut and {Zurm}, Walter and {Wenzel}, Hans-Georg},
        month = jan,
        pages = {345},
       adsurl = {https://ui.adsabs.harvard.edu/abs/1997tiph.conf..345S}
}

@ARTICLE{2018JGRA..123.9286B,
       author = {{Bl{\"o}cker}, Aljona and {Saur}, Joachim and {Roth}, Lorenz and {Strobel}, Darrell F.},
        title = "{MHD Modeling of the Plasma Interaction With Io's Asymmetric Atmosphere}",
      journal = {JGRA},
         year = 2018,
        month = nov,
       volume = {123},
       number = {11},
        pages = {9286-9311},
          doi = {10.1029/2018JA025747},
       adsurl = {https://ui.adsabs.harvard.edu/abs/2018JGRA..123.9286B}
}

@ARTICLE{2013Icar..223..621N,
       author = {{Noyelles}, Beno{\^\i}t},
        title = "{The rotation of Io predicted by the Poincar{\'e}-Hough model}",
      journal = {\icarus},
         year = 2013,
        month = mar,
       volume = {223},
       number = {1},
        pages = {621-624},
          doi = {10.1016/j.icarus.2012.12.008},
archivePrefix = {arXiv},
       eprint = {1203.3867},
 primaryClass = {astro-ph.EP},
       adsurl = {https://ui.adsabs.harvard.edu/abs/2013Icar..223..621N}
}

@ARTICLE{2019Icar..321..126R,
       author = {{Rovira-Navarro}, Marc and {Rieutord}, Michel and {Gerkema}, Theo and {Maas}, Leo R.~M. and {van der Wal}, Wouter and {Vermeersen}, Bert},
        title = "{Do tidally-generated inertial waves heat the subsurface oceans of Europa and Enceladus?}",
      journal = {\icarus},
         year = 2019,
        month = mar,
       volume = {321},
        pages = {126-140},
          doi = {10.1016/j.icarus.2018.11.010},
       adsurl = {https://ui.adsabs.harvard.edu/abs/2019Icar..321..126R}
}

@ARTICLE{2019JGRE..124.2198R,
       author = {{Rekier}, J. and {Trinh}, A. and {Triana}, S.~A. and {Dehant}, V.},
        title = "{Internal Energy Dissipation in Enceladus's Subsurface Ocean From Tides and Libration and the Role of Inertial Waves}",
      journal = {JGRE},
         year = 2019,
        month = aug,
       volume = {124},
       number = {8},
        pages = {2198-2212},
          doi = {10.1029/2019JE005988},
archivePrefix = {arXiv},
       eprint = {1904.02487},
 primaryClass = {physics.geo-ph},
       adsurl = {https://ui.adsabs.harvard.edu/abs/2019JGRE..124.2198R}
}

@ARTICLE{2013P&SS...78....1G,
       author = {{Grasset}, O. and {Dougherty}, M.~K. and {Coustenis}, A. and {Bunce}, E.~J. and {Erd}, C. and {Titov}, D. and {Blanc}, M. and {Coates}, A. and {Drossart}, P. and {Fletcher}, L.~N. and {Hussmann}, H. and {Jaumann}, R. and {Krupp}, N. and {Lebreton}, J.-P. and {Prieto-Ballesteros}, O. and {Tortora}, P. and {Tosi}, F. and {Van Hoolst}, T.},
        title = "{JUpiter ICy moons Explorer (JUICE): An ESA mission to orbit Ganymede and to characterise the Jupiter system}",
      journal = {P\&SS},
         year = 2013,
        month = apr,
       volume = {78},
        pages = {1-21},
          doi = {10.1016/j.pss.2012.12.002},
       adsurl = {https://ui.adsabs.harvard.edu/abs/2013P&SS...78....1G}
}

@ARTICLE{2020NatCo..11.1311H,
       author = {{Howell}, Samuel M. and {Pappalardo}, Robert T.},
        title = "{NASA's Europa Clipper{\textemdash}a mission to a potentially habitable ocean world}",
      journal = {NatCo},
         year = 2020,
        month = mar,
       volume = {11},
          eid = {1311},
        pages = {1311},
          doi = {10.1038/s41467-020-15160-9},
       adsurl = {https://ui.adsabs.harvard.edu/abs/2020NatCo..11.1311H}
}

@ARTICLE{1979Sci...203..892P,
       author = {{Peale}, S.~J. and {Cassen}, P. and {Reynolds}, R.~T.},
        title = "{Melting of Io by Tidal Dissipation}",
      journal = {Sci},
         year = 1979,
        month = mar,
       volume = {203},
       number = {4383},
        pages = {892-894},
          doi = {10.1126/science.203.4383.892},
       adsurl = {https://ui.adsabs.harvard.edu/abs/1979Sci...203..892P}
}

@ARTICLE{2024GeoRL..5107869A,
       author = {{Ayg{\"u}n}, B. and {{\v{C}}adek}, O.},
        title = "{Tidal Heating in a Subsurface Magma Ocean on Io Revisited}",
      journal = {GeoRL},
         year = 2024,
        month = may,
       volume = {51},
       number = {10},
          eid = {e2023GL107869},
        pages = {e2023GL107869},
          doi = {10.1029/2023GL107869},
       adsurl = {https://ui.adsabs.harvard.edu/abs/2024GeoRL..5107869A}
}

@ARTICLE{2021A&A...650A..72K,
       author = {{Kervazo}, M. and {Tobie}, G. and {Choblet}, G. and {Dumoulin}, C. and {B{\v{e}}hounkov{\'a}}, M.},
        title = "{Solid tides in Io's partially molten interior. Contribution of bulk dissipation}",
      journal = {A\&A},
         year = 2021,
        month = jun,
       volume = {650},
          eid = {A72},
        pages = {A72},
          doi = {10.1051/0004-6361/202039433},
       adsurl = {https://ui.adsabs.harvard.edu/abs/2021A&A...650A..72K}
}

@ARTICLE{2018ApJ...857...98R,
       author = {{Renaud}, Joe P. and {Henning}, Wade G.},
        title = "{Increased Tidal Dissipation Using Advanced Rheological Models: Implications for Io and Tidally Active Exoplanets}",
      journal = {AJ},
         year = 2018,
        month = apr,
       volume = {857},
       number = {2},
          eid = {98},
        pages = {98},
          doi = {10.3847/1538-4357/aab784},
archivePrefix = {arXiv},
       eprint = {1707.06701},
 primaryClass = {astro-ph.EP},
       adsurl = {https://ui.adsabs.harvard.edu/abs/2018ApJ...857...98R}
}

@ARTICLE{ROSS1985391,
title = {Tidally forced viscous heating in a partially molten Io},
journal = {\icarus},
volume = {64},
number = {3},
pages = {391-400},
year = {1985},
issn = {0019-1035},
doi = {https://doi.org/10.1016/0019-1035(85)90063-6},
url = {https://www.sciencedirect.com/science/article/pii/0019103585900636},
author = {M.N. Ross and G. Schubert}
}

@ARTICLE{2003JGRE..108.5096M,
       author = {{Moore}, W.~B.},
        title = "{Tidal heating and convection in Io}",
      journal = {JGRE},
         year = 2003,
        month = aug,
       volume = {108},
       number = {E8},
          eid = {5096},
        pages = {5096},
          doi = {10.1029/2002JE001943},
       adsurl = {https://ui.adsabs.harvard.edu/abs/2003JGRE..108.5096M}
}

@ARTICLE{2022Icar..37314737K,
       author = {{Kervazo}, Mathilde and {Tobie}, Gabriel and {Choblet}, Ga{\"e}l and {Dumoulin}, Caroline and {B{\v{e}}hounkov{\'a}}, Marie},
        title = "{Inferring Io's interior from tidal monitoring}",
      journal = {\icarus},
         year = 2022,
        month = feb,
       volume = {373},
          eid = {114737},
        pages = {114737},
          doi = {10.1016/j.icarus.2021.114737},
       adsurl = {https://ui.adsabs.harvard.edu/abs/2022Icar..37314737K}
}

@ARTICLE{2025NatCo..16.6798V,
       author = {{Veenstra}, Allard and {Rovira-Navarro}, Marc and {Steinke}, Teresa and {Davies}, Ashley Gerard and {van der Wal}, Wouter},
        title = "{Lateral melt variations induce shift in Io's peak tidal heating}",
      journal = {NatCo},
         year = 2025,
        month = jul,
       volume = {16},
       number = {1},
          eid = {6798},
        pages = {6798},
          doi = {10.1038/s41467-025-62059-4},
       adsurl = {https://ui.adsabs.harvard.edu/abs/2025NatCo..16.6798V}
}

@ARTICLE{2011CoMP..162..637N,
       author = {{Ni}, Huaiwei and {Keppler}, Hans and {Behrens}, Harald},
        title = "{Electrical conductivity of hydrous basaltic melts: implications for partial melting in the upper mantle}",
      journal = {CoMP},
         year = 2011,
        month = sep,
       volume = {162},
       number = {3},
        pages = {637-650},
          doi = {10.1007/s00410-011-0617-4},
       adsurl = {https://ui.adsabs.harvard.edu/abs/2011CoMP..162..637N}
}

@ARTICLE{2014Natur.509...81S,
       author = {{Sifr{\'e}}, David and {Gard{\'e}s}, Emmanuel and {Massuyeau}, Malcolm and {Hashim}, Leila and {Hier-Majumder}, Saswata and {Gaillard}, Fabrice},
        title = "{Electrical conductivity during incipient melting in the oceanic low-velocity zone}",
      journal = {Nat},
         year = 2014,
        month = may,
       volume = {509},
       number = {7498},
        pages = {81-85},
          doi = {10.1038/nature13245},
       adsurl = {https://ui.adsabs.harvard.edu/abs/2014Natur.509...81S}
}

@ARTICLE{1998Sci...279.1514S,
       author = {{Schenk}, Paul M. and {Bulmer}, Mark H.},
        title = "{Origin of Mountains on Io by Thrust Faulting and Large-Scale Mass Movements}",
      journal = {Sci},
         year = 1998,
        month = mar,
       volume = {279},
        pages = {1514},
          doi = {10.1126/science.279.5356.1514},
       adsurl = {https://ui.adsabs.harvard.edu/abs/1998Sci...279.1514S}
}

@ARTICLE{MEI2002491,
title = {Influence of melt on the creep behavior of olivine–basalt aggregates under hydrous conditions},
journal = {E\&PSL},
volume = {201},
number = {3},
pages = {491-507},
year = {2002},
issn = {0012-821X},
doi = {https://doi.org/10.1016/S0012-821X(02)00745-8},
url = {https://www.sciencedirect.com/science/article/pii/S0012821X02007458},
author = {S. Mei and W. Bai and T. Hiraga and D.L. Kohlstedt}
}

@ARTICLE{1980JGR....85.5173M,
       author = {{Mavko}, G.~M.},
        title = "{Velocity and attenuation in partially molten rocks}",
      journal = {JGRB},
         year = 1980,
        month = oct,
       volume = {85},
        pages = {5173-5189},
          doi = {10.1029/JB085iB10p05173},
       adsurl = {https://ui.adsabs.harvard.edu/abs/1980JGR....85.5173M}
}

@ARTICLE{SCOTT2006177,
title = {The effect of large melt fraction on the deformation behavior of peridotite},
journal = {E\&PSL},
volume = {246},
number = {3},
pages = {177-187},
year = {2006},
issn = {0012-821X},
doi = {https://doi.org/10.1016/j.epsl.2006.04.027},
url = {https://www.sciencedirect.com/science/article/pii/S0012821X06003311},
author = {T. Scott and D.L. Kohlstedt}
}

@incollection{LESHER2015113,
title = {Chapter 5 - Thermodynamic and Transport Properties of Silicate Melts and Magma},
editor = {Haraldur Sigurdsson},
booktitle = {The Encyclopedia of Volcanoes (Second Edition)},
publisher = {Academic Press},
edition = {Second Edition},
address = {Amsterdam},
pages = {113-141},
year = {2015},
isbn = {978-0-12-385938-9},
doi = {https://doi.org/10.1016/B978-0-12-385938-9.00005-5},
url = {https://www.sciencedirect.com/science/article/pii/B9780123859389000055},
author = {Charles E. Lesher and Frank J. Spera}
}

@ARTICLE{https://doi.org/10.1029/GL008i004p00313,
author = {O'Reilly, Thomas C. and Davies, Geoffrey F.},
title = {Magma transport of heat on Io: A mechanism allowing a thick lithosphere},
journal = {GeoRL},
volume = {8},
number = {4},
pages = {313-316},
doi = {https://doi.org/10.1029/GL008i004p00313},
url = {https://agupubs.onlinelibrary.wiley.com/doi/abs/10.1029/GL008i004p00313},
eprint = {https://agupubs.onlinelibrary.wiley.com/doi/pdf/10.1029/GL008i004p00313},
year = {1981}
}

@ARTICLE{10.1093/petrology/25.3.713,
    author = {McKenzie, Dan},
    title = {The Generation and Compaction of Partially Molten Rock},
    journal = {JPet},
    volume = {25},
    number = {3},
    pages = {713-765},
    year = {1984},
    month = {08},
    issn = {0022-3530},
    doi = {10.1093/petrology/25.3.713},
    url = {https://doi.org/10.1093/petrology/25.3.713},
    eprint = {https://academic.oup.com/petrology/article-pdf/25/3/713/4198364/25-3-713.pdf},
}

@ARTICLE{https://doi.org/10.1029/2025JE008940,
author = {Lopes, Rosaly M. C. and Mura, Alessandro and Mouginis-Mark, Peter and Radebaugh, Jani and Tosi, Federico and Zambon, Francesca and Sordini, Roberto and Schenk, Paul and Bolton, Scott},
title = {Thermal Characteristics of Active Lava Flows on Io Observed by the JIRAM Instrument on Juno},
journal = {JGRE},
volume = {130},
number = {11},
pages = {e2025JE008940},
doi = {https://doi.org/10.1029/2025JE008940},
url = {https://agupubs.onlinelibrary.wiley.com/doi/abs/10.1029/2025JE008940},
eprint = {https://agupubs.onlinelibrary.wiley.com/doi/pdf/10.1029/2025JE008940},
year = {2025}
}

@ARTICLE{2004JGRB..109.6201J,
       author = {{Jackson}, Ian and {Faul}, Ulrich H. and {Fitz Gerald}, John D. and {Tan}, Ben H.},
        title = "{Shear wave attenuation and dispersion in melt-bearing olivine polycrystals: 1. Specimen fabrication and mechanical testing}",
      journal = {JGRB},
         year = 2004,
        month = jun,
       volume = {109},
       number = {B6},
          eid = {B06201},
        pages = {B06201},
          doi = {10.1029/2003JB002406},
       adsurl = {https://ui.adsabs.harvard.edu/abs/2004JGRB..109.6201J}
}

@ARTICLE{2025FrASS..1268185T,
       author = {{Tosi}, Federico and {Mura}, Alessandro and {Zambon}, Francesca},
        title = "{Re-evaluating Io's volcanic heat flow: critical limitations in Juno/JIRAM M-band analysis}",
      journal = {FrASS},
         year = 2025,
        month = nov,
       volume = {12},
          eid = {1668185},
        pages = {1668185},
          doi = {10.3389/fspas.2025.1668185},
archivePrefix = {arXiv},
       eprint = {2412.04321},
 primaryClass = {astro-ph.EP},
       adsurl = {https://ui.adsabs.harvard.edu/abs/2025FrASS..1268185T}
}

@article{Mura_Synchronized,
author = {Mura, A. and Lopes, R. and Nimmo, F. and Bolton, S. and Ermakov, A. and Keane, J. T. and Tosi, F. and Zambon, F. and Sordini, R. and Radebaugh, J. and Rathbun, J. and McKinnon, W. and Goossens, S. and Paris, M. and Mirino, M. and Cicchetti, A. and Piccioni, G. and Noschese, R. and Sindoni, G. and Plainaki, C.},
title = {Synchronized Eruptions on Io: Possible Evidence of Interconnected Subsurface Magma Reservoirs},
journal = {JGRE},
volume = {131},
number = {1},
pages = {e2025JE009047},
doi = {https://doi.org/10.1029/2025JE009047},
url = {https://agupubs.onlinelibrary.wiley.com/doi/abs/10.1029/2025JE009047},
eprint = {https://agupubs.onlinelibrary.wiley.com/doi/pdf/10.1029/2025JE009047},
year = {2026}
}

@ARTICLE{2016NatGe...9..429B,
       author = {{Bland}, Michael T. and {McKinnon}, William B.},
        title = "{Mountain building on Io driven by deep faulting}",
      journal = {NatGe},
         year = 2016,
        month = jun,
       volume = {9},
       number = {6},
        pages = {429-432},
          doi = {10.1038/ngeo2711},
       adsurl = {https://ui.adsabs.harvard.edu/abs/2016NatGe...9..429B}
}

@article{https://doi.org/10.1029/2010JE003664,
author = {Castillo-Rogez, Julie C. and Efroimsky, Michael and Lainey, Valéry},
title = {The tidal history of Iapetus: Spin dynamics in the light of a refined dissipation model},
journal = {JGRE},
volume = {116},
number = {E9},
pages = {},
doi = {https://doi.org/10.1029/2010JE003664},
url = {},
eprint = {},
year = {2011}
}

@ARTICLE{2012ApJ...746..150E,
       author = {{Efroimsky}, Michael},
        title = "{Tidal Dissipation Compared to Seismic Dissipation: In Small Bodies, Earths, and Super-Earths}",
      journal = {AJ},
         year = 2012,
        month = feb,
       volume = {746},
       number = {2},
          eid = {150},
        pages = {150},
          doi = {10.1088/0004-637X/746/2/150},
archivePrefix = {arXiv},
       eprint = {1105.3936},
 primaryClass = {astro-ph.EP},
       adsurl = {https://ui.adsabs.harvard.edu/abs/2012ApJ...746..150E}
}

@article{JACKSON2010151,
title = {Grainsize-sensitive viscoelastic relaxation in olivine: Towards a robust laboratory-based model for seismological application},
journal = {PEPI},
volume = {183},
number = {1},
pages = {151-163},
year = {2010},
note = {Special Issue on Deep Slab and Mantle Dynamics},
issn = {0031-9201},
doi = {https://doi.org/10.1016/j.pepi.2010.09.005},
url = {https://www.sciencedirect.com/science/article/pii/S0031920110001871},
author = {Ian Jackson and Ulrich H. Faul}
}

@ARTICLE{2002JGRB..107.2360J,
       author = {{Jackson}, Ian and {Fitz Gerald}, John D. and {Faul}, Ulrich H. and {Tan}, Ben H.},
        title = "{Grain-size-sensitive seismic wave attenuation in polycrystalline olivine}",
      journal = {JGRB},
         year = 2002,
        month = dec,
       volume = {107},
       number = {B12},
          eid = {2360},
        pages = {2360},
          doi = {10.1029/2001JB001225},
       adsurl = {https://ui.adsabs.harvard.edu/abs/2002JGRB..107.2360J}
}

@article{Gyalay2024,
author = {Gyalay, S. and Nimmo, F.},
title = {Io's Long-Wavelength Topography as a Probe for a Subsurface Magma Ocean},
journal = {GeoRL},
volume = {51},
number = {9},
pages = {e2023GL106993},
doi = {https://doi.org/10.1029/2023GL106993},
url = {https://agupubs.onlinelibrary.wiley.com/doi/abs/10.1029/2023GL106993},
eprint = {https://agupubs.onlinelibrary.wiley.com/doi/pdf/10.1029/2023GL106993},
year = {2024}
}

@misc{paris_2026_18877424,
  author       = {Paris, Matteo},
  title        = {Simulated Dataset of Io's Love Numbers and
                   Libration Amplitudes for a Dynamically Coupled
                   Melt-Rheology Framework
                  },
  month        = mar,
  year         = 2026,
  publisher    = {Zenodo},
  doi          = {10.5281/zenodo.18877424},
  url          = {https://doi.org/10.5281/zenodo.18877424},
}

@ARTICLE{Walterova2023,
       author = {{Walterov{\'a}}, Michaela and {Plesa}, Ana-Catalina and {Wagner}, Frank W. and {Breuer}, Doris},
        title = "{Andrade rheology in planetary science}",
      journal = {esoar},
         year = 2023,
        month = jul,
       volume = {992},
          eid = {169008299.99203183},
        pages = {169008299.99203183},
          doi = {10.22541/essoar.169008299.99203183/v1},
       adsurl = {https://ui.adsabs.harvard.edu/abs/2023esoar.99203183W}
}

@article{Pommier2008,
author = {Pommier, A. and Gaillard, F. and Pichavant, M. and Scaillet, B.},
title = {Laboratory measurements of electrical conductivities of hydrous and dry Mount Vesuvius melts under pressure},
journal = {JGRB},
volume = {113},
number = {B5},
pages = {},
doi = {https://doi.org/10.1029/2007JB005269},
year = {2008}
}

@ARTICLE{Lopes2004,
       author = {{Lopes}, Rosaly M.~C. and {Kamp}, Lucas W. and {Smythe}, William D. and {Mouginis-Mark}, Peter and {Kargel}, Jeff and {Radebaugh}, Jani and {Turtle}, Elizabeth P. and {Perry}, Jason and {Williams}, David A. and {Carlson}, R.~W. and {Dout{\'e}}, S. and {the Galileo NIMS} and {SSI Teams}},
        title = "{Lava lakes on Io: observations of Io's volcanic activity from Galileo NIMS during the 2001 fly-bys}",
      journal = {\icarus},
         year = 2004,
        month = may,
       volume = {169},
       number = {1},
        pages = {140-174},
          doi = {10.1016/j.icarus.2003.11.013},
       adsurl = {https://ui.adsabs.harvard.edu/abs/2004Icar..169..140L}
}

@article{Comstock2003,
author = {Comstock, Robert L. and Bills, Bruce G.},
title = {A solar system survey of forced librations in longitude},
journal = {JGRE},
volume = {108},
number = {E9},
pages = {},
doi = {https://doi.org/10.1029/2003JE002100},
url = {https://agupubs.onlinelibrary.wiley.com/doi/abs/10.1029/2003JE002100},
year = {2003}
}

\begin{appendix}
\nolinenumbers

\section{Andrade dissipation model}\label{andrade}
In the Andrade model, the effective forcing frequency ($\omega_P$) is related to the actual forcing frequency via an Arrhenius term (see Eq.~\ref{eq3}) that accounts for the changing response as a function of temperature~\citep{JACKSON2010151,Bierson, park_ios_2025}. This term is taken to be $3.16$~\citep{park_ios_2025}, representing mantle material that is close to the melting point.
\begin{linenomath*}
\begin{equation}
\begin{split}
    \omega_P&=\frac{2\pi}{X}\,,\\
     X&=P\exp\biggl[\frac{E_b}{R_g}\biggl(\frac{1}{T}-\frac{1}{T_r}\biggr)\biggr]\,.  
\end{split}
\label{eq3}
\end{equation}
\end{linenomath*}
Here $R_g$ is the ideal gas constant, $E_b$ is the activation energy, and $T_r$ is a reference temperature of $1374.15\,\mathrm{K}$~\citep{2004JGRB..109.6201J}.

\section{Selection of benchmark models}\label{app1}
\begin{figure}[!ht]
\resizebox{\hsize}{!}
{\includegraphics{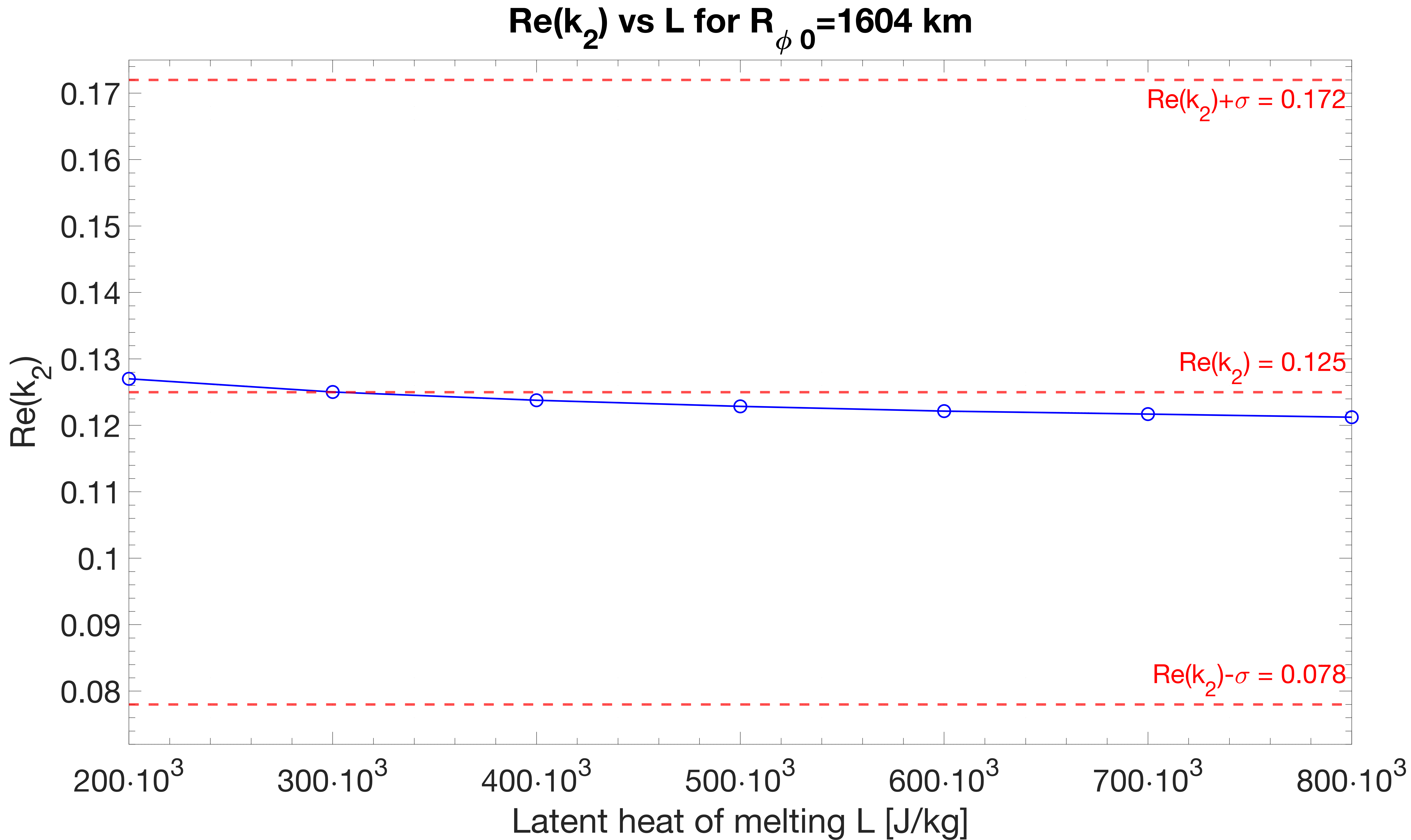}}
\caption{Real part of $k_2$ ($\Re(k_2)$) as a function of latent heat of fusion ($L$). This case assumes a fixed melting onset radius ($R_{\phi 0}$) and a fixed initial $\beta_0 = 10^{-12}\,\mathrm{Pa^{-1}s^{-0.3}}$ . The model results intersect the reference value from~\citet{park_ios_2025} at $L=3\times10^5\,\mathrm{J\,kg^{-1}}$.}
\label{figS2}
\end{figure}

This appendix details the identification of the benchmark mantle configurations discussed in the main text. Within the broader parametric analysis, we chose specific combinations of the melting onset radius ($R_{\phi0}$) and latent heat of fusion ($L$) because they successfully reproduced the real part of the Love number, $\Re(k_2)$, estimated by~\citet{park_ios_2025}. 
We derived that a melting onset radius of $R_{\phi0} = 1604\,\mathrm{km}$ is required to match the observations (using the same reference $\beta_0$). Figure~\ref{figS2} shows the variation in $\Re(k_2)$ as a function of $L$ for this configuration. In this case, the intersection with the reference value occurs at a lower latent heat, specifically $L=3\times10^5\,\mathrm{J\,kg^{-1}}$.

\FloatBarrier

\section{Iterative loop}\label{iterativeloop}
\begin{figure*}[ht]
\centering
\includegraphics[width=\textwidth]{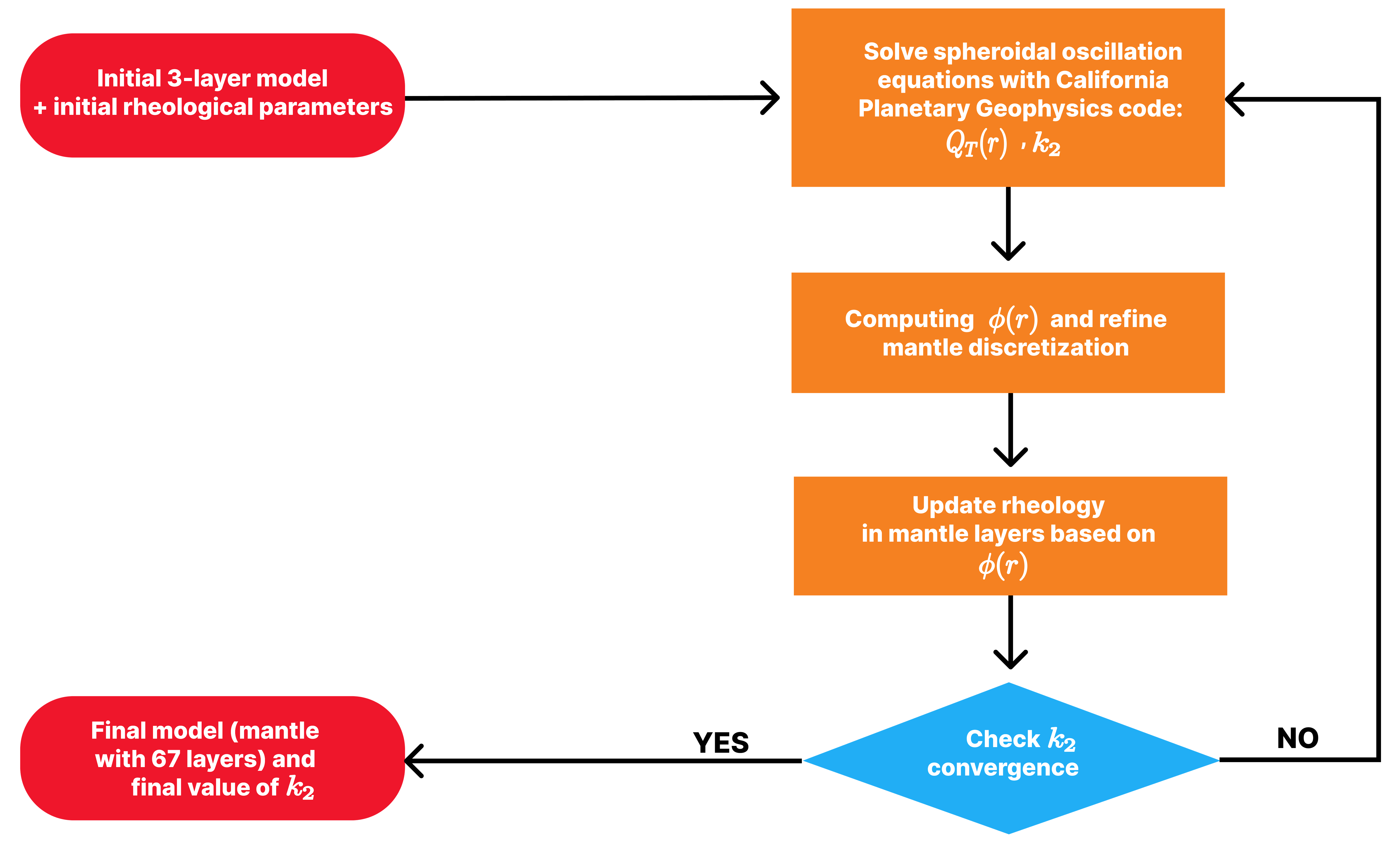}
\caption{Schematic of the iterative refinement loop. The model evolves from an initial three-layer structure to a final, converged configuration where the mantle is resolved into 67 sub-layers. This final model yields the complex Love number $k_2$ and the radial profiles of the volumetric tidal heating rate $Q_T(r)$ and melt fraction $\phi(r)$. This entire procedure is repeated for each pair of ($R_{\phi0}$, $L$) values in the parametric grid search.}
\label{loop}
\end{figure*}

Figure~\ref{loop} illustrates the iterative workflow employed in this study. Each block corresponds to one of the steps described in Sect.~\ref{melt_and_iterative}.
\FloatBarrier

\section{Fixed physical, rheological, and numerical parameters used in the modeling framework}\label{apptab}
While $L$, $\beta_0$, and  $R_{\phi 0}$ are varied in the parametric study, the values listed in Table \ref{tab:parameters} are held constant or represent the baseline constants adopted from literature. The corresponding literature sources are indicated in the table footnotes.
\begin{table}[!ht]
\caption{Physical and rheological parameters used in the simulations.}
\centering
\begin{tabular}{l c c}
\hline\hline
Parameter Description & Symbol & Value \\
\hline
\multicolumn{3}{l}{Melt Parameters} \\
Initial melt velocity & $v_0$ & $0\,\mathrm{m\,s^{-1}}$\;\tablefootmark{a}\\
Initial melt fraction & $\phi_0$ & $1\%$\;\tablefootmark{b}\\
Velocity scale & $\gamma$ & $5 \times 10^{-7}\,\mathrm{m\,s^{-1}}$\;\tablefootmark{a}\\
Permeability exponent & $m$ & $3$\;\tablefootmark{a}\\
Viscosity melt exponent & $\alpha_{melt}$ & $26$\;\tablefootmark{c}\\
Rigidity melt slope & $c$ & $67/15$\;\tablefootmark{d}\\
Beta melt exponent & $n_\beta$ & $20$\;\tablefootmark{d}\\
Critical melt fraction (RCMF) & $\phi_{crit}$ & $\sim 20\%$\;\tablefootmark{e}\\
\hline
\multicolumn{3}{l}{Rheology Parameters} \\
Ref. Lamé coefficient & $\lambda$ & $223.\bar{3}\,\mathrm{GPa}$\;\tablefootmark{f}\\
Reference temperature & $T_r$ & $1374.15\,\mathrm{K}$\;\tablefootmark{g}\\
Activation energy & $E_b$ & $3 \times 10^5\,\mathrm{J\,mol^{-1}}$\;\tablefootmark{b}\\
Alpha parameter & $\alpha$ & $0.3$\;\tablefootmark{h}\\
\hline
\multicolumn{3}{l}{Model Structure} \\
Lithosphere density & $\rho_{Cr}$ & $3259\,\mathrm{kg\,m^{-3}}$\;\tablefootmark{h}\\
Mantle density & $\rho_{M}$ & $3259\,\mathrm{kg\,m^{-3}}$\;\tablefootmark{h}\\
Melt density & $\rho_{melt}$ & $2800\,\mathrm{kg\,m^{-3}}$\;\tablefootmark{a}\\
Core density & $\rho_{C}$ & $5150\,\mathrm{kg\,m^{-3}}$\;\tablefootmark{h}\\
Lithosphere viscosity & $\eta_{Cr}$ & $10^{25}\,\mathrm{Pa\;s}$\;\tablefootmark{h}\\
Mantle initial viscosity & $\eta_{0}$ & $10^{21}\,\mathrm{Pa\;s}$\;\tablefootmark{h}\\
Lithosphere shear modulus & $\mu_{Cr}$ & $40\,\mathrm{GPa}$\;\tablefootmark{h}\\
Mantle initial shear modulus & $\mu_{0}$ & $40\,\mathrm{GPa}$\;\tablefootmark{h}\\
Mantle initial bulk modulus & $K_{0}$ & $250\,\mathrm{GPa}$\;\tablefootmark{f}\\
Surface radius & $R$ & $1820\,\mathrm{km}$\;\tablefootmark{h}\\
Crustal thickness  & $h_{Cr}$ & $50\,\mathrm{km}$\;\tablefootmark{h}\\
Core radius & $R_C$ & $950\,\mathrm{km}$\;\tablefootmark{h}\\
Lithosphere base radius & $R_{top}$ & $1770\,\mathrm{km}$\;\tablefootmark{h}\\
Core-mantle boundary & $R_{bot}$ & $950\,\mathrm{km}$\;\tablefootmark{h}\\
\hline
\end{tabular}
\tablefoottext{a}{\citet{moore_thermal_2001}}\,
\tablefoottext{b}{\citet{Bierson}}\,
\tablefoottext{c}{\citet{MEI2002491}}\,
\tablefoottext{d}{\citet{1980JGR....85.5173M}}\,
\tablefoottext{e}{\citet{Miyazaki_2022}}\,
\tablefoottext{f}{\citet{Tobie}}\,
\tablefoottext{g}{\citet{2004JGRB..109.6201J}}\,
\tablefoottext{h}{\citet{park_ios_2025}}\,
\label{tab:parameters}
\end{table}
\FloatBarrier

\section{Sensitivity analysis and numerical convergence}\label{Sens}
To ensure the reliability of the findings, we tested the sensitivity of the model to auxiliary physical parameters and numerical initialization values.
Specifically, Figs. \ref{figS4} to \ref{figS12} display the results for the benchmark incompressible models identified in Appendix \ref{app1} (i.e., those reproducing the observed real part of $k_2$).
Figures~\ref{figS4} and \ref{figS6} address the uncertainty in the melt transport parameters (velocity scale $\gamma$ ) used in the 1D equations from~\citet{moore_thermal_2001}. Figure~\ref{figS8} validates the choice of the initial melt fraction guess used to initialize the iterative process. Figures~\ref{figS10} and \ref{figS12} demonstrate the numerical stability of the coupled tidal-thermal model by showing the convergence of the Love number $k_2$ over successive iterations.

\begin{figure}[!ht]
\resizebox{\hsize}{!}
{\includegraphics{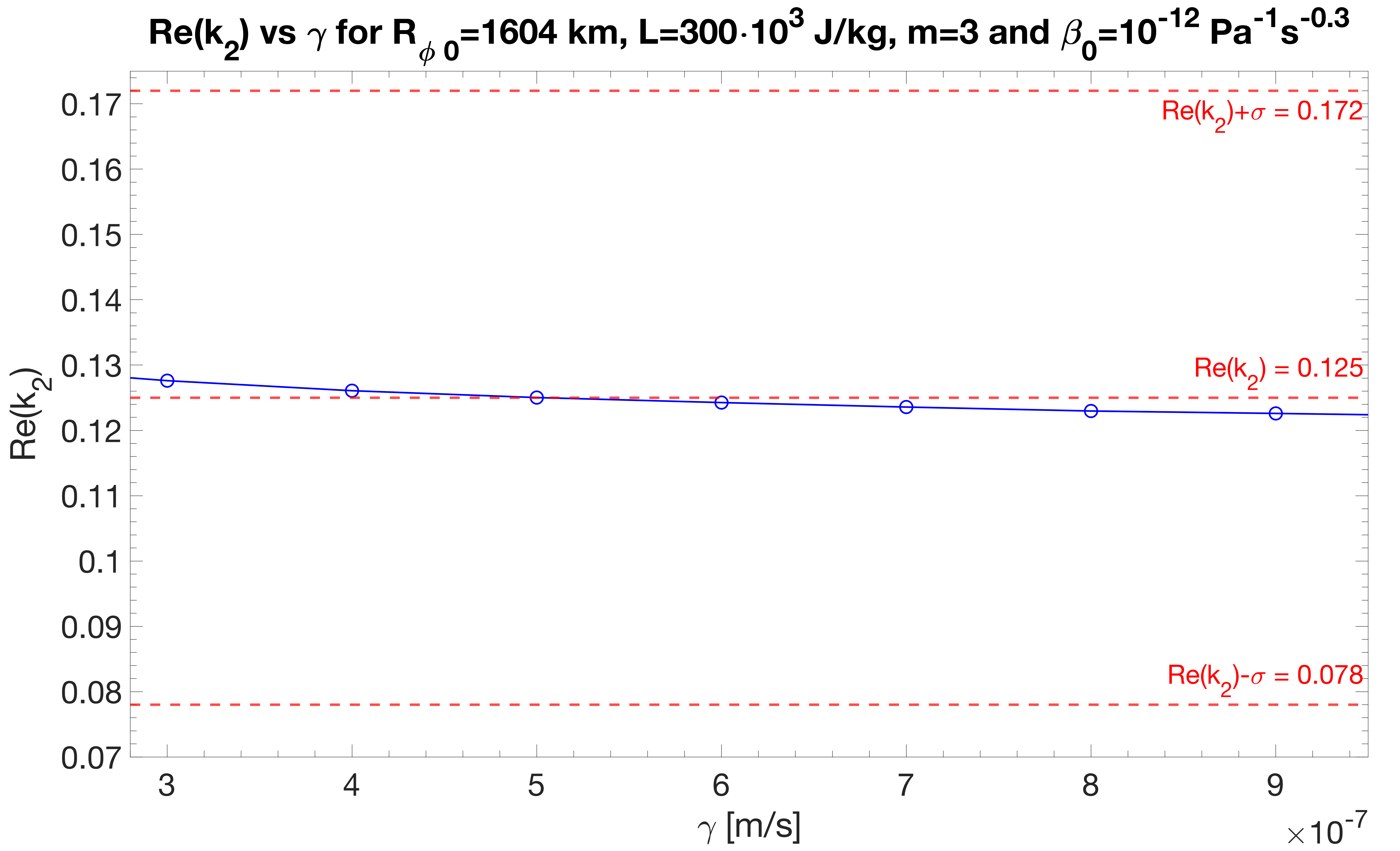}}
\caption{Sensitivity analysis of the real Love number, $\Re(k_2)$, to the melt transport velocity scale $\gamma$. This analysis considers the benchmark incompressible model ($R_{\phi0} = 1604\,\mathrm{km}$). The plot illustrates the variation in $\Re(k_2)$ over a range of $\gamma$ values relative to the standard value used in this study ($\gamma=5\times10^{-7}\,\mathrm{m/s}$) . The limited variation observed confirms the robustness of the model results against uncertainties in percolation efficiency.}
\label{figS4}
\end{figure}

\begin{figure}[!ht]
\resizebox{\hsize}{!}
{\includegraphics{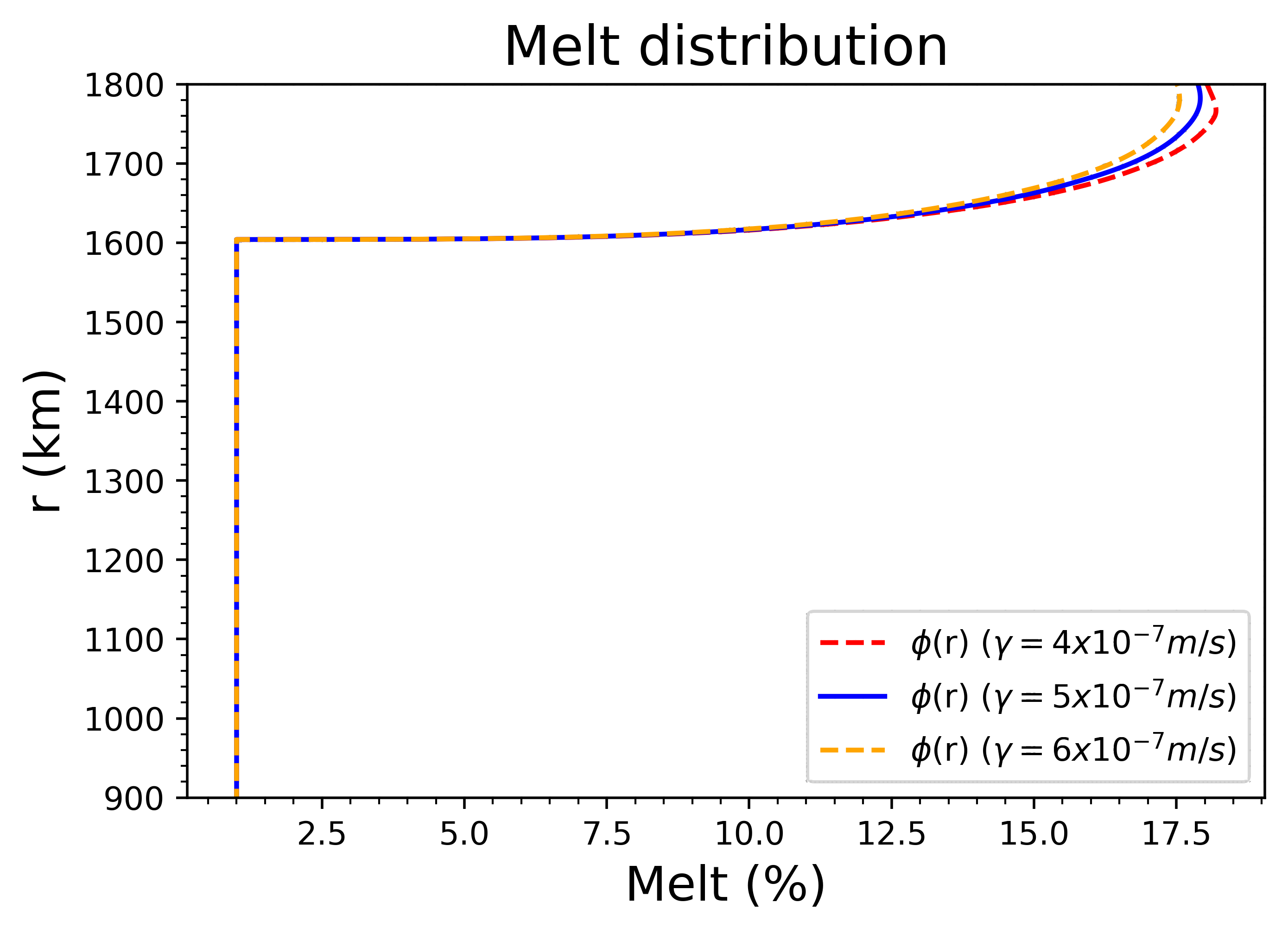}}
\caption{Sensitivity analysis of the computed melt fraction, $\phi(r)$, to the melt transport velocity scale $\gamma$. This analysis considers the benchmark incompressible model ($R_{\phi0} = 1604\,\mathrm{km}$). The plot compares the final radial profiles obtained using the standard parameter set (solid blue lines) with those obtained using alternative values for the velocity scale $\gamma$ (dashed red and orange lines). The near-perfect overlap of the profiles confirms that the model results are robust against uncertainties in melt transport efficiency.}
\label{figS6}
\end{figure}

\begin{figure}[!ht]
\resizebox{\hsize}{!}
{\includegraphics{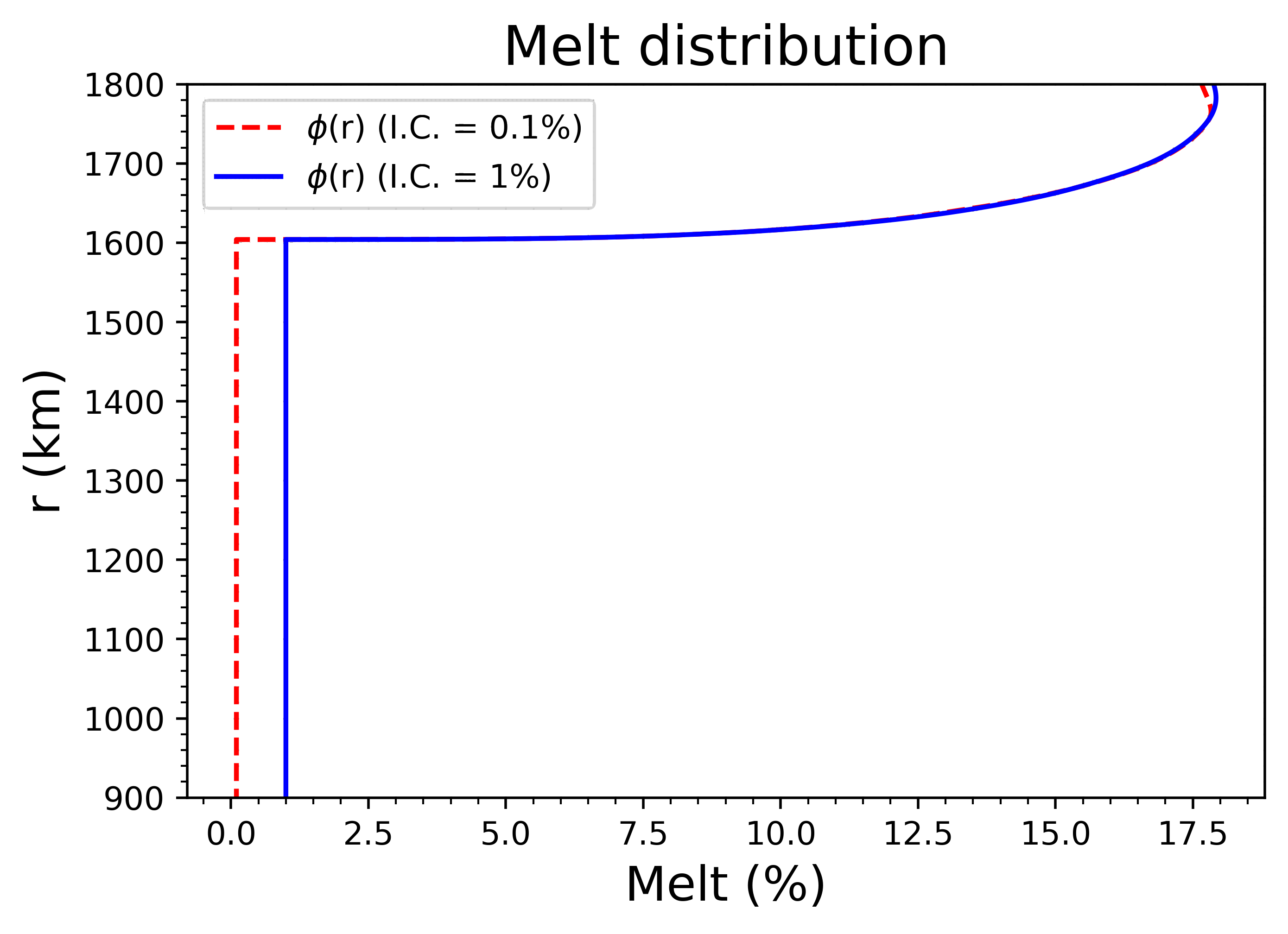}}
\caption{Effect of the initial melt fraction guess on the final converged model. This analysis considers the benchmark incompressible model ($R_{\phi0} = 1604\,\mathrm{km}$). The plot compares the final radial profile of the melt fraction obtained starting from an initial guess of $\phi_0=1\%$ (solid blue curves) with the result obtained starting from $\phi_0=0.1\%$  (dashed red curves). The negligible difference between the profiles justifies the choice of $1\%$ used to reduce computational cost.}
\label{figS8}
\end{figure}

\begin{figure}[!ht]
\resizebox{\hsize}{!}
{\includegraphics{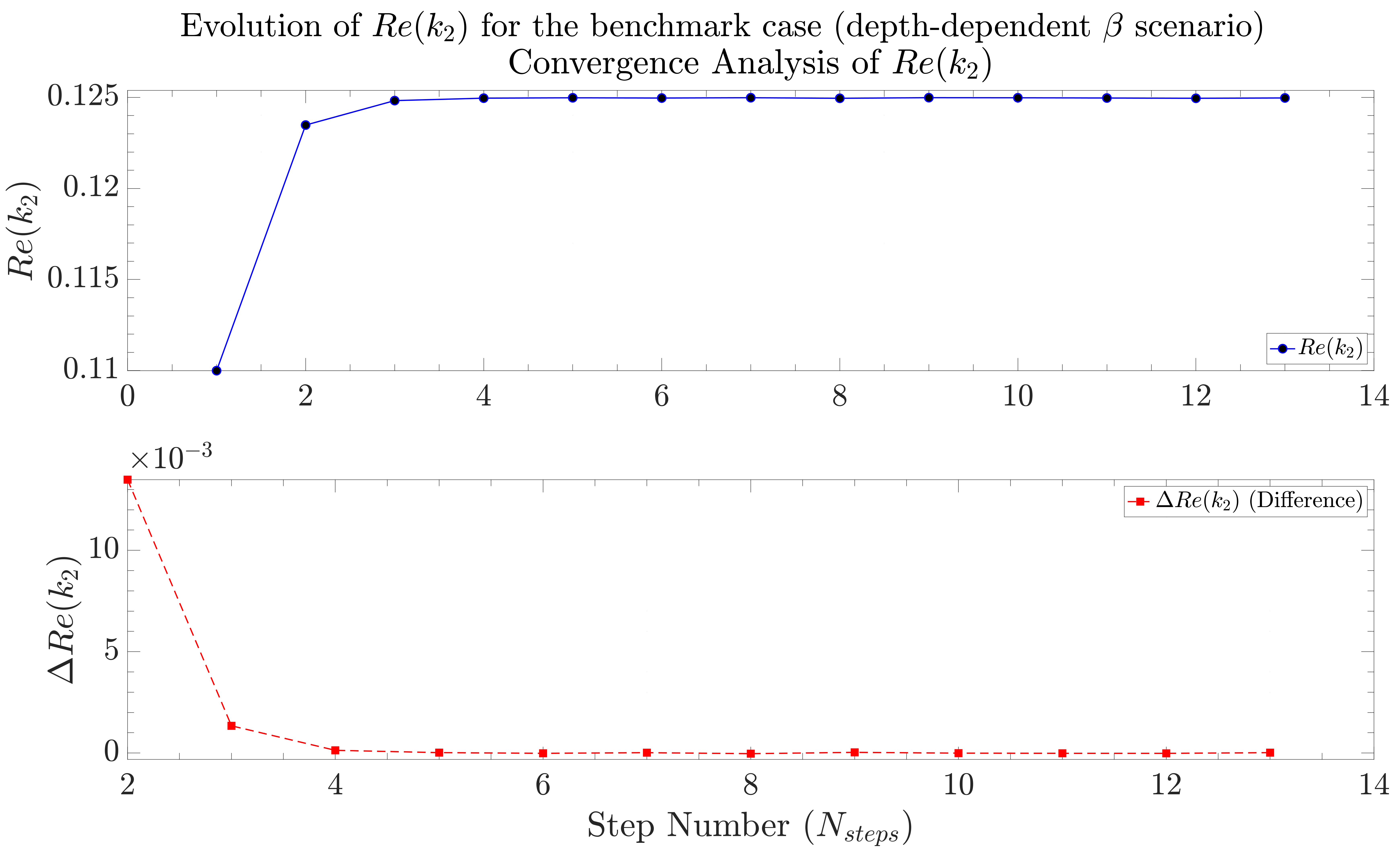}}
\caption{Numerical stability and convergence of the iterative procedure. This analysis considers the benchmark incompressible model ($R_{\phi0} = 1604\,\mathrm{km}$). The plot illustrates the variation of the real part $\Re(k_2)$ as a function of the iteration step. The value stabilizes asymptotically, demonstrating the numerical robustness of the coupled tidal-thermal model implementation.}
\label{figS10}
\end{figure}

\begin{figure}[!ht]
\resizebox{\hsize}{!}
{\includegraphics{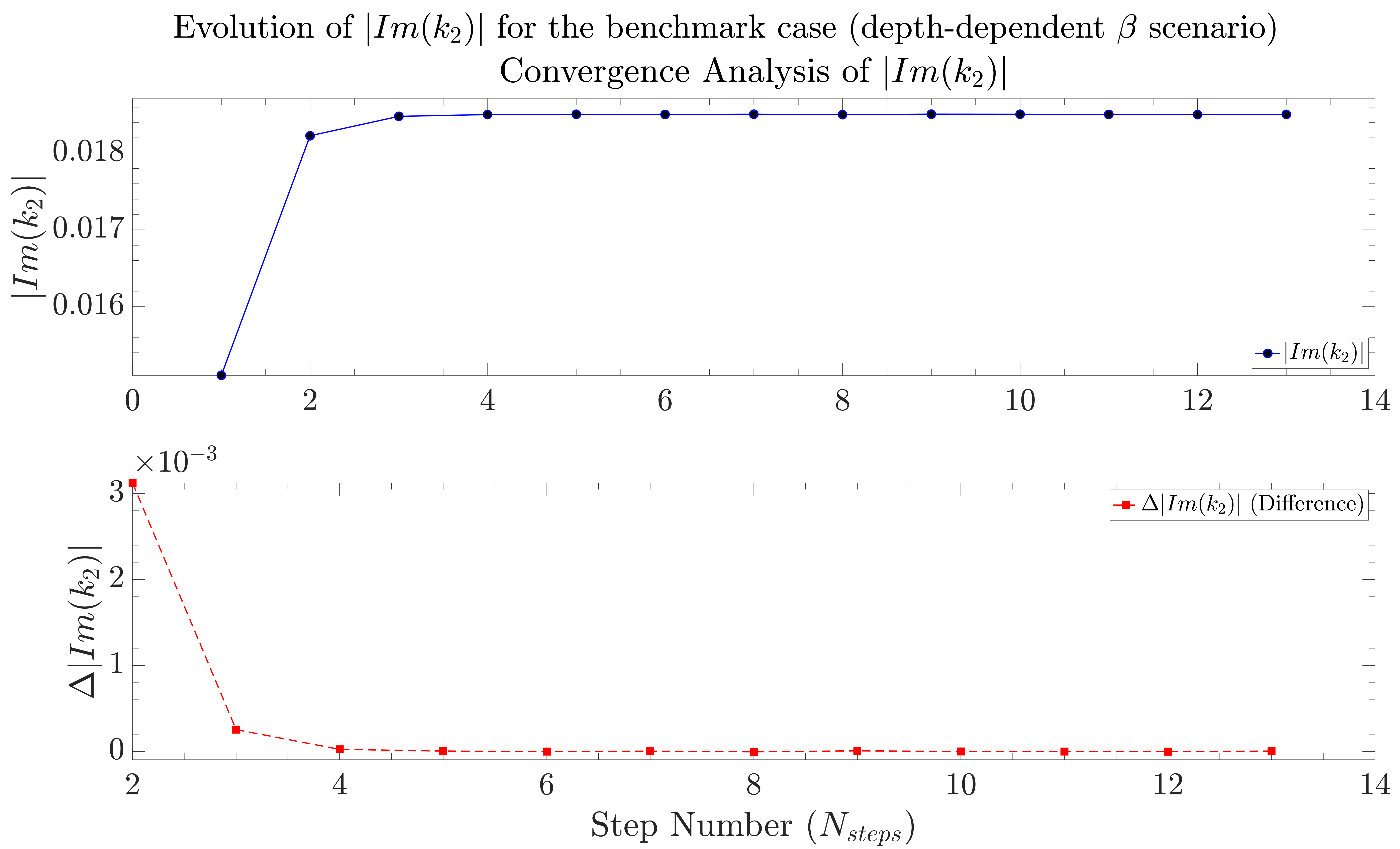}}
\caption{Numerical stability and convergence of the iterative procedure. This analysis considers the benchmark incompressible models ($R_{\phi0} = 1604\,\mathrm{km}$). The plot illustrates the variation of the modulus of the imaginary part  $|\Im(k_2)|$ as a function of the iteration step. The value stabilizes asymptotically, demonstrating the numerical robustness of the coupled tidal-thermal model implementation.}
\label{figS12}
\end{figure}
\FloatBarrier

\section{Incompressible mantle: Sensitivity to initial \texorpdfstring{$\beta_0$}{beta0} (fixed \texorpdfstring{$R_{\phi_0}$}{Rphi0})}\label{sensitivity_study}
In this scenario, we fixed the melting onset radius at $R_{\phi0} = 1604\,\mathrm{km}$, the value that previously reproduced the observed $\Re(k_2)$. We then explored the effect of varying the initial Andrade parameter $\beta_0$ on the tidal response.

\begin{figure}[!ht]
\resizebox{\hsize}{!}
{\includegraphics{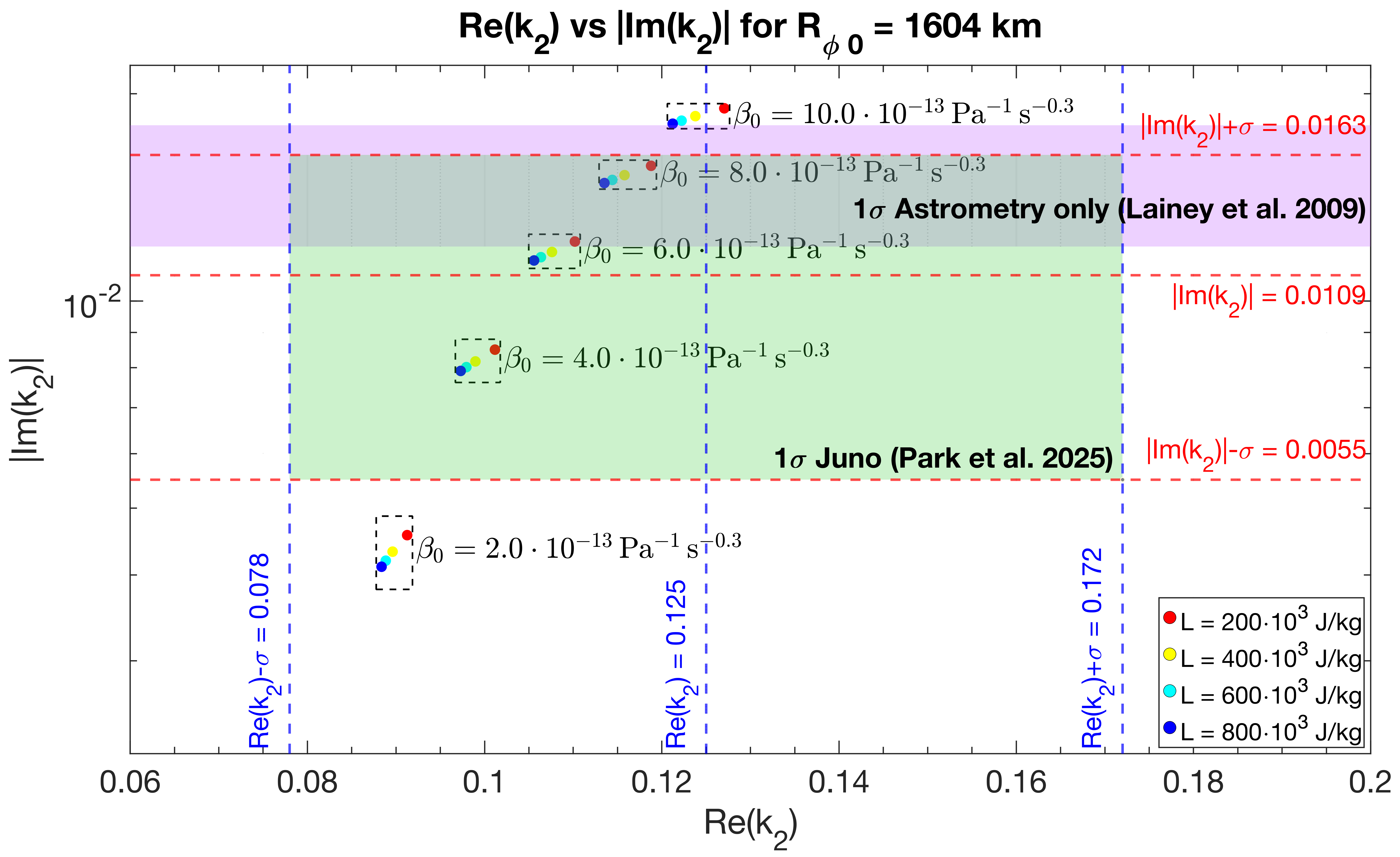}}
\caption{Variation in $\Re(k_2)$ versus $|\Im(k_2)|$. The melting onset radius is fixed at $R_{\phi0} = 1604\,\mathrm{km}$, the value that reproduces the observed $\Re(k_2)$ in the reference model (i.e., for $\beta_0=10^{-12}\,\mathrm{Pa^{-1}s^{-0.3}}$). Models are grouped by the initial $\beta_0$ value and color coded by latent heat ($L$). The green shaded region represents the $1$$\sigma$ confidence interval of the Juno measurements~\citep{park_ios_2025}, while the purple region corresponds to the $|\Im(k_2)|$ range estimated by~\citet{2009Natur.459..957L}.}
\label{fig11}
\end{figure}

We analyzed the sensitivity of $k_2$ to $\beta_0$ (varied over $10^{-13}$ -- $10^{-12}\,\mathrm{Pa^{-1}s^{-0.3}}$) and to the latent heat of fusion $L$ (varied over $2\times10^5$ -- $8\times10^{5}\,\mathrm{J\,kg^{-1}}$).
Figure~\ref{fig11} compares the variation of $\Re(k_2)$ and $|\Im(k_2)|$ with $L$ and with the initial value of $\beta_0$, together with the $1$$\sigma$ confidence interval of the Juno measurements~\citep{park_ios_2025} and the $|\Im(k_2)|$ range estimated by~\citet{2009Natur.459..957L}. The plot thus allows for the identification of models and parameter spaces ($L$, $\beta_0$) that are compatible with both observational estimates.

Table \ref{tab:combined_models_2} summarizes the parameter values for models (from Fig.~\ref{fig11}) that fall within the $1$$\sigma$ observational constraints. The parameters listed are the latent heat of fusion ($L$) and the initial Andrade parameter ($\beta_0$), because $R_{\phi0}$ was fixed at $1604\,\mathrm{km}$ for this entire scenario.

\begin{table}[!ht]
\caption{Model parameters ($L$, $\beta_0$) (fixed $R_{\phi0} = 1604\,\mathrm{km}$).}
\centering
\begin{tabular}{c c l}
\hline\hline
$\beta_0$ ($10^{-13}\,\mathrm{Pa^{-1}s^{-0.3}}$) & $L$ ($10^{5}\,\mathrm{J\,kg^{-1}}$) & Satisfied Constraints \\
\hline
4 & 2--8 & Juno \\
\hline
\multirow{2}{*}{6} & 2 & Juno + Lainey \\
 & 3--8 & Juno \\
\hline
8 & 2--8 & Juno + Lainey \\
\hline
\end{tabular}
\tablefoot{The last column indicates satisfied observational constraints: "Juno" refers to the $1$$\sigma$ interval for $\Re(k_2)$ and $|\Im(k_2)|$ \protect\citep{park_ios_2025}; "Lainey" refers to the $|\Im(k_2)|$ range by \protect\citet{2009Natur.459..957L}.}
\label{tab:combined_models_2}
\end{table}

Within the framework of the adopted Andrade rheology, all models that fall within the $1$$\sigma$ uncertainty of the Juno measurement exhibit a melt fraction below the critical value (RCMF). This finding reinforces the conclusion that, conditional on this specific viscoelastic parameterization, configurations consistent with the $k_2$ value reported by~\citet{park_ios_2025} preclude the presence of a uniform magma layer.
\FloatBarrier

\section{Parametric distributions of companion Love numbers and libration amplitude}
\label{app:companion_parameters}

\begin{figure}[!ht]
\resizebox{\hsize}{!}
    {\includegraphics{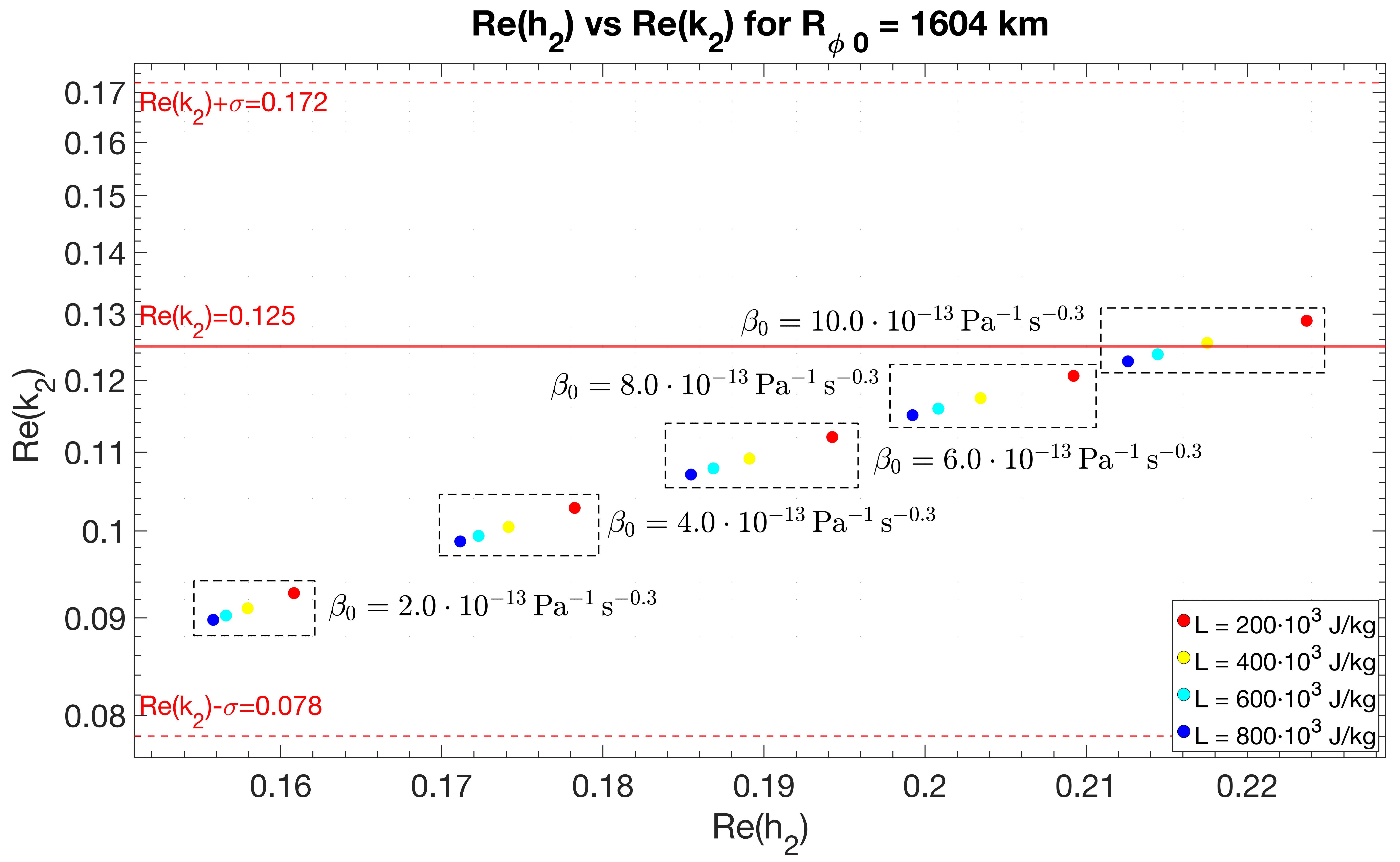}}
    \resizebox{\hsize}{!}
    { \includegraphics{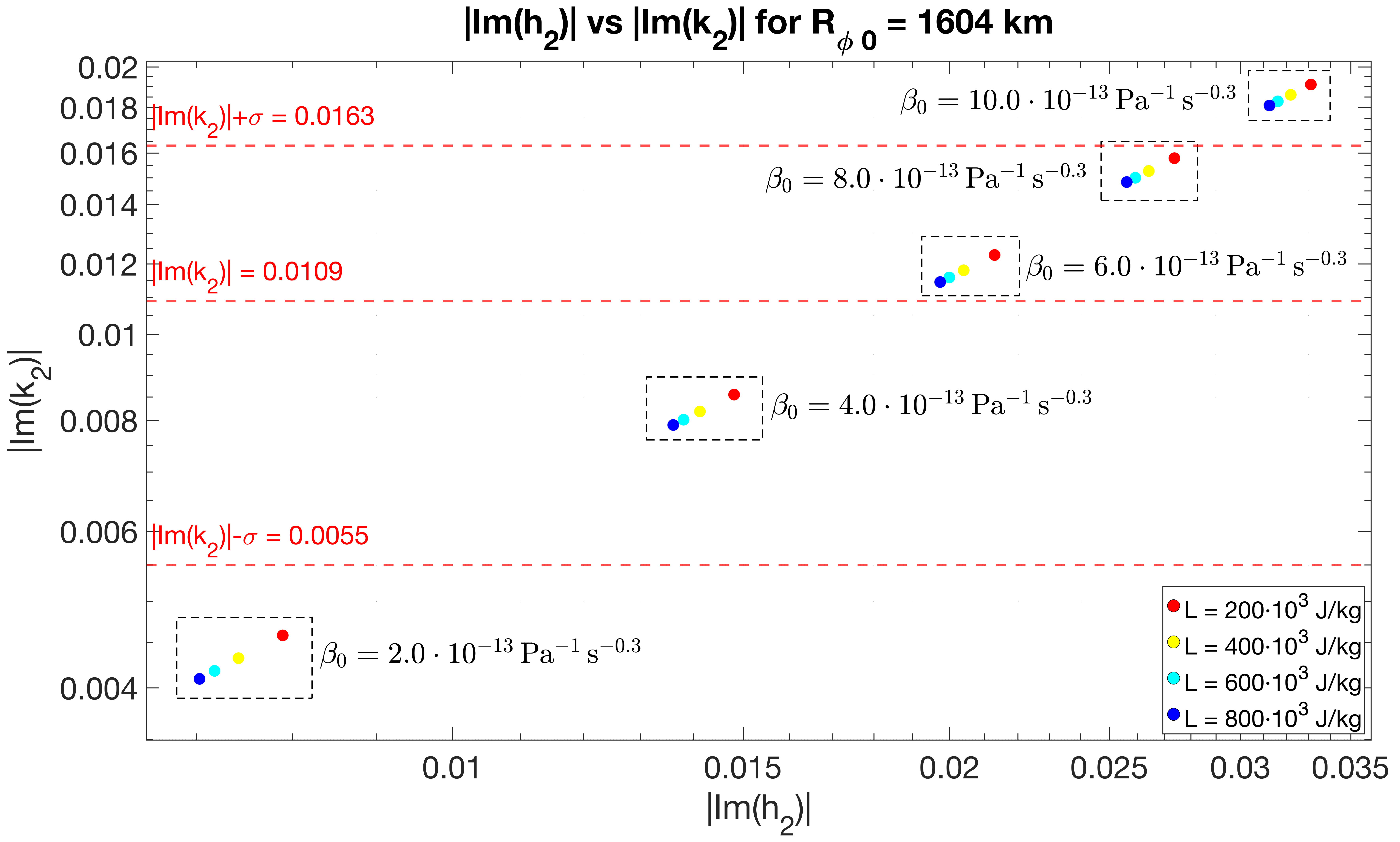}}
\caption{Tidal Love number $h_2$ versus $k_2$ for the compressible scenario, with $R_{\phi0}$ fixed at $1604\,\mathrm{km}$. \textit{Top}: Real components: $\Re(h_2)$ vs. $\Re(k_2)$. \textit{Bottom}: Imaginary components: $|\Im(h_2)|$ vs. $|\Im(k_2)|$. Models are grouped by initial $\beta_0$ and color coded by latent heat ($L$). The horizontal lines indicate the $1$$\sigma$ confidence interval for the Juno $k_2$ measurement~\citep{park_ios_2025}, enabling identification of the corresponding predicted $h_2$ values.}
    \label{fig_h2_2_app}
\end{figure}

\begin{figure}[!ht]
\resizebox{\hsize}{!}
    {\includegraphics{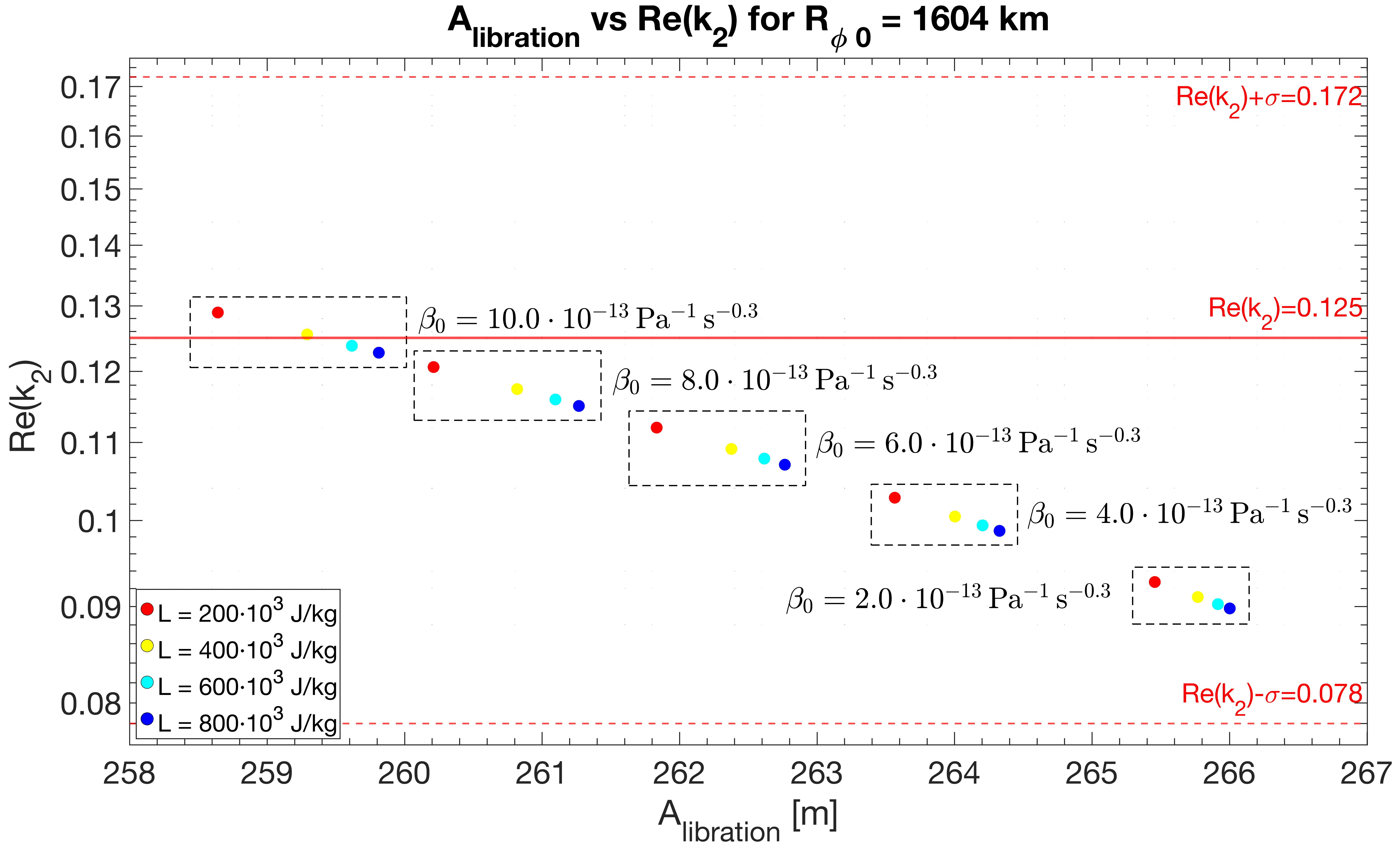}}
    \resizebox{\hsize}{!}
    { \includegraphics{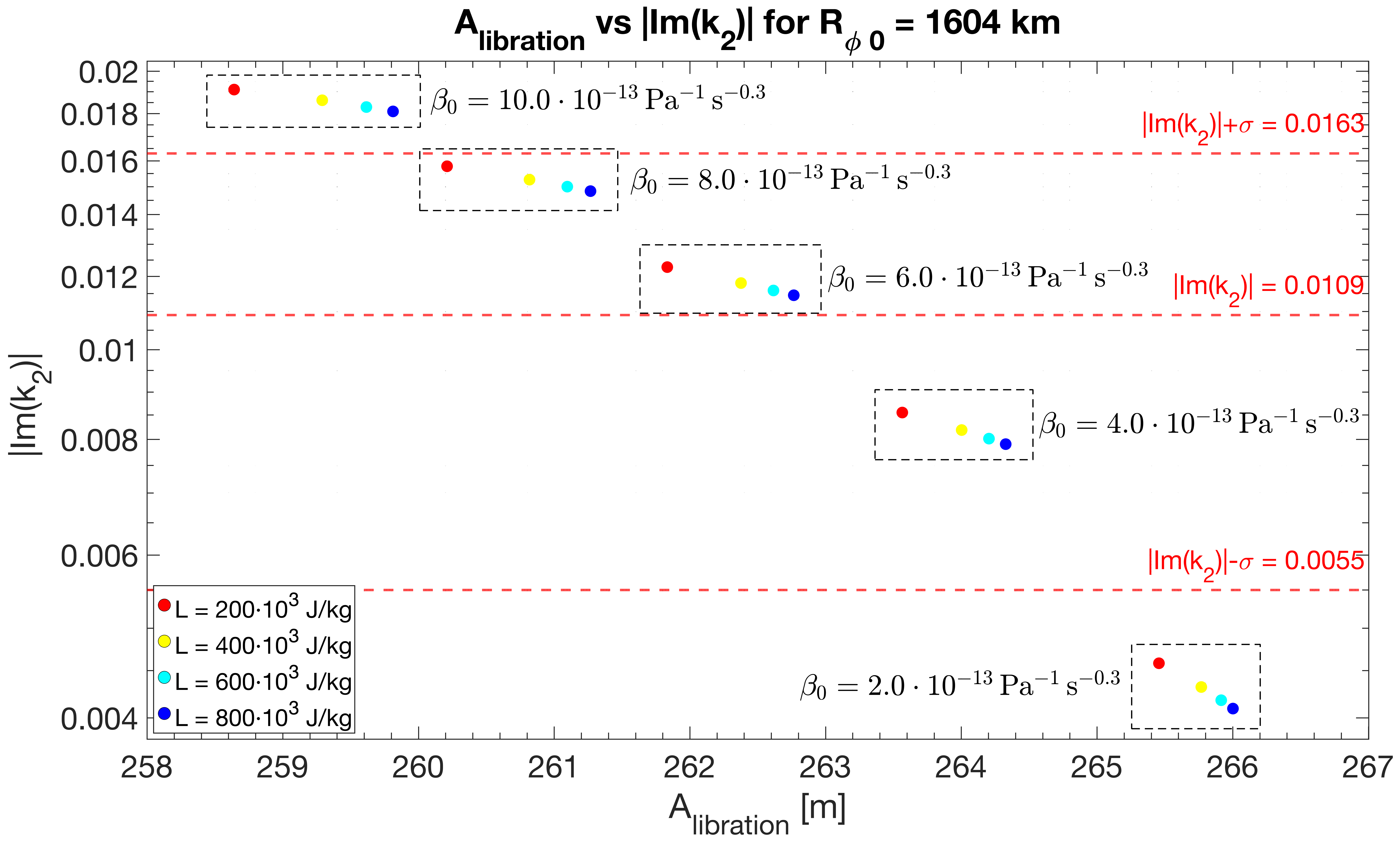}}
    \caption{Libration amplitude versus $k_2$ for the compressible scenario, with $R_{\phi0}$ fixed at $1604\,\mathrm{km}$. \textit{Top}: Libration amplitude vs. $\Re(k_2)$. \textit{Bottom}: Libration amplitude vs. $|\Im(k_2)|$. Models are grouped by initial $\beta_0$ and color coded by latent heat ($L$). The horizontal lines indicate the $1$$\sigma$ confidence interval for the Juno $k_2$ measurement~\citep{park_ios_2025}, enabling identification of the corresponding predicted libration amplitudes.}
    \label{fig_A_2_app}
\end{figure}

\begin{figure}[!ht]
\resizebox{\hsize}{!}
    {\includegraphics{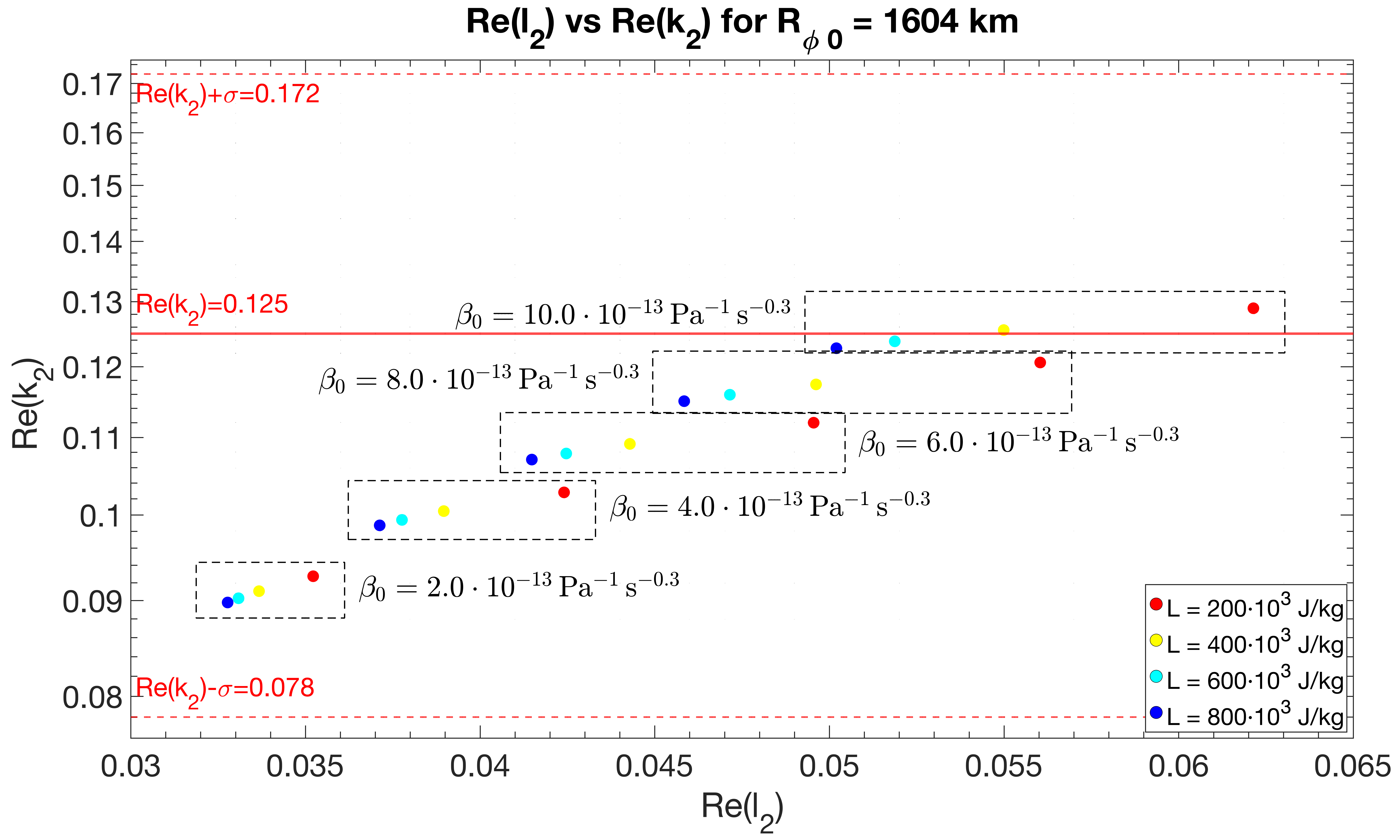}}
    \resizebox{\hsize}{!}
    { \includegraphics{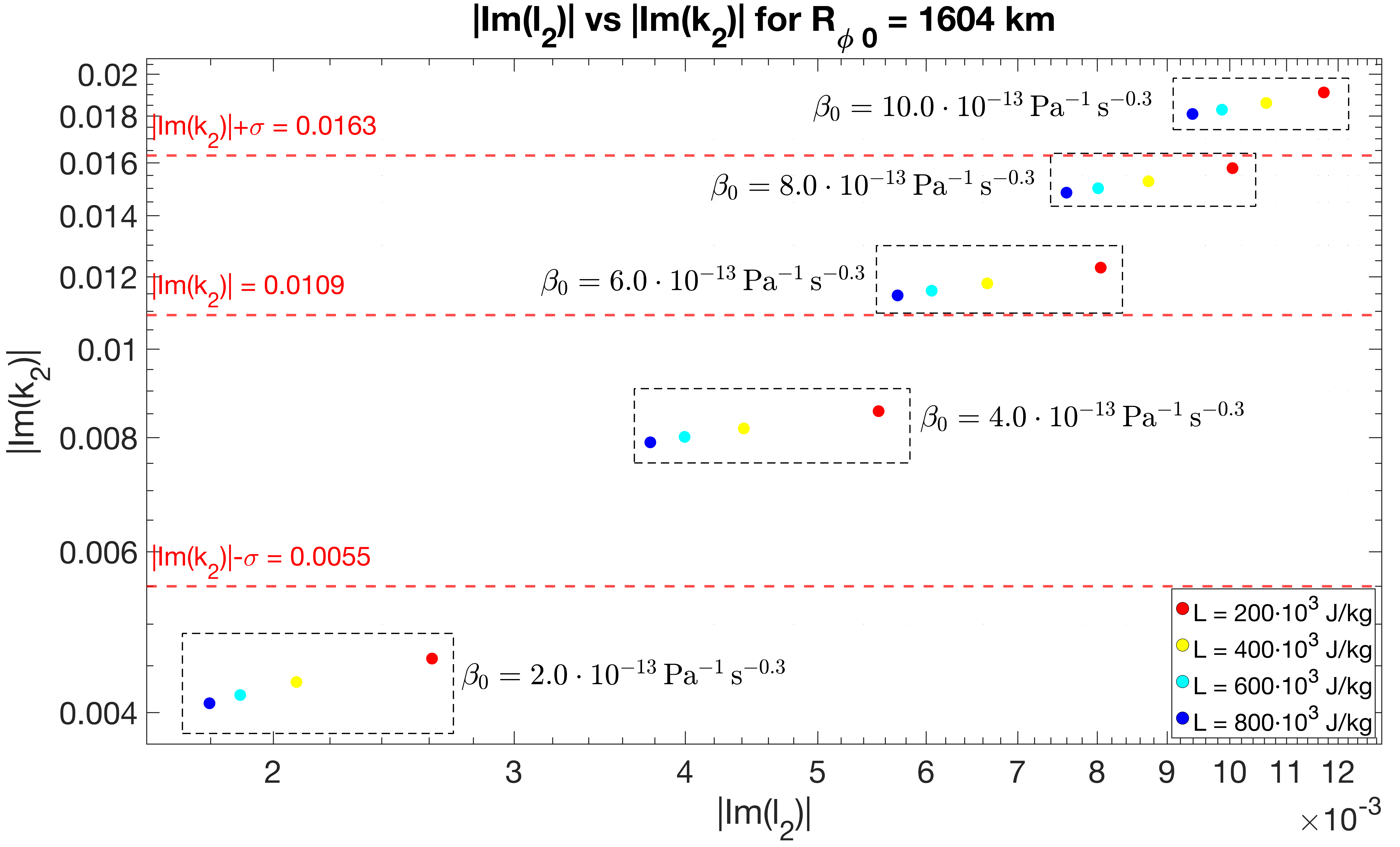}}
    \caption{Tidal Love number $l_2$ versus $k_2$ for the compressible scenario, with $R_{\phi0}$ fixed at $1604\,\mathrm{km}$. \textit{Top}: Real components: $\Re(l_2)$ vs. $\Re(k_2)$. \textit{Bottom}: Imaginary components: $|\Im(l_2)|$ vs. $|\Im(k_2)|$. Models are grouped by initial $\beta_0$ and color coded by latent heat ($L$). The horizontal lines indicate the $1$$\sigma$ confidence interval for the Juno $k_2$ measurement~\citep{park_ios_2025}, enabling identification of the corresponding predicted $l_2$ values.}
    \label{fig_l2_2_app}
\end{figure}

This appendix provides the extended graphical representation of the companion Love numbers ($h_2$, $l_2$) and the libration amplitude. Figures~\ref{fig_h2_2_app}, \ref{fig_A_2_app}, and \ref{fig_l2_2_app} illustrate the behavior of these parameters evaluated as a function of the complex degree-2 Love number $k_2$. 

The results are presented for a compressible scenario, with the melting onset boundary fixed at $R_{\phi0} = 1604\,\mathrm{km}$. Consistent with the structural representation adopted in the main text, the modeled configurations are grouped according to $\beta_0$ and to the applied latent heat of melting ($L$). This parametric mapping visually demonstrates the clustering of the theoretical models and their intersection with the $1$$\sigma$ Juno confidence interval for $k_2$~\citep{park_ios_2025}, which defines the bounded ranges summarized in Table~\ref{tab:predictions}.

\FloatBarrier

\section{Spatially uniform profile (constant-\texorpdfstring{$\beta$}{beta} case)}
To isolate the effect of structural weakening from intrinsic variations in transient anelasticity, we additionally investigated a constant-$\beta$ reference configuration. In this formulation, the Andrade parameter was kept globally fixed at $\beta = \beta_0$, while viscosity and shear modulus were iteratively updated according to Eq. \ref{eq1}. While a spatially constant $\beta$ represents a simplified effective mantle rheology rather than a fully realistic, structure-dependent model, it provides a baseline for comparison. The physical parameters (Table \ref{tab:parameters}), the initial interior structure (Sect.~\ref{internal_structure}), and the parametric ranges for $R_{\phi0}$ and $L$ (Sect.~\ref{melt}) were strictly consistent with the primary coupled analysis.

This investigation focused on three key parameters that influence the Love number: the radial position at which mantle melting begins ($R_{\phi0}$), the latent heat of fusion ($L$), and the initial Andrade parameter ($\beta_0$).
The initial $\beta_0$ value was fixed at $10^{-12}\,\mathrm{Pa^{-1}s^{-0.3}}$. This choice is consistent with~\citep{park_ios_2025}.

With this fixed initial $\beta_0$ value, we investigated the variation of $k_2$ as a function of the melting onset radius ($R_{\phi0}$) and the latent heat of fusion ($L$). We then examined the corresponding radial profiles of the volumetric tidal heating rate ($Q_T(r)$) and the melt fraction ($\phi(r)$), focusing on models found to be consistent with the Juno measurements. Furthermore, we treated the mantle and its sub-layers as incompressible~\citep{1988Icar...75..187S, 2013E&PSL.361..272H, Bierson, 2024GeoRL..5107869A, park_ios_2025, 2025NatCo..16.6798V}.

\begin{figure}[!ht]
\resizebox{\hsize}{!}{\includegraphics{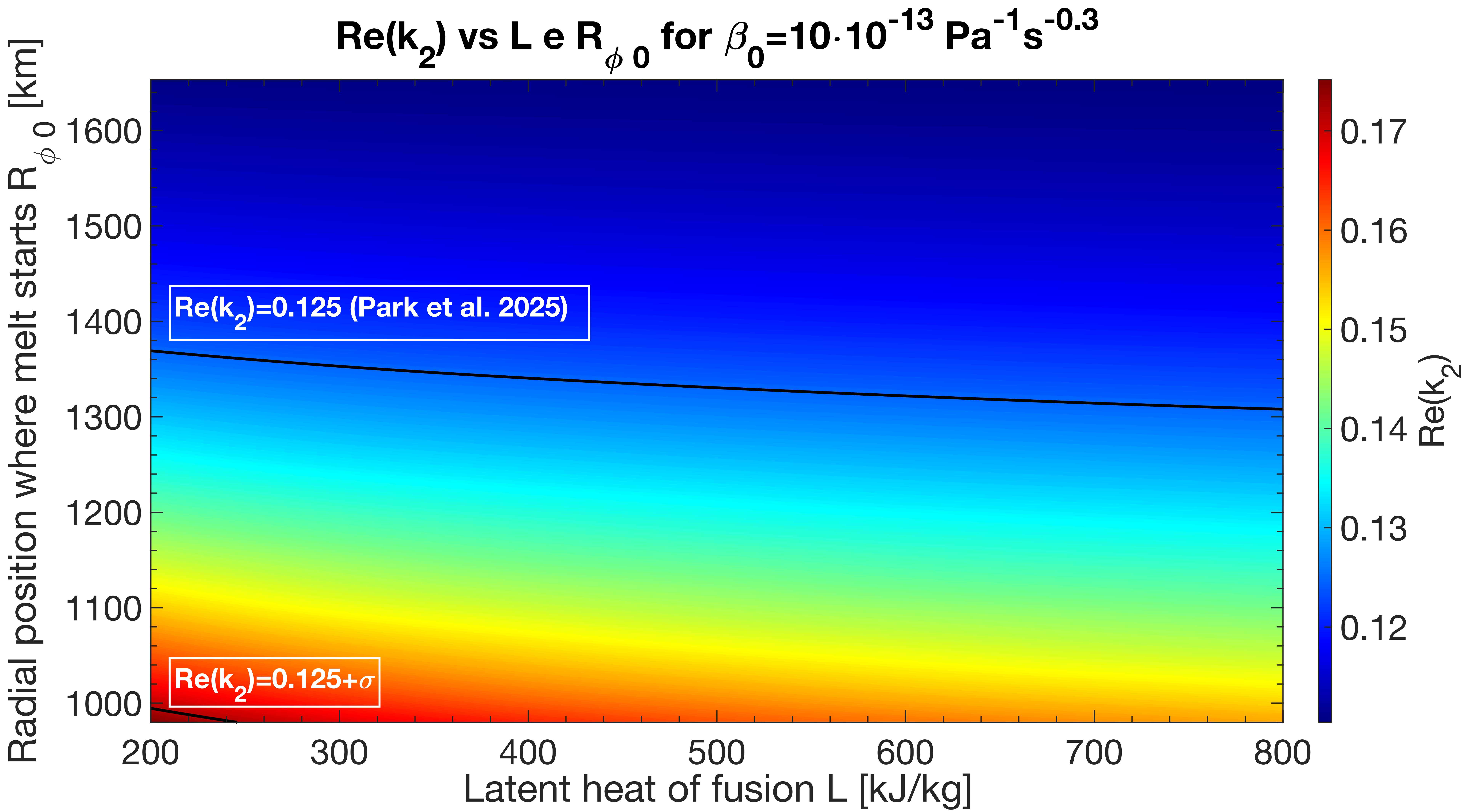}}
    \hspace{0.6 mm}
    \resizebox{\hsize}{!}
{\includegraphics{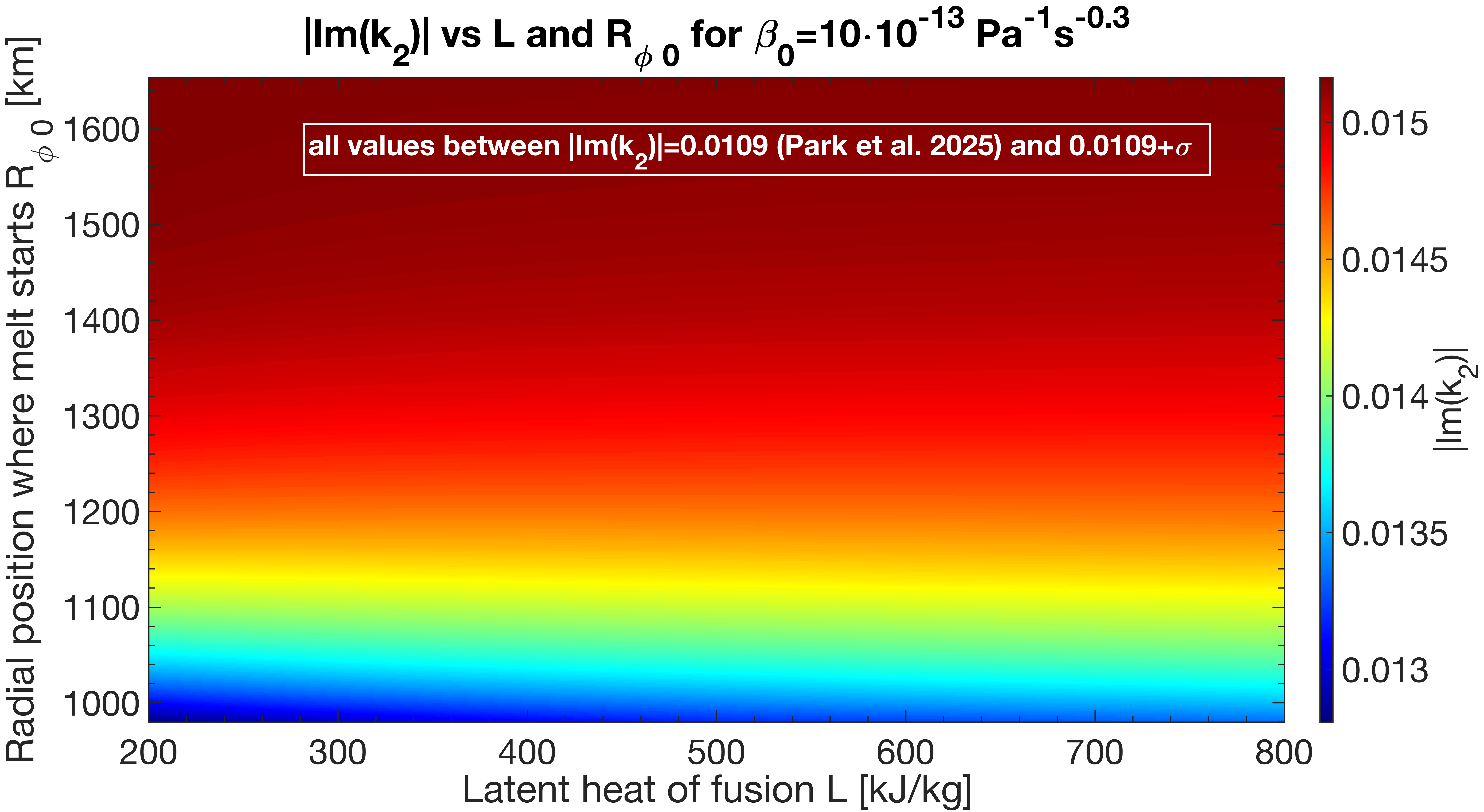}}
    \caption{Love number $k_2$ as a function of melting radius ($R_{\phi 0}$) and latent heat of fusion ($L$), assuming constant $\beta$. In the top panel, the estimated value of $\Re(k_2)$ from~\citet{park_ios_2025} ($0.125 \pm 0.047$) is shown as a reference curve. The bottom panel omits the reference for $|\Im(k_2)|$, as all computed values exceed $0.0109$ but remain within $1$$\sigma$, i.e., below $0.0163$.
}
    \label{fig1}
\end{figure}

Figure~\ref{fig1} shows the behavior of $k_2$ as a function of the melting radius ($R_{\phi0}$) and the latent heat of fusion ($L$) for a constant $\beta$ ($\beta_0=10^{-12}\,\mathrm{Pa^{-1}s^{-0.3}}$). Several combinations reproduce the real part of $k_2$  ($\Re(k_2 )  = 0.125 \pm 0.047$,~\citealt{park_ios_2025}), while the modulus of the imaginary part slightly exceeds the corresponding estimate ($|\Im(k_2 |= 0.0109 \pm 0.0054$) but remains within one standard deviation.
Multiple combinations of $R_{\phi0}$ and $L$ yield a real part of the Love number ($\Re(k_2)$) within one sigma of the value measured by Juno (Fig.~\ref{fig1}). However, a systematic shift is observed for the imaginary part ($|\Im(k_2)|$), which consistently produces values higher than those estimated by~\citet{park_ios_2025} and closer to the estimate by~\citet{2009Natur.459..957L}.

\begin{figure}[!ht]
\resizebox{\hsize}{!}
{\includegraphics{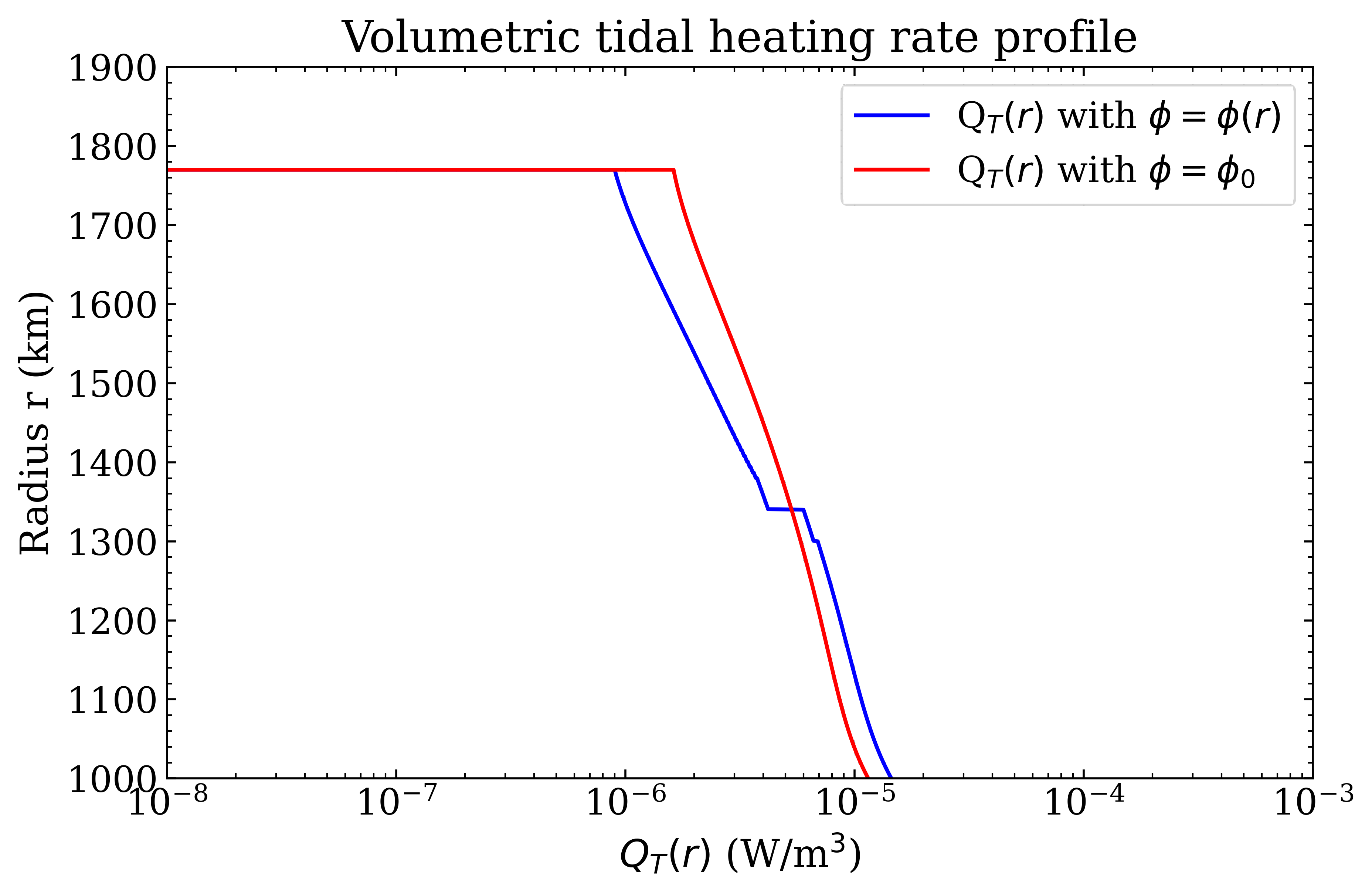}}
\caption{Radial profile of the volumetric tidal heating rate, $Q_T(r)$, for the benchmark case ($R_{\phi0} = 1340\,\mathrm{km}$, $L = 4\times10^{5}\,\mathrm{J\,kg^{-1}}$) that matches the observed $\Re(k_2)$ $\approx 0.125$~\citep{park_ios_2025}. The red curve shows the initial (first iteration) profile, while the blue curve represents the final, converged profile. This model assumes the constant $\beta$ scenario, which results in tidal heating being concentrated primarily in the deep mantle.}
\label{fig3}
\end{figure}

\begin{figure}[!ht]
\resizebox{\hsize}{!}{
\includegraphics{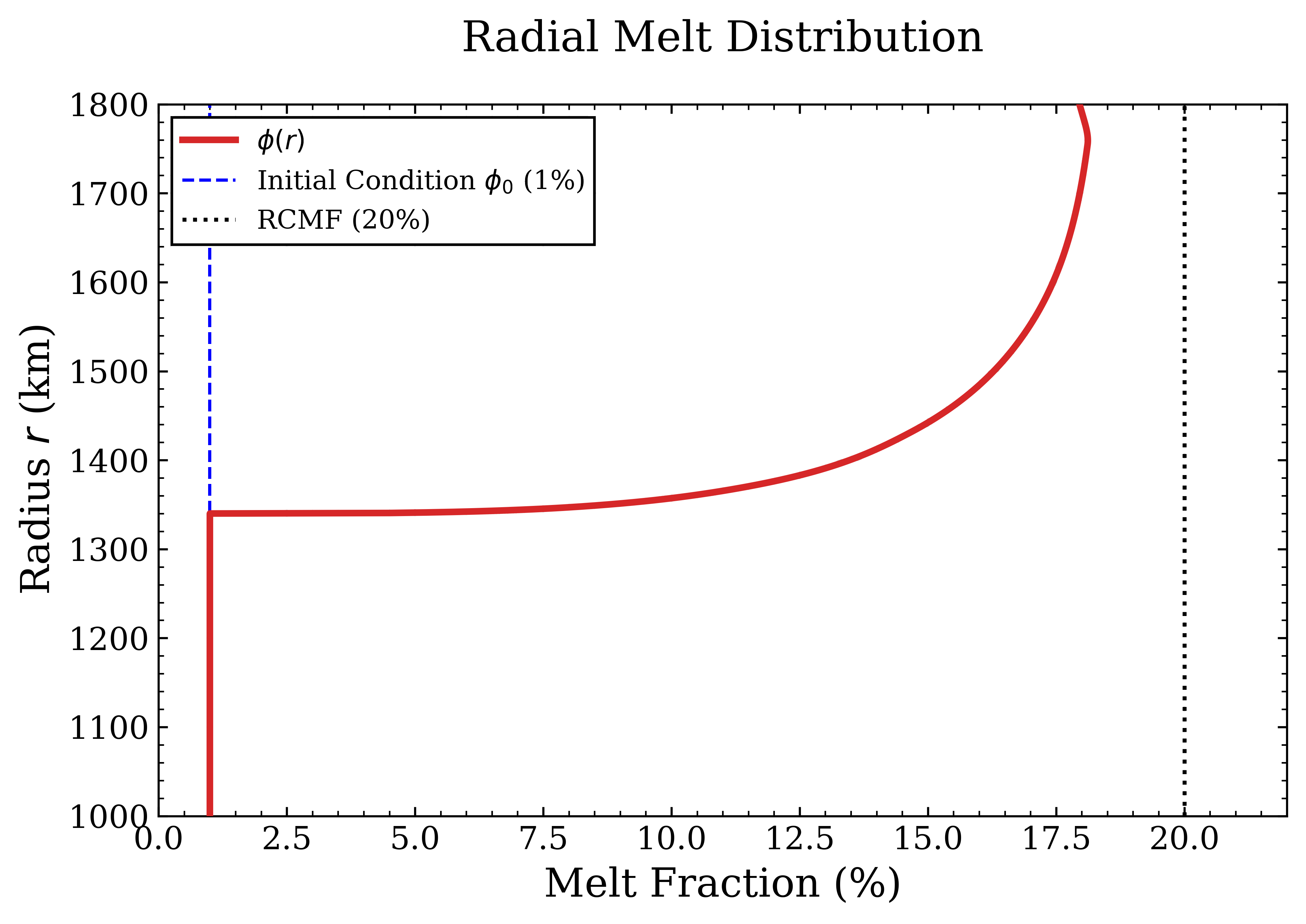}}
\caption{Final radial profile of the melt fraction, $\phi(r)$, for the same benchmark model shown in Fig.~\ref{fig3} ($R_{\phi0} = 1340\,\mathrm{km}$, $L = 4\times10^{5}\,\mathrm{J\,kg^{-1}}$, and $\beta_0=10^{-12}\,\mathrm{Pa^{-1}s^{-0.3}}$). The melt fraction remains below the RCMF associated with the formation of a laterally uniform magma layer~\citep{Miyazaki_2022}. An initial melt fraction of $1\%$ was assumed; lower initial values did not significantly affect the results but increased computational time, justifying the choice of $1\%$.}
\label{fig4}
\end{figure}

As a specific example, the observed value of $\Re(k_2)$ $\approx 0.125$ is recovered for a model with $R_{\phi0} = 1340\,\mathrm{km}$ and $L = 4\times10^{5}\,\mathrm{J\,kg^{-1}}$. For this benchmark case, the corresponding radial profiles of the volumetric tidal heating rate ($Q_T(r)$) and the melt fraction ($\phi(r)$) are shown in Figs.~\ref{fig3} and~\ref{fig4}, respectively.

As this is a parametric study, the $R_{\phi0}$ and $L$ values are treated as exact model inputs and are therefore not reported with associated uncertainties. Consequently, when $\Re(k_2)$ is stated to be 0.125 for a given ($R_{\phi0}$, $L$) pair, this signifies the deterministic value produced by the iterative simulation for those specific inputs.

\begin{figure}[!ht]
\resizebox{\hsize}{!}
{\includegraphics{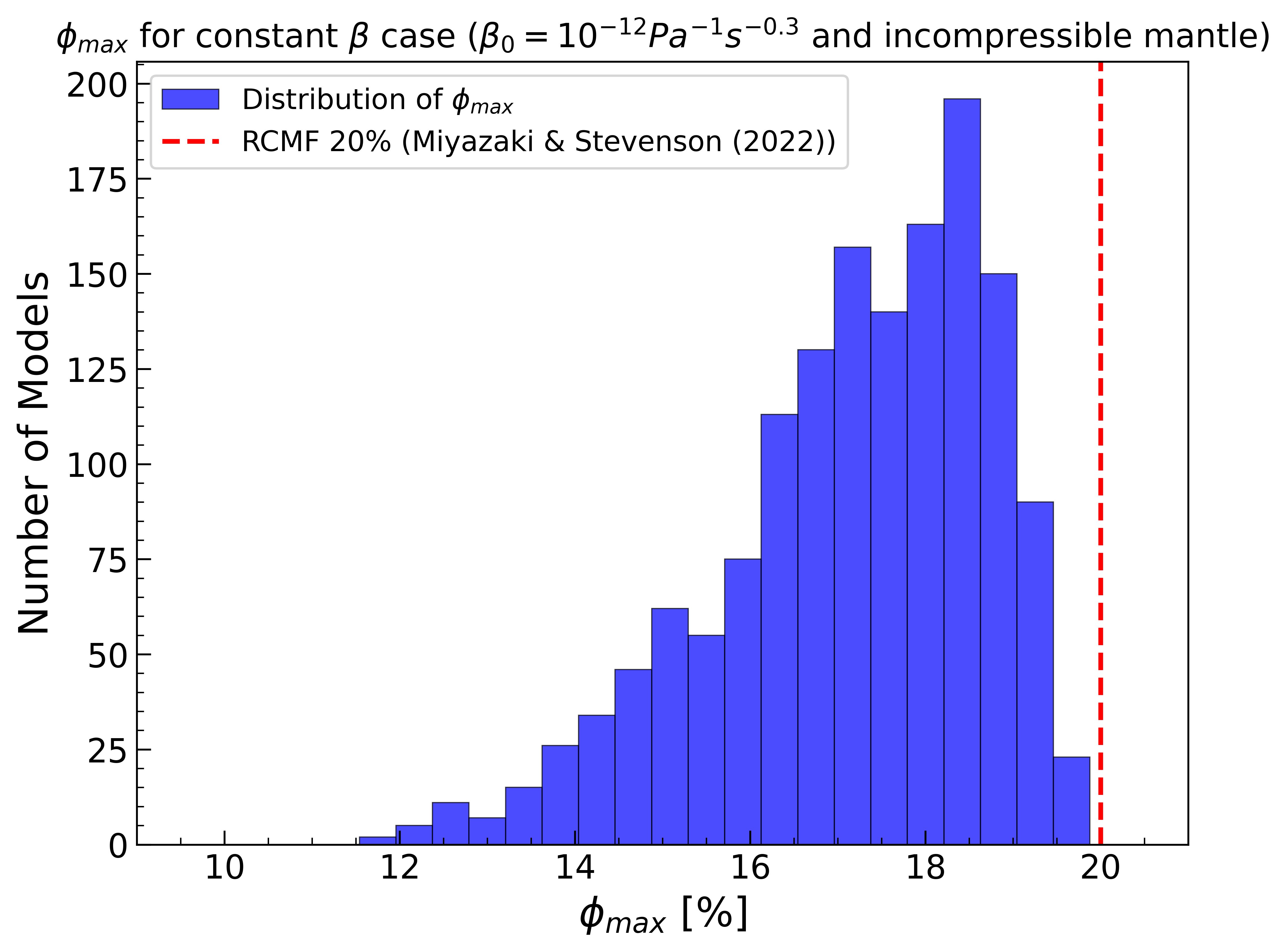}}
\caption{Distribution of peak melt fractions $\phi_{max}$ for models consistent with Juno observations. This analysis focuses on the constant $\beta$ case. The figure displays the maximum melt fraction obtained for all model configurations that reproduce the real part of the Love number, $\Re(k_2)$, within the $1$$\sigma$ confidence interval estimated by~\citet{park_ios_2025}. The vertical red dashed line indicates the RCMF of $20\%$~\citep{Miyazaki_2022}.}
\label{figS13}
\end{figure}

As shown in Figs.~\ref{fig3} and \ref{fig4}, tidal heating is concentrated primarily in the deep mantle. The benchmark model that reproduces the $\Re(k_2)$ value estimated by~\citet{park_ios_2025} exhibits a peak melt fraction of approximately $18.5\%$. This value lies below both the RCMF (\citealt{Hamilton}) and the $\sim20\%$ threshold at which a partially molten "magmatic sponge" structure is expected to become unstable~\citep{Miyazaki_2022}.
This behavior is common to all models reproducing $\Re(k_2)$ values within the $1$$\sigma$ Juno confidence interval, as such models maintain melt fractions below this critical threshold. Therefore, within this rheological parameterization, only models with melt fractions below the RCMF successfully reproduce the Juno constraint (see Fig.~\ref{figS13}).

In addition, for this specific parametric configuration, an a posteriori evaluation of the mass-flux framework yields a column-integrated thermodynamic melt production rate of $\dot{M}^{\text{max}}_{\text{gen}} \approx 1.32 \times 10^8 \, \text{kg s}^{-1}$ and a maximum melt percolation capacity of $\dot{M}^{\text{max}}_{\text{migr}} \approx 4.07 \times 10^8 \, \text{kg s}^{-1}$. The comparison reveals that $\dot{M}^{\text{max}}_{\text{migr}} > \dot{M}^{\text{max}}_{\text{gen}}$ by a factor of approximately $3.08$.

This inequality confirms that the transport-efficient, drainage-controlled regime identified for the benchmark case persists under these alternative parametric assumptions, verifying the internal fluid-dynamics plausibility of the model. The calculated steady-state production rate ($1.32 \times 10^8\,\mathrm{kg\,s^{-1}}$) remains fully compatible with the empirical planetary limits, falling near the upper bound of the global mass discharge range scaled from JIRAM observations. For a detailed description of the algebraic scaling methodology and the corresponding comparative volcanic frameworks, the reader is referred to Sect.~\ref{comparison}.

Ultimately, we conducted this constant-$\beta$ analysis primarily to establish a comparative baseline. The results obtained in this simplified configuration are largely analogous to those presented in the main body of the article. However, because the dynamically coupled framework more accurately captures the self-consistent thermomechanical feedback within Io's mantle, we derive our primary physical interpretations and geological implications exclusively from the main case.

\end{appendix}
\end{document}